\documentclass[12pt,a4paper]{article}
\usepackage{graphicx}
\usepackage{amsmath,amssymb,mathrsfs, url,cite,ifpdf,slashed,multirow,tabularx,type1cm}

\usepackage{subfigure}

\usepackage{caption}
\usepackage[colorlinks=true,linkcolor=blue,citecolor=blue,urlcolor=magenta]{hyperref} 

\allowdisplaybreaks

\newcommand{\TeV}{\,{\rm TeV}}
\newcommand{\GeV}{\,{\rm GeV}}

\newcommand{\Slash}[1]{{\ooalign{\hfil \hspace*{-5pt}~#1\hfil\crcr\raise.167ex\hbox{/}}}}

\def\be{\begin{equation}}
\def\ee{\end{equation}}
\def\beq{\begin{eqnarray}}
\def\eeq{\end{eqnarray}}

\def\({\left(}
\def\){\right)}
\def\<{\langle}
\def\>{\rangle}
\newcommand{\non}{\nonumber \\ }
\newcommand{\matl}{\left( \begin{array}}
\newcommand{\matr}{\end{array} \right)}

\newcommand{\eq}[1]{Eq.~(\ref{#1})}

\usepackage{subfigure, url, ulem}

\def\thefootnote{\ifnum\c@footnote>\z@\textasteriskcentered\@arabic\c@footnote\fi}
\makeatletter
\renewcommand{\footnoterule}{%
\kern-3\p@
\hrule width 0.4\columnwidth
\kern 2.6\p@}
\def\thefootnote{\ifnum\c@footnote>\z@\@arabic\c@footnote\fi}
\makeatother
\makeatletter
\newcommand{\@authornote}[2]{{\def\thefootnote{\fnsymbol{footnote}}\setcounter{footnote}{#1}#2\setcounter{footnote}{0}}}
\newcommand{\authornotemark}[1]{\@authornote#1{\addtocounter{footnote}{-1}\footnotemark}}
\newcommand{\authornotetext}[2]{\@authornote#1{\footnotetext{#2}}}
\makeatother

\begin{document}

\begin{titlepage}
\begin{flushright}
\hfill KEK--TH--1843\\
\hfill IPMU--15--0101\\
\hfill July, 2015\\
\end{flushright}

\vskip 1.5 cm
\begin{center}

{\Large \bf 
Prospects for Spin-1 Resonance Search  \\ \vspace{0.4cm}
at 13 TeV  LHC and  the ATLAS Diboson Excess
}

\vskip .55in
{\large
\textbf{Tomohiro Abe}$^{\rm (a)}$\footnote[0]{${}${\it E-mail:} \textcolor{magenta}{abetomo@post.kek.jp}},
\textbf{Teppei Kitahara}$^{\rm (a)}$\footnote[0]{${}${\it E-mail:} \textcolor{magenta}{kteppei@post.kek.jp}},
and
\textbf{Mihoko M. Nojiri}$^{\rm (a,b,c)}$\footnote[0]{${}${\it E-mail:} \textcolor{magenta}{nojiri@post.kek.jp}}
}
\vskip 0.4in
$^{\rm (a)}${\it KEK Theory Center, IPNS, KEK, Tsukuba, Ibaraki 305-0801, Japan}
\vskip 0.1in
$^{\rm (b)}${\it The Graduate University of Advanced Studies (Sokendai),\\
Tsukuba, Ibaraki 305-0801, Japan}
\vskip 0.1in
$^{\rm (c)}${\it Kavli IPMU (WPI), University of Tokyo, Kashiwa, Chiba 277--8583, Japan}
\vskip 0.1in

\end{center}
\vskip .25in

\begin{abstract}
Motivated by ATLAS diboson excess around 2 TeV, we investigate a
 phenomenology of spin-1 resonances in a model where electroweak sector
 in the SM is weakly coupled to  strong dynamics.
The spin-1 resonances,  $W'$ and $Z'$, are introduced as 
effective degrees of freedom of the dynamical sector. 
We explore  several theoretical constraints  by investigating the scalar
 potential of the model
as well as the current bounds from the LHC and precision
 measurements.
It is found that the main decay modes are $V' \to VV$ and $V' \to Vh$,
 and the $V'$ width 
is  narrow enough so that the ATLAS diboson excess can be explained.
In order to  investigate future prospects, we  also perform collider
 simulations at $\sqrt{s} = 13$ TeV LHC, and obtain a model independent
 expected exclusion limit for $\sigma(pp \to W' \to WZ \to JJ)$.
We find a parameter space where 
the diboson excess  can be explained, and are within a reach of  the LHC  at $\int dt
 \mathcal{L}=10$ fb$^{-1}$ and $\sqrt{s}= 13 $ TeV.
\end{abstract}

\end{titlepage}
\renewcommand{\thefootnote}{\#\arabic{footnote}}
\setcounter{page}{1}
\hrule
\tableofcontents
\vskip .2in
\hrule
\vskip .4in

\section{Introduction}

The ATLAS  experiment recently reported an excess of the events in the search
for the diboson resonance,
in the $pp \to$
($WW$, $WZ$, and $ZZ$)
 $\to JJ$ channels, 
where $J$ is a  fat-jet formed by boosted $W$ or $Z$
boson~\cite{Aad:2015owa}. 
The largest local significance is 3.4~$\sigma$ around 2~TeV  in the $WZ$ channel, and the
global significance is 2.5 $\sigma$. 
The CMS experiment  also studied the same channels. The largest deviation
they found is 1.4 $\sigma $ at $\sim$
1.9~TeV~\cite{Khachatryan:2014hpa}.
Although we cannot conclude that there is a
new particle with the mass around 2~TeV from this data,
it is worthwhile to consider models which can explain this excess, and
many papers have already appeared discussing 
interpretations of the excess~\cite{1506.03751, 
1506.03931, 1506.04392, 1506.05994, 1506.06064, 1506.06736, 1506.06739
,Alves:2015mua, 1506.07511, Thamm:2015csa, Brehmer:2015cia, Cao:2015lia, Cacciapaglia:2015eea, 1507.01185, 1507.01923,
1508.04129}.
As discussed in these references, 
a simple candidate is a spin-1 particle.

New vector resonances often appear in the models with
dynamical symmetry breaking.
Such spin-1 resonances appear in  
the composite Higgs scenario \cite{HEP-PH/0412089, Panico:2011pw, Contino:2011np,
Bellazzini:2012tv} with the dynamics at TeV scale to account for the naturalness
problem.
Since the models are based on the non-Linear sigma models,
the effective theory involves many operators 
whose coefficients are unknown.
They possibly affect to the couplings of the new spin-1 particles
to the standard model (SM) particles, and thus there 
is uncertainty in the prediction of the $W'$ couplings.

Another way to include spin-1 particles is to extend the
electroweak gauge symmetry. 
We can easily introduce new spin-1 particles in renormalizable manner.
In that case, the models are calculable and
we can avoid the operators whose coefficients are unknown,
in contrast to the models based on the non-linear sigma models. 
Besides, some renormalizable models with extended gauge sector can be
regarded as the low energy effective theory of UV theory with some dynamics.

Such renormalizable models have been discussed in the context of
the left-right (LR) symmetric model~\cite{1506.06736,1507.01923,
1508.04129} and
the leptophobic G221 model~\cite{1506.07511}.
These models contain the right-handed SM fermions
which are not singlet under the new gauge symmetry.
In such case, the couplings of the SM fermions to the new gauge boson
are not suppressed, 
and the new gauge bosons mainly decay into the SM fermions.

It is also possible to use linear sigma model, 
instead of non-linear sigma models, 
for models emerged from the dynamics at TeV scale. An
example was proposed in Ref.~\cite{Abe:2013jga}.
This model, called the
 partially composite standard model, has three Higgs fields.
Two of them are regarded  as effective degrees of freedom below the dynamical scale.
The other one is an elementary field.
Spin-1 resonances  are  introduced as new gauge bosons
{\it a la} Hidden local symmetry~\cite{Bando:1987ym, Bando:1984ej,
Bando:1984pw, Bando:1985rf}. 
A feature of the model is that the SM fermion are singlet under the new
gauge symmetry, and thus
all the fermion couplings to the new gauge bosons are suppressed by
the mixing angle in the gauge 
sector. As a result, the new gauge bosons 
mainly decay into the SM gauge bosons. 
This is an important feature of this model.

In this paper,
we investigate the possibility to explain the diboson excess
by the partially composite standard model,
and also the future prospects of 
$W'$ and $Z'$ bosons searches at 
the LHC Run-2, where $\sqrt{s} = 13$~TeV.
We perform a comprehensive study 
to find the parameter space
which has not been excluded 
from current experimental data.
The constraints on the model parameters come from
the LHC searches and 
the electroweak precision measurements.
We also require
theoretical constraints such as
perturbativity condition, bounded below condition, global
minimum vacuum condition, and stability condition of the scalar
potential.  
We find a parameter space where the diboson excess can be explained.

We further investigate a model-independent  sensitivity 
at the LHC Run-2
by generating both 
signal $pp\rightarrow W'\rightarrow WZ$ and dijet background events, and performing  detector simulations.

We organize the rest of this paper as follows.
We briefly review the partially composite standard model in Sec.~\ref{sec:model}.
In Sec.~\ref{sec:Phenomenology of Spin-1 Resonances},  
we show the constraints to the model, and find that there are  parameter
regions  where are consistent with the ATLAS excess.
In Sec.~\ref{sec:LHC}, we perform the collider simulations for the
signal and the background, and obtain the sensitivity at the LHC Run-2.
In Sec.~\ref{sec13TeV}, 
we investigate the future prospects of the spin-1 resonances search
using our simulation results. 
Section~\ref{sec:conclutions} is devoted for conclusion.

\section{The partially composite standard model}
\label{sec:model}

\subsection{The model setup}
\label{subsec:model}

\begin{figure}[t]
\begin{center}
\includegraphics[width =6.5cm, bb=0 0 291 202]{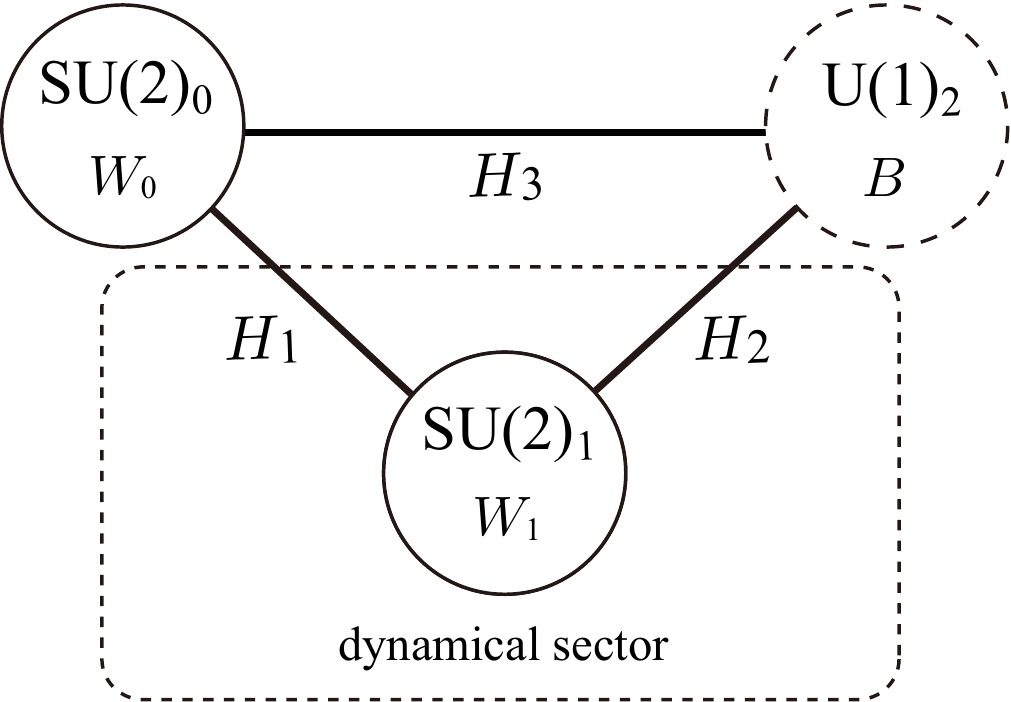}
 \vspace{.2cm}
\caption{The moose diagram of this setup:
the circles represent the gauge groups, and the thick lines that connect two circles are the Higgs fields. 
The Higgs fields $H_1$, $H_2$, and SU(2$)_1$ gauge group
can be regarded as the operators originated from the dynamical sector.
 }  
  \label{moose}
\end{center}
\end{figure}
In the partially composite standard model,
the gauge symmetry of the electroweak sector is 
SU(2)$_0 \times$SU(2)$_1 \times$U(1)$_2$,  
and three Higgs fields ($H_1$, $H_2$, $H_3$) are introduced for the
symmetry breaking,
SU(2)$_0$ $\times$SU(2)$_1 \times$U(1)$_2$  $\to $ U(1)$_{EM}$.
We denote the gauge couplings $g_0$, $g_1$, and $g_2$, respectively. 
The three gauge couplings are related to the 
QED coupling as
\begin{align}
 \frac{1}{e^2}
=&
 \frac{1}{g_0^2} + \frac{1}{g_1^2} + \frac{1}{g_2^2}.
\end{align}
Here, we assume that 
the SU(2)$_1$ gauge symmetry belongs to a dynamical sector,
and  $g_1 \gg g_{0}, g_{2}$. Under this assumption,
$g_0$ and $g_2$ are approximately $g_W$ and $g_Y$ which are the gauge couplings of 
SU(2)$_L$ and U(1)$_Y$, respectively.
We regard the gauge field
associated with SU(2)$_1$ as the vector resonance originated from unknown dynamics  above TeV scale. 
This implies that fields transformed under the SU(2)$_1$ gauge symmetry
also belong to the dynamical sector. We take $H_1$
and $H_2$ as such fields, and regard $H_3$ as an elementary field.
All the fermions are also elementary,
and 
they are singlet under SU(2)$_1$. 
We  schematically show the model structure in
the moose notation~\cite{Georgi:1985hf} in Fig.~\ref{moose}, and 
also summarize the field contents and  their charge assignments in Table~\ref{charge}.

The scalar potential is given as\footnote{
We omit the term $i \kappa' \text{tr}(H_1 H_2 H_3^{\dagger}
\tau^3)$ in this paper because 
this term can be eliminated by the field
redefinition~\cite{Abe:2013jga}.
}
\beq
V(H_1, H_2, H_3) 
&=& 
   \mu^2_1 \textrm{tr} (H_1 H_1^{\dag})  
+  \mu^2_2 \textrm{tr} (H_2H_2^{\dag}) 
+  \mu^2_3 \textrm{tr} (H_3 H_3^{\dag})       \non
& &+ \kappa  \textrm{tr} (H_1 H_2 H_3^{\dag})      \non
& &+ \lambda_1  (\textrm{tr} (H_1 H_1^{\dag})  )^2+ \lambda_2  (\textrm{tr} (H_2 H_2^{\dag})  )^2+ \lambda_3  (\textrm{tr} (H_3 H_3^{\dag})  )^2 \non
& &+  \lambda_{12}  \textrm{tr} (H_1 H_1^{\dag})  \textrm{tr} (H_2 H_2^{\dag}) +    \lambda_{23}  \textrm{tr} (H_2 H_2^{\dag})  \textrm{tr} (H_3 H_3^{\dag}) \non
& &+   \lambda_{31}  \textrm{tr} (H_3 H_3^{\dag})  \textrm{tr} (H_3 H_3^{\dag}).
\label{scapote}
\eeq
Here all the Higgs fields are represented by two-by-two 
matrices, and they are real, namely
\begin{align}
 \epsilon H_{i}^{\ast} \epsilon = -H_i,
 ~~~\text{where }
 \epsilon = 
 \begin{pmatrix} 
  0 & 1 \\ 
  -1 & 0 
 \end{pmatrix}
.
\end{align}
All parameters in the Higgs potential are also real.
\begin{table}[tp]
\begin{center}
\caption{The charge assignment of the partially composite standard model.
  Only $H_1$ and $H_2$  are the representations of the SU(2)$_1$ gauge symmetry.
}
\label{charge}
\begin{tabular}{l | c c c |c } \hline \hline
Fields & SU(2$)_0$ & U(1$)_2$ & SU(3$)_c$ & SU(2$)_1$   \\ \hline 
$H_1$& $\mathbf{2}$  & 0 & $\mathbf{1}$ & $\mathbf{2}$ \\ 
$H_2$& $\mathbf{1}$  & 1/2 & $\mathbf{1}$ & $\mathbf{2}$ \\ \hline  
$H_3$ & $\mathbf{2}$  & 1/2 & $\mathbf{1}$ & $\mathbf{1}$ \\ 
$Q$ & $\mathbf{2}$ & 1/6 & $\mathbf{3}$ & $\mathbf{1}$ \\ 
$u_R$ & $\mathbf{1}$  & 2/3 & $\mathbf{3}$& $\mathbf{1}$ \\
$d_R$ & $\mathbf{1}$ & -1/3 & $\mathbf{3}$ & $\mathbf{1}$  \\ 
$L$ & $\mathbf{2}$& -1/2 & $\mathbf{1}$   & $\mathbf{1}$ \\
$e_R$ & $\mathbf{1}$  & -1 & $\mathbf{1}$ & $\mathbf{1}$\\ \hline   \hline
\end{tabular}
\end{center}
\end{table}
We assume that all the vacuum expectation values (VEVs) of the Higgs
fields are diagonal,
real and positive to realize  desired  electroweak symmetry breaking.
The Higgs fields are expanded around their VEVs, $v_1, ~v_2$ and $v_3$,
\beq
H_i = \frac{v_i}{2} + \frac{1}{2} \left( h_i + i \tau^a \pi^a_i \right),
\eeq
where $\tau^a$ is the Pauli 
matrices, and $h_i,
~\pi^a_i$ are 
the four real scalar component fields.
The covariant derivatives of the Higgs fields
 are given as
\beq
D_{\mu} H_1 &=& \partial_{\mu} H_1 + i g_0 \frac{\tau^a}{2} W^a_{0 \mu} H_1 - i g_1 H_1 \frac{\tau^a}{2} W^a_{1 \mu}, \\
D_{\mu} H_2 &=& \partial_{\mu} H_2 + i g_1 \frac{\tau^a}{2} W^a_{1 \mu}  H_2 - i g_2 H_2 \frac{\tau^3}{2} B_{\mu}, \\
D_{\mu} H_3 &=& \partial_{\mu} H_3 + i g_0 \frac{\tau^a}{2} W^a_{0 \mu}
H_3 - i g_2 H_3 \frac{\tau^3}{2} B_{\mu}
.
\eeq
By calculating the muon life time in this model at the tree level, we 
find the relation 
between the Fermi constant and the VEVs in this model as
\beq
v_3^2 + \frac{1}{\frac{1}{v_1^2} + \frac{1}{v_2^2}} 
= 
v^2 
\equiv 
\left( \sqrt{2} G_F \right)^{-1},
\eeq
where $v \simeq 246$ GeV.
For the later convenience, we introduce a new parameter $r$, 
\beq
r \equiv \frac{v_2}{v_1}
.
\eeq
Thus $v_1$ and $v_2$ are expressed by $r$, $v_3$, and $v$,
\beq
v_1^2 = (1 + r^{-2})(v^2 - v_3^2),~~~~~
v_2^2= (1 + r^2) (v^2 - v_3^2).
\eeq

There are twelve scalars in this model,
and 
six of them are eaten by the gauge bosons. 
 Thus this model has six physical scalars: 
three CP-even Higgs bosons ($h,~H,~H'$), one CP-odd Higgs boson ($A$),
and 
two charged Higgs bosons
($H^{\pm}$). We identify $h$ as the
SM-like 125~GeV Higgs bosons.
The masses of the CP-odd and the charged Higgs bosons
are the same at the tree level
and given by
\begin{align}
 m_{A}^2 = m_{H^{\pm}}^2
=
 -\frac{1}{4} \frac{\kappa}{v_3} \frac{1+r^2}{r} v^2.
\end{align}
The mass eigenstates of the CP-even Higgs bosons are related to
 the gauge
eigenstates through the mixing angles $\theta_1,~\theta_2,$ and
$~\theta_3$ as 
follows,  
\beq
\matl{c} H' \\ H \\ h  \matr   = \matl{ccc} s_1 s_2 - c_1 c_2 s_3 &     -s_1 c_2 - c_1 s_2 s_3      & 
     c_1 c_3    \\
 -c_1 s_2 - 
   s_1 c_2 s_3 & 
  c_1 c_2 - s_1 s_2 s_3 & s_1 c_3   
       \\
     c_2 c_3&    s_2 c_3& s_3 \matr \matl{c} h_1 \\ h_2 \\ h_3 \matr  \label{Higgsangle}
\eeq
where $s_i~(c_i)$ stands for
$\sin \theta_i$~($\cos \theta_i$) for $ i = 1,~2,~3$.

The Yukawa interactions are given as
\begin{align}
  \mathcal{L}^{\textrm{Yukawa}} 
=& 
 - \bar{Q}^i H_3 
\begin{pmatrix}
  y_u^{ij} & 0 \\ 
  0 & y_d^{ij}
\end{pmatrix}
\begin{pmatrix}
  u_R^{j} \\
  d_R^{j}
\end{pmatrix}
- 
\bar{L}^i H_3 
\begin{pmatrix}
  0 & 0 \\ 
  0 & y_e^{ij}
\end{pmatrix}
\begin{pmatrix}
  0 \\
  e_R^{j}
\end{pmatrix}
+
H.c.
,
\end{align}
where $i$ and $j$ are the generation indices.  
We introduce a parameter $\kappa_F$
which is the ratio of the couplings between
the lightest CP-even Higgs boson and the fermions to its SM value,  
\begin{align}
 \kappa_F 
 \equiv& 
 \frac{g_{hff}}{m_f/v} = \frac{v}{v_3} s_3
.
\end{align}
Since $|s_3| \leq 1$ and $v/v_3  > 1$, $\kappa_F$ can be larger than
one. We will discuss  the viable range of $\kappa_F$ in the next subsection.
The fermion masses are given as
\begin{align}
m_f 
=& 
y_f \frac{v_3}{2}
=
y_f \frac{v_3}{v} \frac{v}{2}
,
\label{eq:fermionMass}
\end{align}
and the Yukawa couplings are enhanced by $v/v_3$ compared to their SM
values.
Large Yukawa couplings 
could make the Higgs potential unstable 
above the electroweak scale. 
We discuss this point in Sec.~\ref{sec:constraint}.

In addition to the SM gauge bosons, we have extra three vector bosons,
$W'^{\pm}$ and $Z'$.  
In the $g_1 \gg g_0$ regime, 
the mass eigenvalues of the gauge  bosons are given  as 
\beq
m_W^2 &\simeq &\frac{1}{4} g_0^2 v^2 \left( 1 - \frac{g_0^2}{g_1^2} \frac{1}{(1 + r^2)^2} \right), \\
m_{W'}^2 &\simeq& \frac{1}{4} g_1^2 (v_1^2 + v_2^2) \left( 1+\frac{g_0^2}{g_1^2} \frac{1}{(1+r^2)^2} \right),  \\
m_Z^2 &\simeq &\frac{1}{4} (g_0^2 + g_2^2) v^2 \left( 1 - \frac{(g_0^2 - g_2^2 r^2 )^2}{g_1^2 (g_0^2 + g_2^2)}\frac{1}{(1 + r^2)^2} \right), \\
m_{Z'}^2 &\simeq& \frac{1}{4} g_1^2  (v_1^2 + v_2^2) \left( 1 + \frac{g_0^2 + g_2^2 r^4 }{g_1^2 }\frac{1}{(1 + r^2)^2} \right). 
\eeq 
We find $m_{W'} \simeq m_{Z'}$ except in the large $r$ regime.
We need to find the relation between the gauge eigenstates and
the mass eigenstates to evaluate the couplings, and they are 
given as
\begin{align}
 W^{\pm}_{\mu} 
\simeq&
 \left( 1 - \frac{1}{2 (1 + r^2)^2} \frac{g_0^2}{g_1^2} \right) 
 W^{\pm}_{0 \mu}
+
 \left( \frac{1}{1 + r^2} \frac{g_0}{g_1} \right)  
 W^{\pm}_{1 \mu}
,
\\ 
 W'^{\pm}_{\mu} 
\simeq&
-\left( \frac{1}{1 + r^2} \frac{g_0}{g_1} \right)  
 W^{\pm}_{0 \mu}
+
 \left( 1 - \frac{1}{2 (1 + r^2)^2} \frac{g_0^2}{g_1^2} \right) 
 W^{\pm}_{1 \mu}
,
\\ 
  A_{\mu} 
 =&
  \frac{e}{g_0}   W^{3}_{0 \mu}
+ \frac{e}{g_1}   W^{3}_{1 \mu}
+ \frac{e}{g_2}   W^{3}_{2 \mu}
,
\\ 
 Z_{\mu} 
\simeq&
 ~c_W \left( 1 - \frac{1 - 2 r^2 t_W^2}{2 ( 1 + r^2)^2}
 \frac{g_0^2}{g_1^2} \right) 
 W^{3}_{0 \mu}
+
c_W \left( \frac{1 - r^2 t_W^2}{(1 + r^2)} \frac{g_0}{g_1} \right)  
 W^{3}_{1 \mu}
 \non
& -
s_W \left( 1 -  \frac{r^4 t_W^2}{2 ( 1+ r^2)^2} \frac{g_0^2}{g_1^2} \right)  
 W^{3}_{2 \mu}
,
\\ 
  Z'_{\mu} 
 \simeq&
  - \frac{1}{1 + r^2} \frac{g_0}{g_1}   W^{3}_{0 \mu}
 +
 \left( 1 - \frac{1 + r^4 t_W^2}{2 ( 1+r^2)^2 }
  \frac{g_0^2}{g_1^2} \right)   W^{3}_{1 \mu}
 -
 \frac{r^2 t_W}{( 1+ r^2)} \frac{g_0}{g_1}
  W^{3}_{2 \mu}
,
\end{align}
where $c_W = m_W/m_Z$, $s_W = \sqrt{1 - c_W^2}$, and $t_W = s_W/c_W$.
The typical size of the mixing angles is $\mathcal{O}(g_0 / g_1)$,  
and the gauge filed $W_0$ ($W_1$) is the main component of the mass
eigenstate $W$ ($W'$). 
It is worth noting that the mixing angles for $W'$ and $Z'$ become the
same in the small $r$  regime, which means the custodial symmetry is
enhanced.

The approximate expressions for some  couplings of the $W'$ and $Z'$ to 
the SM particles are given as
\begin{align}
 \frac{g_{W'ff}}{g_{Wff}^{\text{SM}}}
\simeq&
 - \frac{m_W}{m_{W'}}
 \sqrt{1 - \frac{v_3^2}{v^2}}
\frac{1}{r}
\label{eq:W'ffcoup}
,
\\ 
\frac{g_{W'WZ}}{g_{WWZ}^{SM}}
\simeq&
- \frac{m_W}{m_{W'}}
 \sqrt{1 - \frac{v_3^2}{v^2}}
\frac{1}{c_W^2}
\frac{r}{1 + r^2}
,
\\ 
\frac{g_{WWZ'}}{g_{WWZ}^{SM}}
\simeq&
- \frac{m_W}{m_{W'}}
 \sqrt{1 - \frac{v_3^2}{v^2}}
\frac{1}{c_W}
\frac{r}{1 + r^2}
.
\label{eq:WpWZcoupling}
\end{align}
Compare to the benchmark model (sequential standard model (SSM) \cite{Quigg:1983gw, Eichten:1984eu, Altarelli:1989ff}) 
used by the ATLAS/CMS, the $W'$ couplings
 to the SM fermions are smaller.
All couplings have a suppression factor of
 $ (m_W / m_{W'} ) \sqrt{1 - v_3^2 / v^2}  $, 
because the $W'$ boson couples to the elementary fermion through $W_0$-$W_1$ mixing,
so that the width is narrow. 
Since $g_{W'ff}$ is proportional to $r^{-1}$, the production cross section of  the Drell-Yan process is proportional to $r^{-2}$.
The $W'$ boson could not be produced in the large $r$ region.  
In Sec.~\ref{sec:Phenomenology of Spin-1 Resonances}, we will 
find a parameter space in small $r$ regions where a signal rate is consistent with the ATLAS diboson excess. 
Some numerical results  are given in
Sec.~\ref{8TeVstatus}

The gauge boson couplings to the scalars are also important in
our analysis. We give their approximated expressions here.
Due to the mixing between
 the two SU(2) gauge eigenstates, the
$WWh$ and $ZZh$ couplings differ from the SM values. 
We denote these coupling ratios to the SM values by $\kappa_{W,Z}$,
\begin{align}
 \kappa_W 
 \equiv 
 \frac{g_{hWW}}{2 m_W^2/v} 
,
\quad
 \kappa_Z 
 \equiv& 
 \frac{g_{hZZ}}{2 m_Z^2/v} 
,
\label{eq:def_of_kappa}
\end{align}
and their approximated formulae in the
$g_1 \gg g_{0}, g_2$ regime are
\begin{align}
 \kappa_W \simeq
 \kappa_Z \simeq
\frac{r^3}{(1+r^2)^{3/2}}\sqrt{1 - \frac{v_3^2}{v^2}} c_2 c_3 
 +
 \frac{1}{(1+r^2)^{3/2}}\sqrt{1 - \frac{v_3^2}{v^2}} s_2 c_3 
 +
 \frac{v_3}{v} s_3
.
\end{align}
The couplings relevant to the $W'/Z'$ decay are
\begin{align}
 g_{WW'h} \simeq
 g_{ZZ'h} \simeq&
\frac{2 m_W m_{W'}}{v}
\left(
- \frac{r^2}{(1+r^2)^{3/2}} c_2 c_3
+ \frac{r}{(1+r^2)^{3/2}} s_2 c_3
- \frac{1}{r} \frac{m_W^2}{m_{W'}^2} 
   \frac{v_3}{v} \sqrt{ 1- \frac{v_3^2}{v^2}} 
   s_3
\right)
,
\\ 
 g_{WW'H} \simeq
 g_{ZZ'H} \simeq
&
\frac{2 m_W m_{W'}}{v}
\Biggl(
 - \frac{r^2}{(1+r^2)^{3/2}} (-c_1 s_2 - s_1 c_2 s_3)
  \nonumber\\
 & \quad \quad \quad \quad \quad
 + \frac{r}{(1+r^2)^{3/2}} (c_1 c_2 - s_1 s_2 s_3) 
 - \frac{1}{r} \frac{m_W^2}{m_{W'}^2} 
 \frac{v_3}{v} \sqrt{ 1- \frac{v_3^2}{v^2}} 
 s_1 c_3
\Biggr)
\label{eq:WW'H}
,
\\ 
 g_{WW'H'} \simeq
 g_{ZZ'H'} \simeq
&
\frac{2 m_W m_{W'}}{v}
\Biggl(
 - \frac{r^2}{(1+r^2)^{3/2}} (s_1 s_2 - c_1 c_2 s_3)
 \nonumber\\
& \quad \quad \quad \quad \quad
 + \frac{r}{(1+r^2)^{3/2}} (-s_1 c_2 - c_1 s_2 s_3) 
 - \frac{1}{r} \frac{m_W^2}{m_{W'}^2} 
 \frac{v_3}{v} \sqrt{ 1- \frac{v_3^2}{v^2}} 
 c_1 c_3
\Biggr)
.
\end{align}
%

\subsection{Model parameters}
\label{sec:modelParams}

In the electroweak sector of the model, there are 13
real parameters, 
\beq
\mu_1^2, ~\mu_2^2, ~\mu_3^2, ~\kappa, ~\lambda_1,~\lambda_2, ~\lambda_3, ~\lambda_{12},~\lambda_{23}, ~\lambda_{31}, ~g_0, ~g_1, ~g_1.
\eeq
It is convenient to use a different set of the parameters instead of these parameters.
We use the following 13 parameters to fix the parameters in the
electroweak sector,
\begin{align}
 r, ~v_3, ~v, ~\alpha, ~m_Z, ~m_{Z'}, 
 ~m_h, ~ m_{H'}, ~m_{H}, ~m_{A}, ~\kappa_F, ~\kappa_Z, ~g_{WW'H'}.
\label{eq:inputParams}
\end{align}
Here, we use $r$, $v_3$, $v$ instead of the three $\mu$ parameters.
$\kappa$ is fixed by the charged Higgs mass or the CP-odd Higgs mass.
The six $\lambda$'s have the same information as the three CP-even Higgs
masses and their mixing angles ($m_h$, $m_H$, $m_{H'}$, $\theta_1$,
$\theta_2$, $\theta_3$). 
We can use $m_Z$, $m_{Z'}$, and $\alpha
(=e^2/4\pi)$ instead of the gauge couplings. 
In addition, we can replace
($\theta_1$, $\theta_2$, $\theta_3$) with
($\kappa_F$, $\kappa_Z$, $g_{WW'H'}$).

The values of the four parameters,
$v,~m_Z,$ $\alpha$, and $m_h$ are already known very precisely.
We take $m_h = 125$ GeV. 
We find $\kappa_F$ is severely constrained  close to $1$ in the $m_A
\gg m_h$ regime.
When the heavy Higgs masses are universal, $m_A = m_{H'} = m_{H}$, 
$\lambda_3$ is simply expressed as
\beq
\lambda_3 (\mu = m_{Z'}) = \frac{\kappa_F^2}{2}\frac{m_h^2}{v^2} + \frac{1 - \kappa_F^2}{2}\frac{m_A^2}{v^2}.
\label{lambda3_solved}
\eeq
Thus, $\lambda_3$ is very large except for $\kappa_F$ =1.
Similarly,  
the coupling ratio $\kappa_Z$ is also severely
constrained  close to its maximal value in $m_A \gg m_h$ regime.
Typically, the allowed value of $\kappa_Z$ is $1 - \mathcal{O}(1)\%$.
The detailed description is given in Appendix~\ref{app:kappaz}.
In the following discussion, we take
$\kappa_F \simeq 1$,
$m_A = m_{H} = m_{H'} ={\cal O}(1)$~TeV,
$g_{WW'H'}=0$,
and we always choose $\kappa_Z$ to be its maximal value. 
For most of our numerical analysis, $\kappa_Z$  is set to $0.95$--$1.00$.

\section{Phenomenology of spin-1 resonances}
\label{sec:Phenomenology of Spin-1 Resonances}

In this section, we discuss the properties of the $W'$ and $Z'$ bosons such as 
the production cross sections and
the decay branching ratios.
And we also discuss both the theoretical and experimental constraints.
 In the following, we show some formulae with approximation, which help to  understand the parameter dependence.
However, we  use the exact formulae in our numerical calculations.

\subsection{Properties of the extra vector bosons }
\label{sec31}

\begin{figure}[tp]
\begin{center}
\includegraphics[width =6.cm,bb = 0 0 360 335]{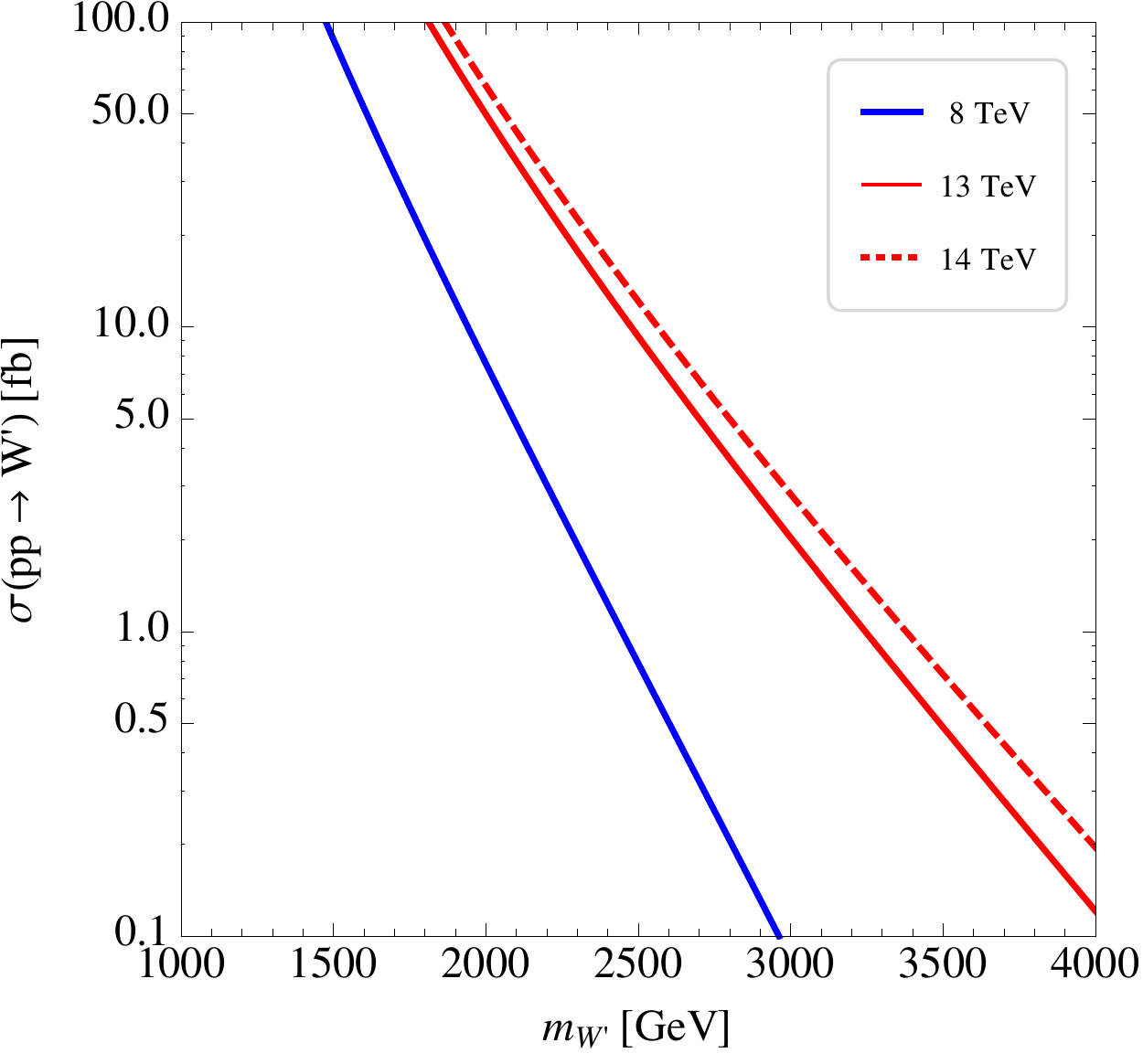}~~~~
\includegraphics[width =6.cm,bb = 0 0 360 333]{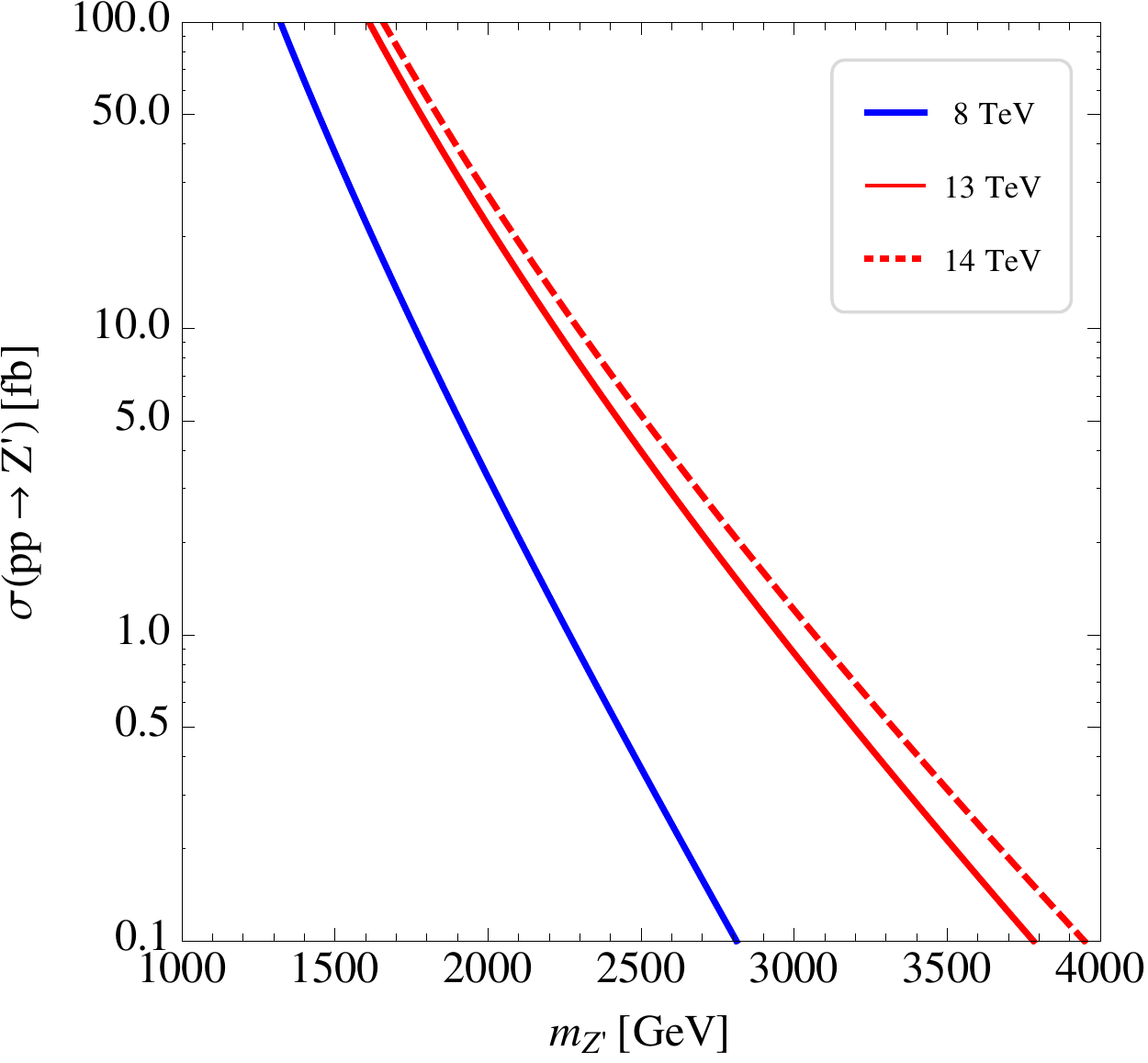}
 \vspace{.2cm}
\caption{The production cross sections of the extra vector bosons, $W'$ and $Z'$.
We take $v_3 = 200 \GeV$, $r = 0.13$. 
}
\label{fig:prop2}
\end{center}
\end{figure}

The main production mode of $W'$ and $Z'$ in $pp$ collisions is the
Drell-Yan process. 
They are proportional to $r^{-2}$ as we can see from
Eq.~(\ref{eq:W'ffcoup}).\footnote{For more details on the formulae, see
Appendix~\ref{App:xsec}.} 
We consider small $r$ in order to make the cross section large.
In Fig.~\ref{fig:prop2}, we show the production cross sections for
$W' ~(W'^{+} + W'^{-})$ and $Z'$. 
Here we take
 $v_3 = 200 \GeV$, $r = 0.13$ and $\kappa_F = 1.00$.
We use the {\sc CTEQ6L} parton distribution functions~\cite{Pumplin:2002vw}.
The production cross section of $W'$ is approximately twice that of $Z'$
for small $r$ because the custodial SU(2) symmetry is recovered in the region.

The partial decay widths of $W'$ and $Z'$ into the SM particles are given as
\begin{align}
 \Gamma(W' \to WZ) 
 \simeq& 
 \frac{1}{48 \pi} \frac{m_{W'}^3}{v^2} \frac{r^2}{(1 + r^2)^2} \left( 1
 - \frac{v_3^2}{v^2} \right)
\label{eq:W'toWZ}
,\\
 \Gamma(W' \to Wh) 
 \simeq& 
 \frac{1}{48 \pi} \frac{m_{W'}^3}{v^2} \frac{r^2}{(1+r^2)^3} \left( - r c_2 c_3 + s_2 c_3 \right)^2,\\
 \Gamma(W' \to f\bar{f}) 
 \simeq& 
 \frac{N_c}{48\pi}
 \frac{m_W^2}{m_{W'}}\frac{e^2}{s_W^2}\frac{1}{r^2}\left( 1 -
 \frac{v_3^2}{v^2}\right), \\ 
 \Gamma(Z' \to WW) 
 \simeq& 
 \frac{1 }{48\pi} \frac{m_{W'}^3}{v^2} \frac{r^2}{(1 + r^2)^2}\left( 1 -
 \frac{v_3^2}{v^2}\right), \\ 
 \Gamma(Z' \to Zh) 
 \simeq&
 \frac{1}{48 \pi} \frac{m_{W'}^3}{v^2} \frac{r^2}{(1+r^2)^3} \left( - r
 c_2 c_3 + s_2 c_3 \right)^2,\\ 
 \Gamma(Z' \to f\bar{f}) 
 \simeq& 
 \frac{N_c}{24 \pi}
 \frac{m_W^2}{m_{W'}}\frac{e^2}{s_W^2}\frac{1}{r^2}\left( 1 -
 \frac{v_3^2}{v^2}\right) \non 
  & \times \left(  \left( \left( 1 - r^2 \frac{s_W^2}{c_W^2} \right)
 T^3_f + r^2 \frac{s_W^2}{c_W^2} Q_f \right)^2 + \left( r^2
 \frac{s_W^2}{c_W^2} Q_f \right)^2  \right), \label{Zpff} 
\end{align}
where $T^3_f ~= 1/2 ~(-1/2)$ for the up-type (down-type) fermions,
$Q_f$ is the electric charge of the fermion, 
and  $N_c$ is the color factor, $N_c = 3$. 
Note that the bosonic channels are dominant among their decay channels
due to an enhancement factor from the wave function of the  longitudinally polarized
gauge bosons in the final states. Therefore the dominant decay modes of
 $W'$ ($Z'$) are $W'\to WZ$ and $W' \to Wh$ ($Z' \to WW$ and $Z' \to Zh$).

There are also decay modes with a heavy scalar in the final states.
For example, $W' \to W H$ and $W' \to h H$ exist and their 
approximated formulae are
\begin{align}
 \Gamma(W' \to W H)
\simeq&
 \frac{1}{48 \pi} 
\left(
\frac{g_{WW'H}}{2 m_W m_{W'}/v}
\right)^2
\frac{m_{W'}^3}{v^2}
\left(
8 \frac{m_W^2}{m_{W'}^2}
+ 
\left(
1 - \frac{m_H^2}{m_{W'}^2} + \frac{m_W^2}{m_{W'}^2} 
\right)^2
\right)
\left( 1 - \frac{m_H^2}{m_{W'}^2}\right)
 ,\\
 \Gamma(W' \to h H)
\simeq&
 \frac{1}{48 \pi} 
\frac{m_{W'}^4}{v^2 - v_3^2}
\frac{v_3^2}{v^2}
\frac{2}{(1 + r^2)^3}
\left(
-r c_2 c_3
+
s_2 c_3
\right)^2
\left(
1 -  \frac{m_{H^{+}}^2 }{m_{W'}^2}
\right)^{3}
,
\end{align}
where $g_{WW'H}$ is given in Eq.~(\ref{eq:WW'H}), and 
$g_{WW'H}/(2 m_W m_{W'}/v)$ is ${\cal O}(1)$ in large regions of the
parameter space. 
These decay widths are comparable to  those of the SM final states. 
The decay channels including only the heavy states, such as $W' \to
H^{\pm} H$,  also have the same feature.
Once these modes are open, they would be dominant decay modes.

\begin{figure}[tp]
\begin{center}
\includegraphics[width =6.cm,bb=0 0 300 280]{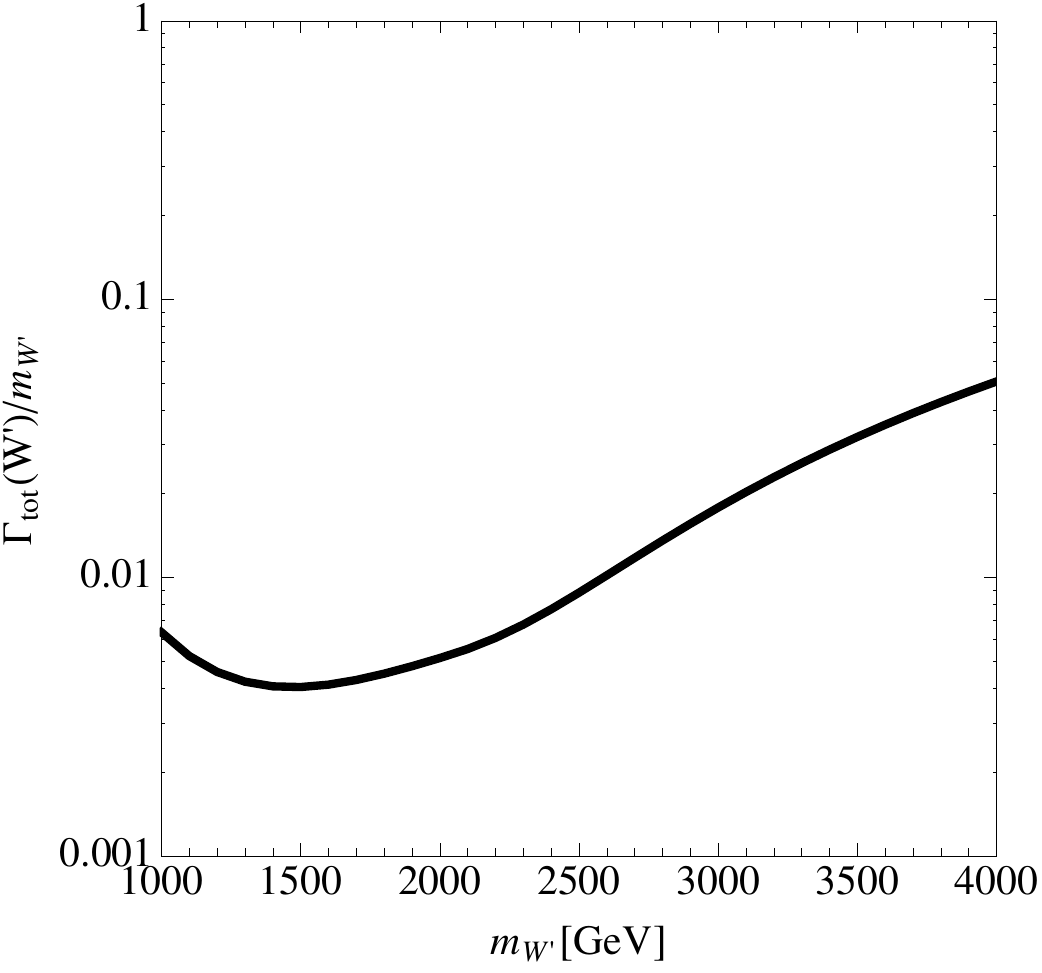}~~~~
\includegraphics[width =6.cm,bb=0 0 300 280]{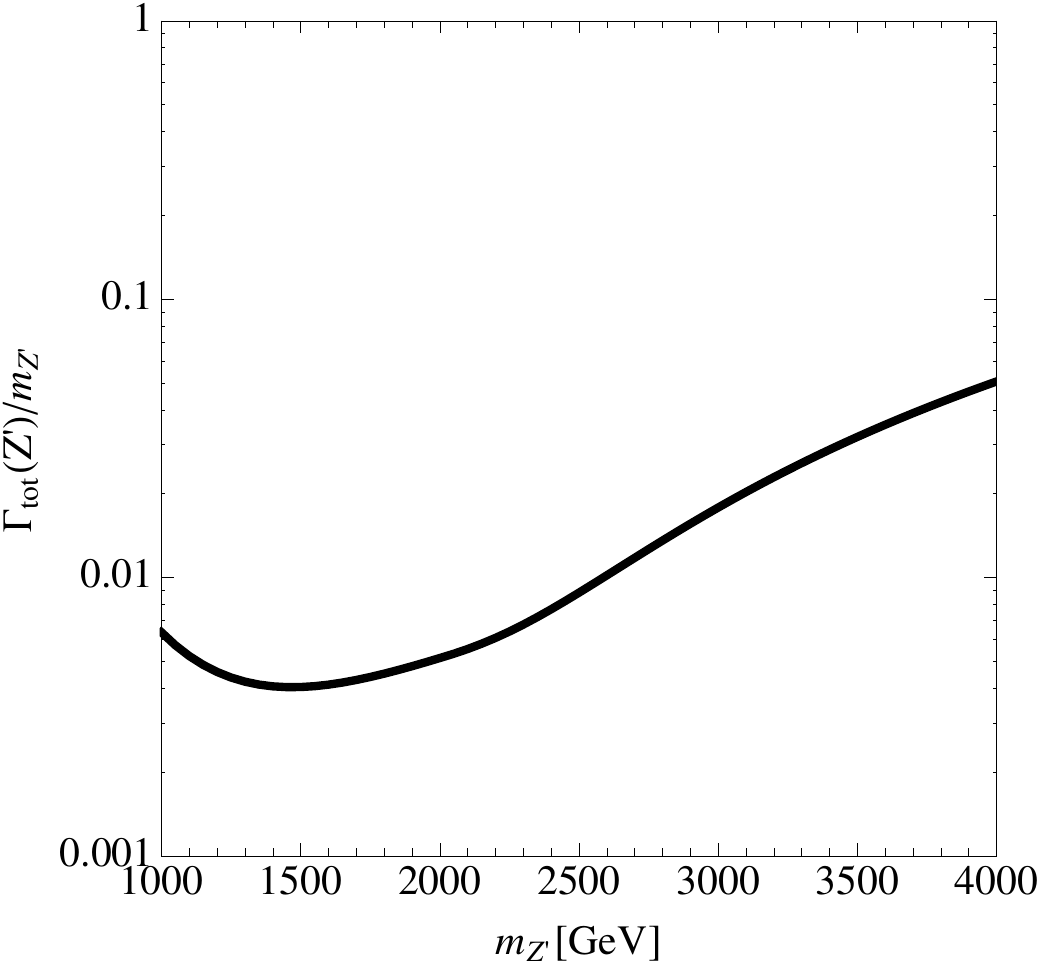}\\
 \vspace{.4cm}
 \includegraphics[width =6.cm,bb=0 0 300 280]{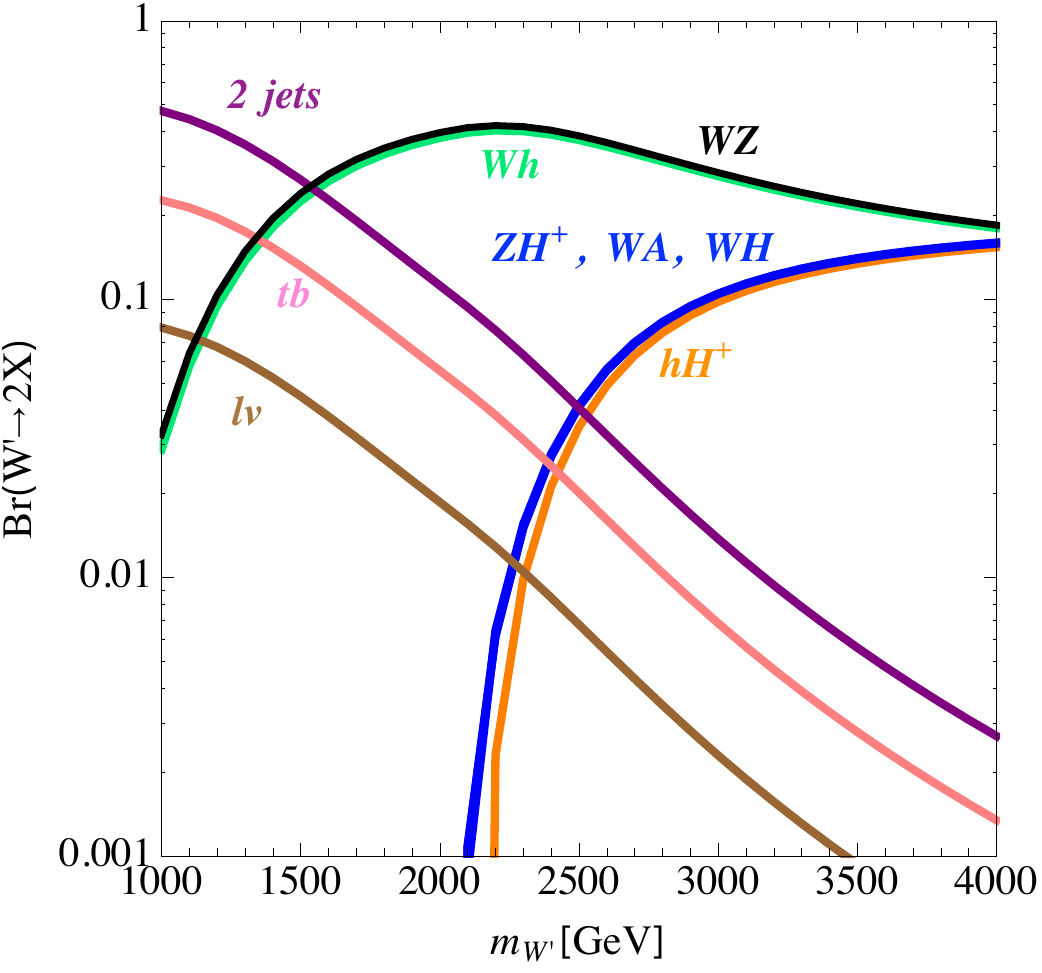}~~~~
\includegraphics[width =6.cm, bb = 0 0 300 280]{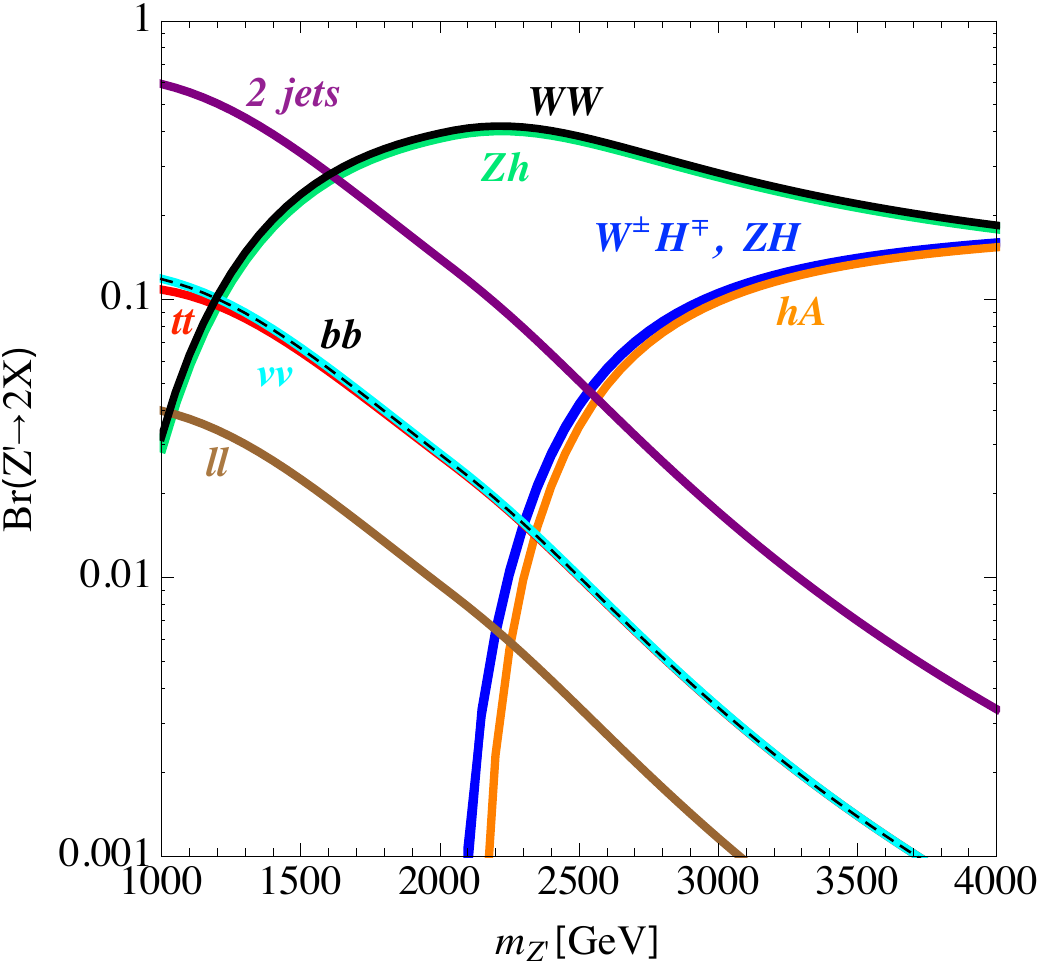}
 \vspace{.2cm}
\caption{The total widths and the branching ratios of the extra vector bosons, $W'$ and $Z'$.
We take $v_3 = 200 \GeV$, $r = 0.13$, $\kappa_F = 1.00$, 
and  $m_A = m_{H'} = m_{H} = 2$ TeV.
Here, 2jets means 
Br($W' \to ud) + $Br($W' \to s c$)
or 
Br($Z' \to u u) + $Br($Z' \to dd) + $Br($Z' \to ss) + $Br($Z' \to cc$), 
$\ell \nu$ means 
Br($W' \to e \nu_e)$   ( $= $Br($W' \to \mu \nu_{\mu})  = $Br($W' \to \tau \nu_{\tau}$)),
$\ell \ell$ means 
Br($Z' \to ee)$   ( $= $Br($Z' \to \mu \mu)  = $Br($Z' \to \tau \tau$)),
$\nu \nu$ means 
Br($Z' \to \nu_e \nu_e) + $Br($Z' \to \nu_{\mu} \nu_{\mu}) + $Br($Z' \to
\nu_{\tau} \nu_{\tau}$), 
and
$W^{\pm} H^{\mp}$ means 
Br($W' \to W^{+} H^{-}) = $Br($W' \to W^{-} H^{+}$).
}
 \label{fig:prop}
\end{center}
\end{figure}
In Fig.~\ref{fig:prop}, we show
the total widths and the branching ratios of $W'$ and $Z'$. 
Here we take $v_3 = 200 \GeV$, $r = 0.13$ and $\kappa_F = 1.00$, $m_A = m_{H'} = m_{H} = 2$ TeV. 
We also take $g_{WW'H'}=0$ as we mentioned in
Sec.~\ref{sec:modelParams}, thus $W' \to WH'$ and $Z' \to ZH'$ are
absent in the figure.
We find that the dominant decay channels are $V' \to VV$
and $V' \to Vh$ for large $m_{V'}$. 
However, the decay branching ratio reduces once $V' \to H X$ decay modes are open.

In Fig.~\ref{fig:prop}, we find 
Br($W' \to WZ$) = Br($W' \to Wh$) $\simeq$ 40~\% 
at $m_{W'} = $  2~TeV.
The relation is easily understood by the equivalence theorem, Br($W'\to W_L Z_L) =$Br($W'\to \pi_{SM} \pi_{SM}$) in the heavy $W'$ mass limit, where $\pi_{SM}$ is the SM Nambu-Goldstone boson. 
In the SM limit, Br($W'\to \pi_{SM} \pi_{SM}$) is equivalent to Br($W'\to \pi_{SM} h$). 
Thus, Br($W'\to WZ)\simeq$Br($W'\to Wh$) is realized.
In addition, when one takes $g_{WW'H'}=0$, the heavy Higgses $H,~A,~H^{\pm}$ form a multiplet, and Br($W' \to ZH^{\pm})\simeq$Br($W'\to WA)\simeq$Br($W'\to WH$) is realized.

We also find the total widths are 
narrow, namely $\Gamma /m \sim
1$--$5 ~\%$, because the resonances decay into the SM particles with suppressed couplings due to the mixing between the elementary and the composite sectors.
Another remark is that the decay properties of $W'$ and $Z'$ are 
similar  due to the enhanced SU(2)
custodial symmetry  
in small $r$ regime.\footnote{In the large $r$  region, the branching ratio $Z' \to f 
\bar{f}$ and the production cross section of $Z'$  become larger (see
Eqs.~(\ref{Zpff}, \ref{prod:Zp})),  
while the production cross section of $W'$ is suppressed.}

\begin{figure}[tp]
\begin{center}
\includegraphics[width =6cm,bb=0 0 300 280]{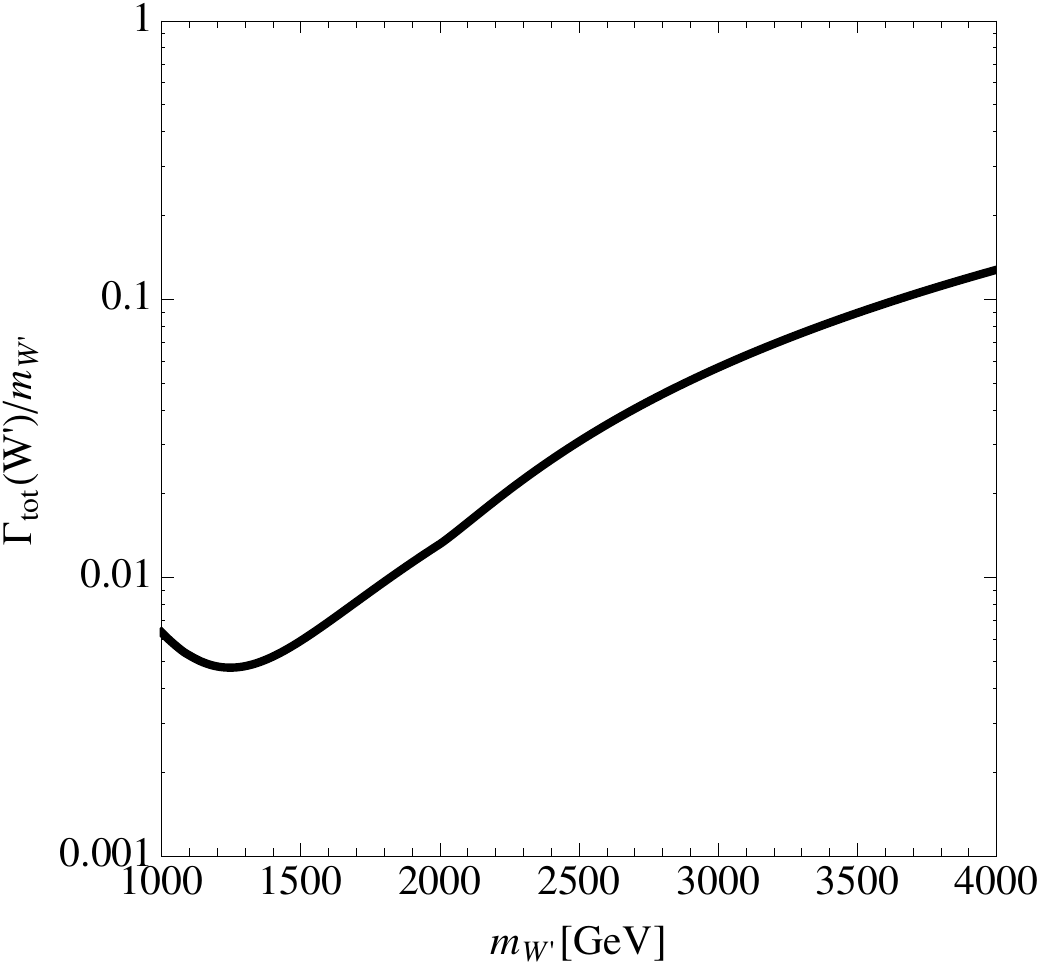}~~~~
\includegraphics[width =6cm,bb=0 0 300 280]{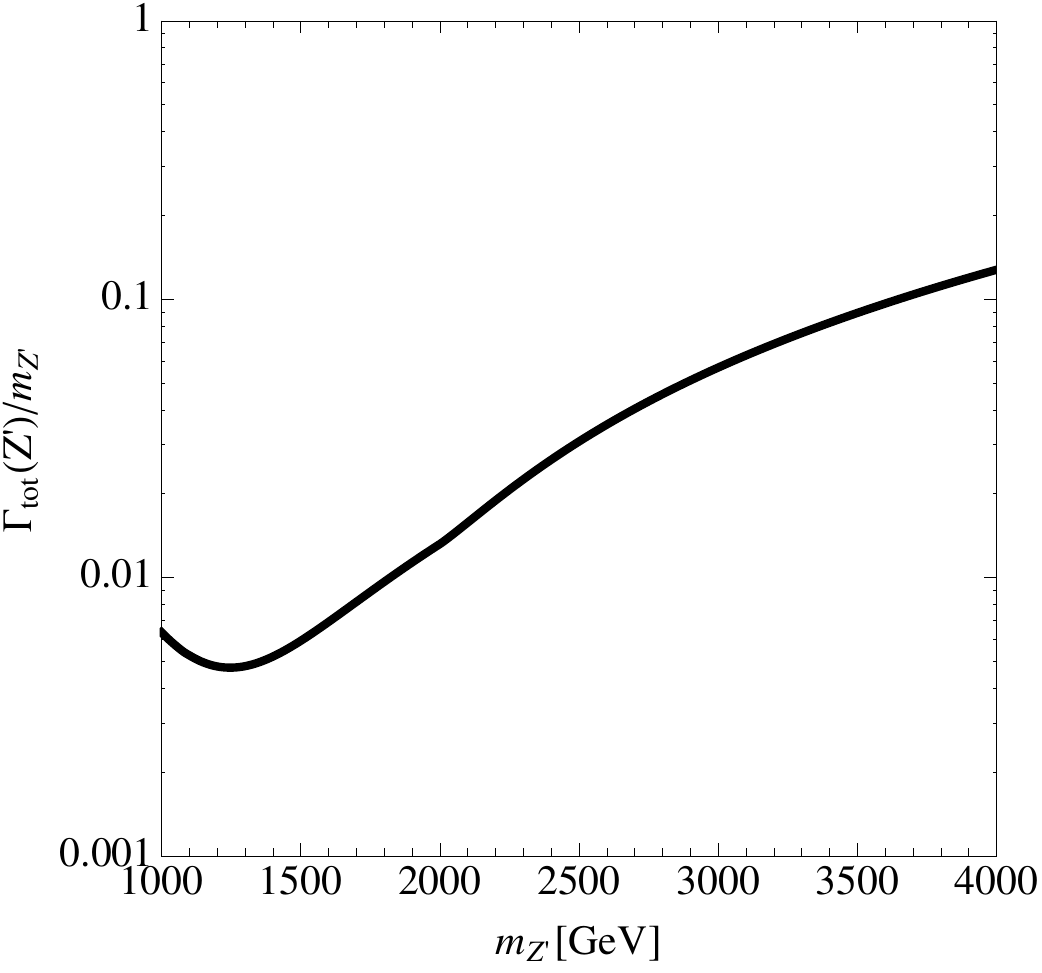}\\
 \vspace{.4cm}
 \includegraphics[width =6cm,bb=0 0 300 280]{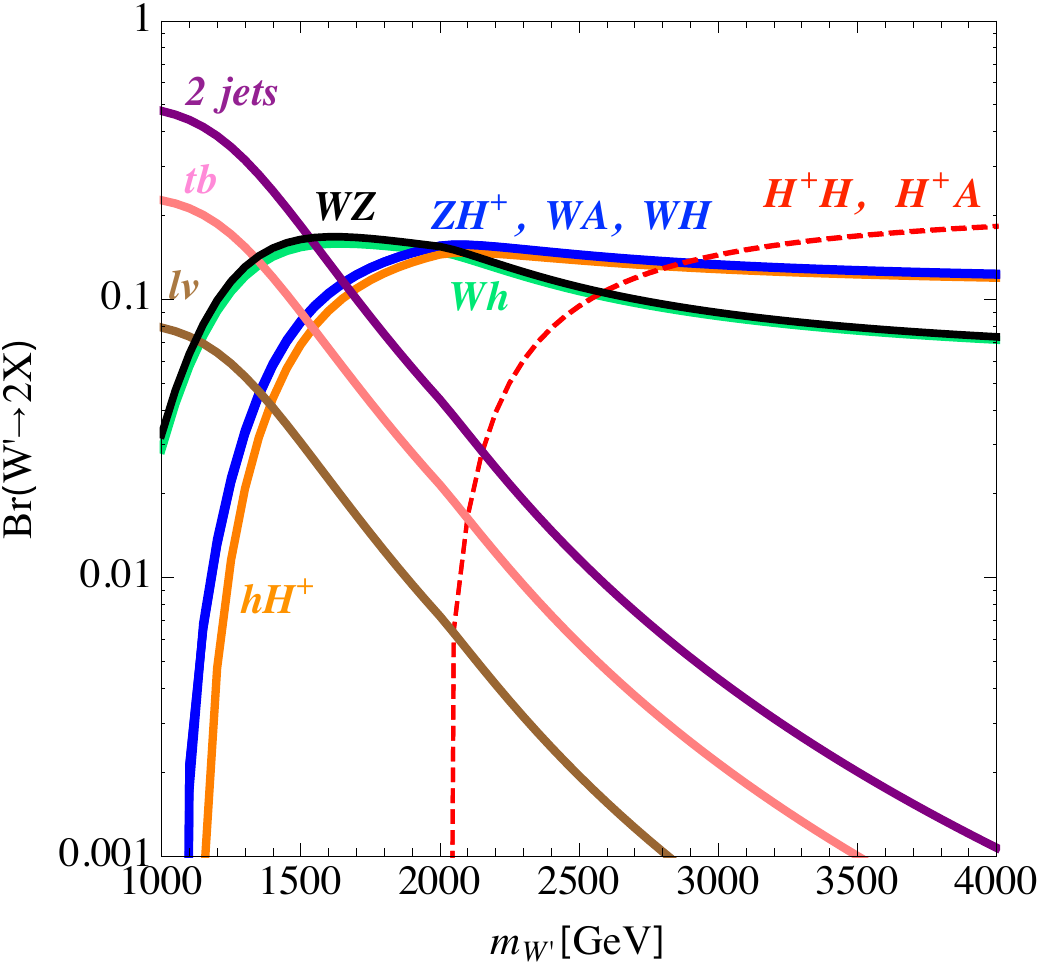}~~~~
\includegraphics[width =6cm, bb = 0 0 300 280]{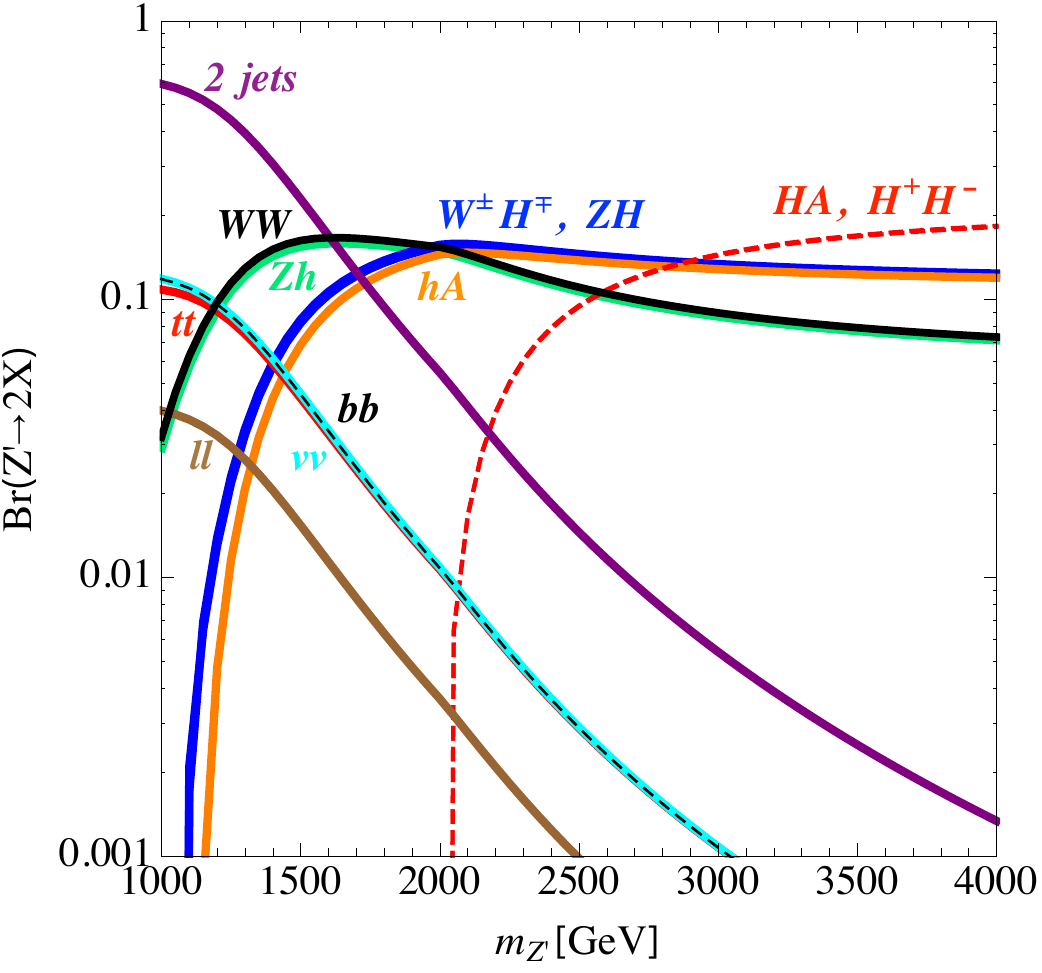}
 \vspace{.2cm}
\caption{The total widths and the branching ratios of $W'$ and $Z'$ for
$m_A = m_{H'} = m_{H} = 1$ TeV. 
Other parameter choice and the notations are the same as in Fig.~\ref{fig:prop}. 
}
\label{fig:prop3}
\end{center}
\end{figure}
In Fig.~\ref{fig:prop3}, we take different scalar masses, 
$m_A = m_{H'} = m_{H} = 1$~TeV. The other parameter choice is
the same as in Fig.~\ref{fig:prop}.
The decay channels to the non-SM particles, 
namely Br($W' \to H^{+}H$), Br($W' \to H^{+} A$), Br($Z' \to H A$), 
and Br($Z' \to H^{+} H^{-}$), are open.    
We took the parameter so that the channels to $H'$ is absent,
$g_{WW'H'}=0$.

As mentioned above, these channels are comparable to the decay modes to the SM particles. 
As a result, the decay modes $V' \to VV$ and $V' \to Vh$ searched at
the ATLAS and the CMS experiments are
suppressed compared to Fig.~\ref{fig:prop}. 
This implies that the decay channels with heavy scalars should not be open if we try to explain
the excess at the ATLAS experiment. Hereafter, we consider the situation
that the heavy scalars are as heavy as the extra gauge bosons.

\subsection{Constraints on the model}
\label{sec:constraint}

In this subsection, we show theoretical and experimental
constraints on the model parameters. 
In order to perform a reliable perturbative calculation,
we demand perturbativity condition, bounded below condition, global minimum
vacuum condition, and stability condition. 
In addition, we take into account the LHC bounds and the electroweak
precision measurements.

\subsubsection{Theoretical constraints}
\label{sec:cons}

\subsubsection*{Perturbativity  condition for the gauge coupling $g_1$}
We require that all the absolute values of the gauge couplings and the Higgs
quartic couplings are
smaller than $4\pi$ and $\left(4\pi \right)^2$ in order to keep the reliability of our analysis based on the
perturbative calculation.

For the gauge  couplings, $g_0$ and $g_2$ are almost the same value as
the SM gauge couplings, but $g_1$ can be very large. Since SU(2)$_1$ is
asymptotic free in our setup, $g_1$ becomes smaller at high energy due
to the quantum effects. 
Thus the maximum value of the $g_1$ is given
at $Z'$ mass scale, and we require $|g_1(\mu = m_{Z'})| < 4\pi$. The
regions where $|g_1(m_{Z'})| > 4\pi$ are filled with yellow in
Figs.~\ref{fig:const} and \ref{constriant2}.

\subsubsection*{Perturbativity  condition for the Higgs quartic
   couplings}

The scalar quartic couplings are large when the mass
differences between the SM-like Higgs and the other heavy 
scalar bosons are large.
Since we take $m_{h} \ll$ $m_{A}=m_{H}=m_{H'} \sim {\cal
O}(1)$~TeV in our analysis,
the quartic couplings tend to be large.
In addition, due to the renormalization group effects, they 
can be even larger at high energy.
The renormalization group equations for this model are 
given in Appendix \ref{App:RGE}. 
We define a cutoff scale $\Lambda$ by 
\begin{align}
 |\lambda_{i}(\mu = \Lambda)| = \left(4\pi\right)^2,
\end{align}
and require $|\lambda_{i} (\mu < \Lambda)| < \left( 4\pi\right)^2$.

This constraint highly depends on our choice of $\Lambda$. We  require 
that
$\Lambda$ should be significantly  higher than  a few TeV, otherwise we have to
take account of interaction terms from higher dimensional operators
whose coefficients are unknown, and some uncertainty is introduced to our analysis.
For example, there are operators which modifies the $W'$ coupling to the
SM fermions such as $(c / \Lambda^2) \bar{Q} i \gamma^{\mu} (H_1
iD_{\mu}H_1^{\dag}) Q$. This operator brings unknown parameter
$c$, and thus brings uncertainty to our calculations such as the
production cross section of $W'$.
Typically, such higher dimensional operators with $c\sim 1$
bring $1\%$ ($10 \%$) uncertainty if $\Lambda = 100 \TeV$($10 \TeV$).
To avoid such uncertainty from unknown parameters, we restrict ourselves for the case
$\Lambda >$ 100~TeV.
The parameter regions where $|\lambda(\Lambda)|> \left( 4\pi \right)^2$ are filled with the
lighter (darker) gray for $\Lambda = 100 (10)$~TeV in
Figs.~\ref{fig:const} and \ref{constriant2} for reference.

\subsubsection*{Bounded below  condition}

Here we consider the conditions that the Higgs potential at the tree
level is bounded below. 
For the purpose, it is enough to check that
the potential value at the large field values, and thus we consider
only the quartic terms in the potential. 
 We rewrite the quartic terms as
\beq
V_{\textrm{quartic} } = R^4 \left(  \lambda_1 n_1^2 + \lambda_2 n_2^2 +\lambda_3 n_3^2+  \lambda_{12} n_1 n_2 + \lambda_{23} n_2 n_3  + \lambda_{31} n_3 n_1 \right),
\eeq
with
\beq
R^2 n_i \equiv \textrm{tr}(H_i H_i^{\dag}) = \frac{1}{2} (h_i^2 + \pi_i^a \pi_i^a), 
\eeq
where $n_i$ satisfies $0 \leq n_{1,2,3} \leq 1$, and $n_1^2 + n_2^2 + n_3^2 = 1$. 
In order to avoid run-away vacua, we demand the following conditions
 for the Higgs quartic couplings at $m_{Z'}$ scale ($\lambda_i(\mu = m_{Z'})$),
 \vspace{0.15cm} 
\beq
\textrm{Min} \left[  \lambda_1 n_1^2 + \lambda_2 n_2^2 +\lambda_3 n_3^2+  \lambda_{12} n_1 n_2 + \lambda_{23} n_2 n_3  + \lambda_{31} n_3 n_1  \right] > 0, \textrm{~~for~} 0 \leq n_{1,2,3} \leq 1. \label{positivity}
\eeq
 \vspace{0.15cm}
The parameter regions where this condition is not satisfied are filled with
cyan in Figs.~\ref{fig:const} and \ref{constriant2}.
Especially, one can solve the above inequality analytically in specific directions as
\beq
\lambda_1 >0 ~~\textrm{for~}n_2=n_3=0,&~~~&\lambda_2>0~~\textrm{for~}n_1=n_3=0,\non
\lambda_3>0~~\textrm{for~}n_1=n_2=0,&~~~&\lambda_{23}  + 2 \sqrt{\lambda_2 \lambda_3} > 0~~\textrm{for~}n_1=0,\non
 \lambda_{31}  + 2 \sqrt{\lambda_3 \lambda_1} > 0~~\textrm{for~}n_2=0,&~~~&\lambda_{12}  + 2 \sqrt{\lambda_1 \lambda_2} > 0~~\textrm{for~}n_3=0.
 \label{bounded}
\eeq 

\subsubsection*{Global minimum vacuum condition}

We demand the  electroweak vacuum to be a global
minimum of the Higgs potential at $\mu =m_{Z'}$.
The parameter regions where this condition is not satisfied are filled with
green in Figs.~\ref{fig:const} and \ref{constriant2}.

\subsubsection*{Stability condition}

Among the quartic couplings, $\lambda_3$ can be very small when
$\kappa_F$ is very close to $1$, see Eq.~(\ref{lambda3_solved}), and
especially it takes the same value as the quartic coupling in the SM for
$\kappa_F = 1$. In that case, $\lambda_3$ can be negative at a high
energy scale due to the contribution from the Yukawa interaction to the
renormalization group equations, and the Higgs potential becomes unstable.
The VEV giving masses to the fermions are 
 $v_3 $, see Eq.~(\ref{eq:fermionMass}). Thus the Yukawa coupling in our setup is
larger than the coupling in the SM by $v/v_3$, and the
Higgs potential can become unstable at a few TeV scale for the small $v_3$ region.
We define the scale $\bar{\Lambda}$ at which $\lambda_3$ becomes negative,
 \beq
 \lambda_3 (\mu = \bar{\Lambda}) = 0, \label{stability}
 \eeq
and we demand $\bar{\Lambda} \gtrsim 100 \TeV$,  as we demand for the perturbativity condition. 
We fill the regions where this condition is not satisfied with
magenta in Figs.~\ref{fig:const} and \ref{constriant2}.
   
 This bound is conservative because we do not 
 allow a meta-stable vacuum.  
Note that we  do not take into account  higher loop corrections.  
 In the SM, the constraints become significantly weaker if higher loop corrections are taken into account \cite{Degrassi:2012ry, Zoller:2014cka}.

\subsubsection{Experimental constraints}

\subsubsection*{Constraints from the direct search for $W'$ and $Z'$}

Since the production cross sections
of the extra vector bosons are relatively large, this model is constrained from current results 
of the exotic resonance searches of various decay channels 
at the $\sqrt{s} = 8 \TeV$ LHC.
We take  account  of  the following constraints:  
$W' \to \ell \nu$ searches  \cite{ATLAS:2014wra, Khachatryan:2014tva}, $Z' \to \ell \ell$
searches \cite{Aad:2014cka,  Khachatryan:2014fba}, $V h$ resonance searches ($V' \to Vh$) 
\cite{Khachatryan:2015ywa, Aad:2015yza, CMS:2015gla, Khachatryan:2015bma}, and the
diboson searches ($V' \to VV$)
 using dijets \cite{Khachatryan:2014hpa, Aad:2015owa}, $\ell \ell jj$ \cite{Aad:2014xka}, $\ell \nu jj$ \cite{Aad:2015ufa}, and   $\ell \nu \ell \ell$ channel \cite{Aad:2014pha}.
The searches for the other channels do not constrain this model. 
We fill the excluded regions with 
blue in Figs.~\ref{fig:const} and \ref{constriant2}.
Among the constraints, $V' \to Vh$ and $V' \to WZ \to \ell \nu jj$
give severe bound, and exclude a part of the  parameter regions in which we can
explain the diboson excess reported by the ATLAS experiment.

\subsubsection*{Constraints from the electroweak precision measurements}

The electroweak precision parameters, $\hat{S},~\hat{T},~W$ and $Y$,
defined in Ref.~\cite{Barbieri:2004qk},  are
severely restricted from the electroweak precision observables. 
Since the interactions of $W'$ and $Z'$ to the light fermions affect 
the low energy observables, 
the light $W'$ and $Z'$ are severely constrained.  
They are calculated at the tree level in Ref.~\cite{Abe:2013jga}, 
\begin{align}
 \hat{S}
=&
 \frac{g_0^2 v_1^2 v_2^2}{g_1^2 (v_1^2 + v_2^2)^2 + g_0^2 v_1^4}
\simeq
\frac{m_W^2}{m_{W'}^2}
\left(
1 - \frac{v_3^2}{v^2}
\right)
,
\\ 
 \hat{T}
=&
0
,
\\ 
 W
=&
4 m_W^2 \frac{g_0^2}{g_1^2} \frac{1}{v_1^2 + v_2^2}
 \frac{v_1^4}{g_1^2 (v_1^2 + v_2^2)^2 + g_0^2 v_1^4}
\simeq
\frac{m_W^4}{m_{W'}^4}
\frac{1}{r^2}
\left(1 - \frac{v_3^2}{v^2} \right)
,
\\ 
 Y
=&
4 m_W^2 \frac{g_2^2}{g_1^2} \frac{1}{v_1^2 + v_2^2}
 \frac{v_2^4}{g_1^2 (v_1^2 + v_2^2)^2 + g_0^2 v_1^4}
\simeq
r^2 t_W^2
\frac{m_W^4}{m_{W'}^4}
\left(1 - \frac{v_3^2}{v^2} \right)
.
\end{align}
We
use them to find the constraint on the parameter space.
The parameter regions  constrained at 95\% C.L. are
filled with red in Figs.~\ref{fig:const} and \ref{constriant2}.
 The small $v_3$, the small $r$  as well as the light $W'$ regions are constrained.

We also consider constraints from flavor physics, and find they are
weaker than the constraints from $\hat{S}$ and/or the current LHC bound.
For example, $K^{0}$-$\bar{K}^{0}$ mixing in this model is almost the same
as that in the SM. This is because the contributions from $W'$ is
sufficiently suppressed due to the suppression of the couplings to the
SM fermions, and also the the modification of the $W$
couplings to the fermions is very small, ${\cal
O}(m_W^4/m_{W'}^4)$. Therefore, we do not show the constraints
in the figures.

\subsubsection{Summary of the constraints}
\begin{figure}[tp]
\begin{center}
 \vspace{-2cm}
 \subfigure[ ]{
  \hspace{-1.3cm}
\includegraphics[width =8.5cm, bb=0 0 289 204]{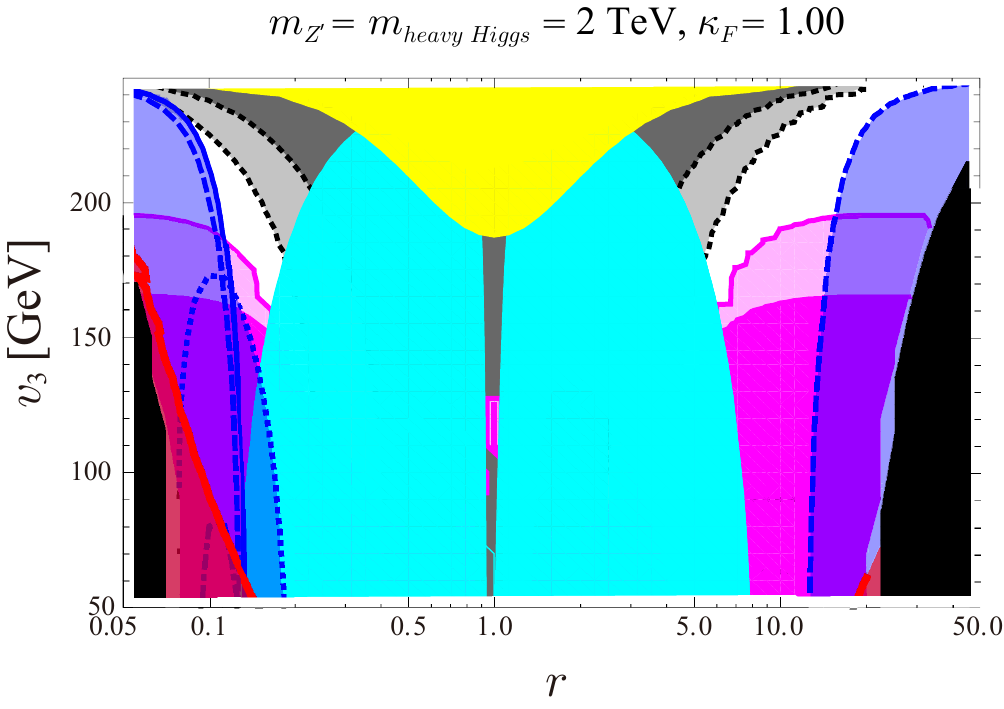}}\\
 \vspace{.2cm}
  \subfigure[ ]{
  \hspace{-1.3cm}
\includegraphics[width =8.5cm, bb=0 0 289 199]{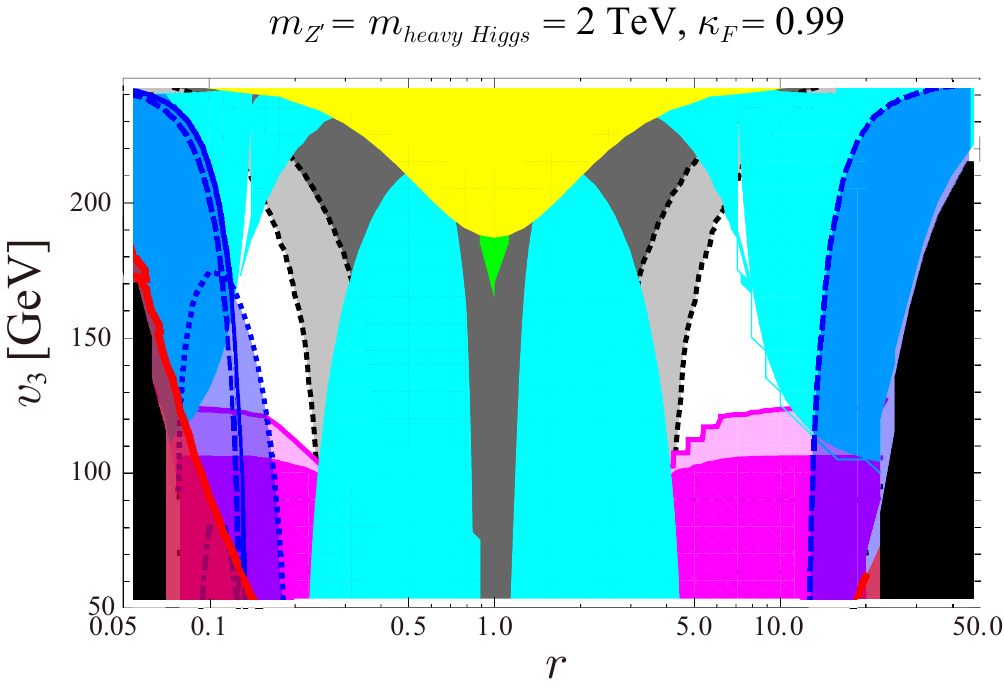}}\\
  \subfigure[ ]{
 \vspace{.2cm}
   \hspace{-1.3cm}
\includegraphics[width =8.5cm, bb=0 0 289 199]{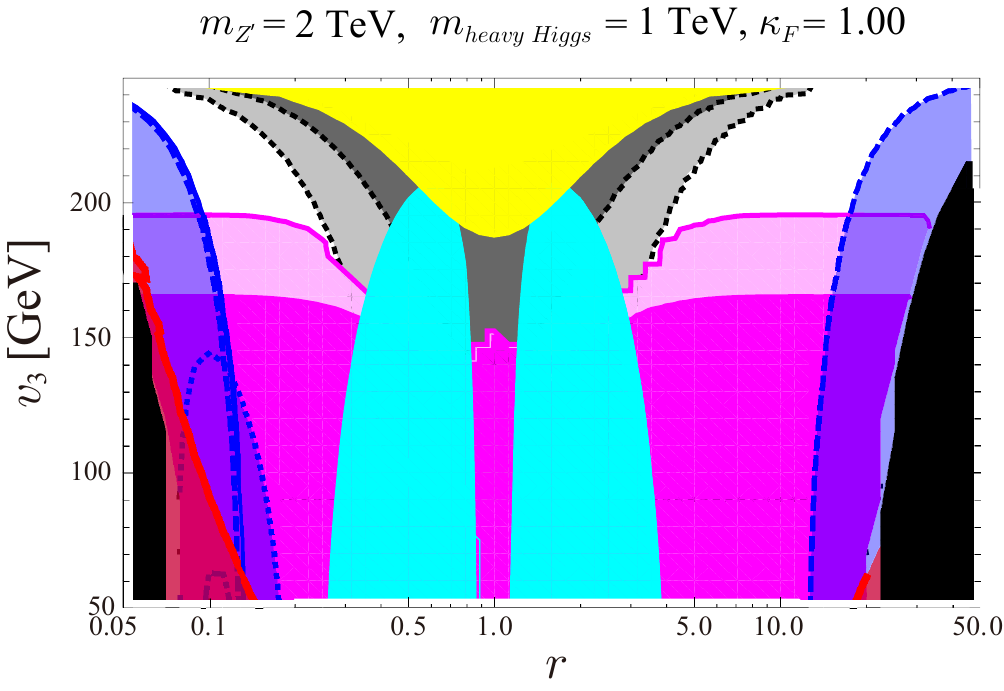}}
 \vspace{-.cm}
 \caption{The theoretical and experimental constraints in  $r$--$v_3$
 planes.
We take three different parameter choices for ($m_{Z'}$, $m_{A}$, $\kappa_F$).
The colored regions are constrained,
\textbf{Gray}: the perturbativity conditions for the Higgs quartic couplings,
\textbf{Yellow}: the perturbativity conditions for $g_1$,
\textbf{Cyan}: the bounded below condition,
\textbf{Green}: the global minimum vacuum condition,
\textbf{Magenta}: the stability condition,
\textbf{Blue}: the LHC bounds,
\textbf{Red}: constraints from the electroweak precision measurements,
\textbf{Black}: no physical solutions.
See also the explanations in the text.
}
 \label{fig:const}
\end{center}
\end{figure}
In Fig.~\ref{fig:const}, we plot all the constraints in $r$--$v_3$ planes
with three different parameter sets,
($m_{Z'}$, $m_{A}$, $\kappa_F$)
= (2~TeV, 2~TeV, 1.00), ~(2~TeV, 2~TeV, 0.99), and (2~TeV, 1~TeV, 1.00). 
Here we take all the heavy scalar bosons are degenerate, $m_A = m_{H'}= m_{H}$.
 The colored regions are excluded or constrained, 
 and  the white regions are allowed  from all constraints.
The gray regions surrounded by the black dotted lines represent the
 perturbativity condition of the Higgs quartic couplings
for $\Lambda = 10$ (darker) 
 and 100 TeV (lighter),  
and the yellow regions are that of $g_1$.
The bounded below  condition 
excludes the cyan region.
The global minimum vacuum condition excludes the green region.
The magenta regions  are excluded
 by the stability condition 
 for $\bar{\Lambda} = 10$ (darker) and 100 TeV (lighter). 
The LHC results exclude blue regions,
where the solid blue lines represent
 $W' \to \ell \nu$ searches
\cite{ATLAS:2014wra, Khachatryan:2014tva},  the dashed lines represent
 $Z' \to \ell \ell$ searches
\cite{Aad:2014cka,  Khachatryan:2014fba},  the dotted lines represent
$V h$
resonance searches \cite{Khachatryan:2015ywa, Aad:2015yza, CMS:2015gla, Khachatryan:2015bma},
and the dot-dashed lines 
represent diboson searches ($V' \to VV$)
\cite{Aad:2015owa, Khachatryan:2014hpa, Aad:2014xka, Aad:2015ufa, Aad:2014pha}.
The regions filled with the red color are excluded by the electroweak precision
measurements.  
No physical solutions are found  in the black region, namely the gauge
couplings and/or the VEVs become complex numbers there.

By comparing all the panels,
we find the experimental bounds (the LHC and the electroweak precision measurements)
are almost insensitive to the heavy Higgs bosons. 
The theoretical bounds are very sensitive to the $\kappa_F$, as we
can see from the panels (a) and (b).
This is because the deviation of  $\kappa_F$  from 1 leads to the large
$\lambda_3$  (cf. $\lambda_3 \sim 0.13$ (0.78) at  $\kappa_F = 1.00$
(0.99)) (see \eq{lambda3_solved}), which   
changes the regions excluded by the bounded below condition (see Eq.~(\ref{bounded})) and
stability condition (see Eq.~(\ref{stability})).
We find that  perturbativity  condition is weaker  in the panel (c)
$(m_A = 1$ TeV) compared with the panel (a) $(m_A = 2$ TeV).
This is because the lighter $m_A$  leads to the smaller quartic couplings.

\begin{figure}[p]
\begin{center}
  \subfigure[ ]{
  \hspace{-1.3cm}
\includegraphics[width =6cm, bb=0 0 263 265]{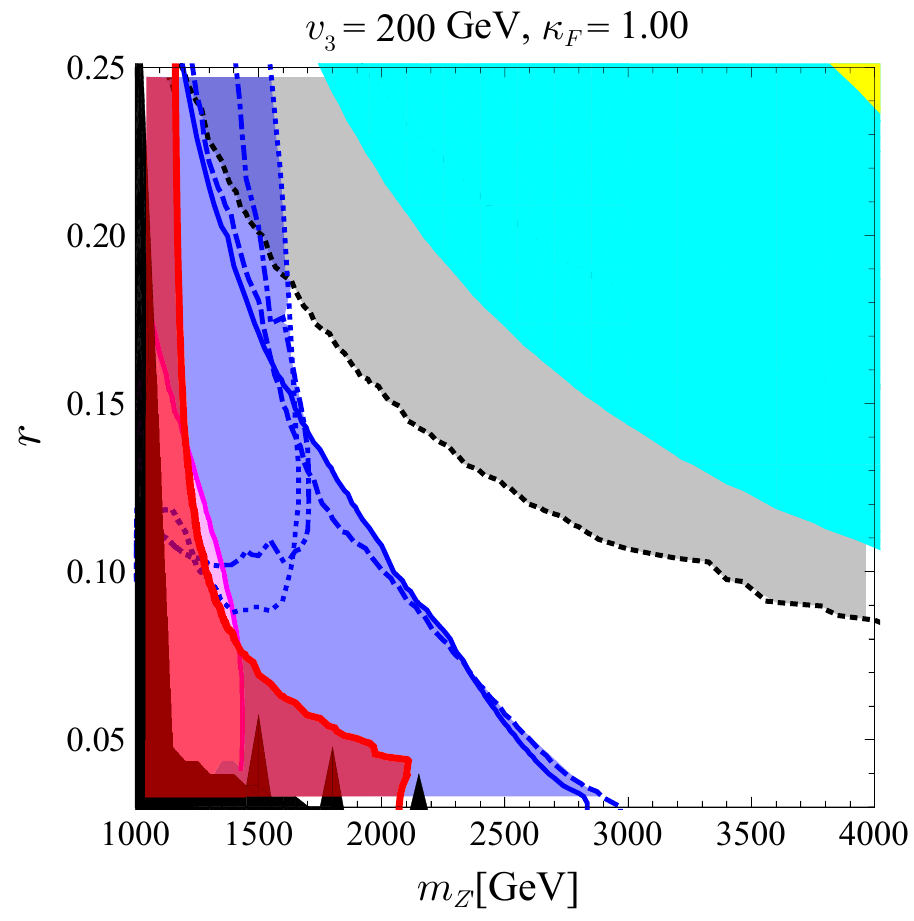}}~~~~~~~~~~
 \subfigure[ ]{
  \hspace{-1.3cm}
\includegraphics[width =6cm, bb=0 0 263 265]{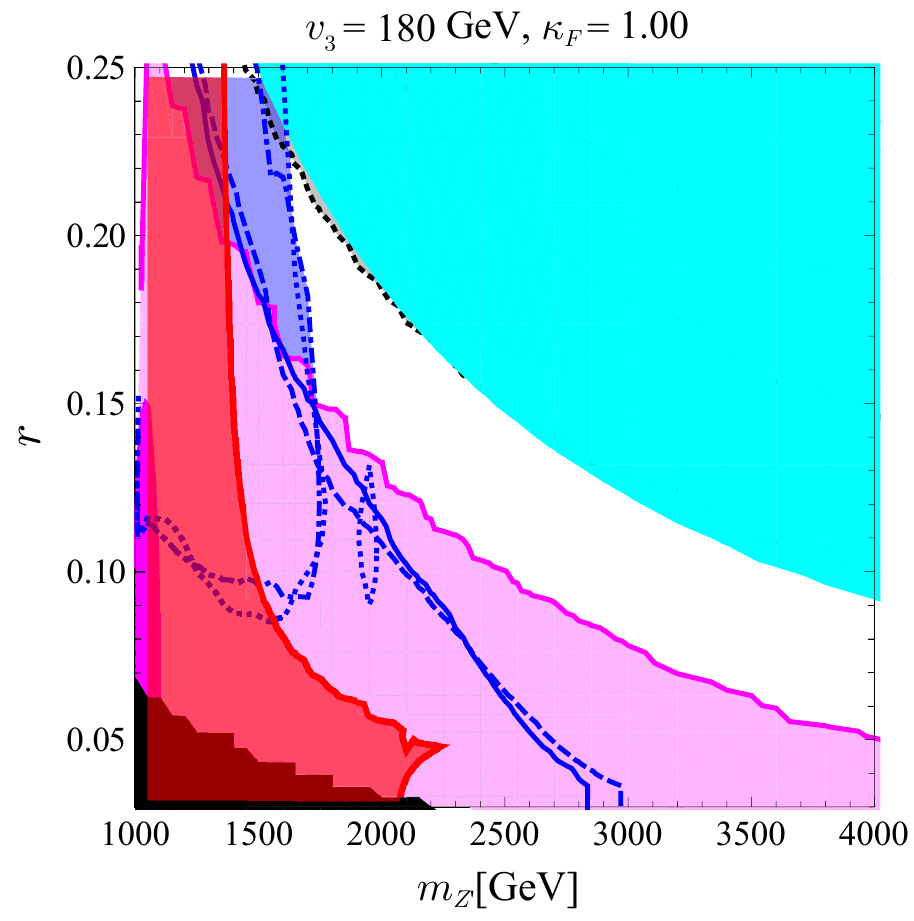}}\\
~~
\\
 \subfigure[ ]{
  \hspace{-1.3cm}
\includegraphics[width =6cm, bb=0 0 263 266]{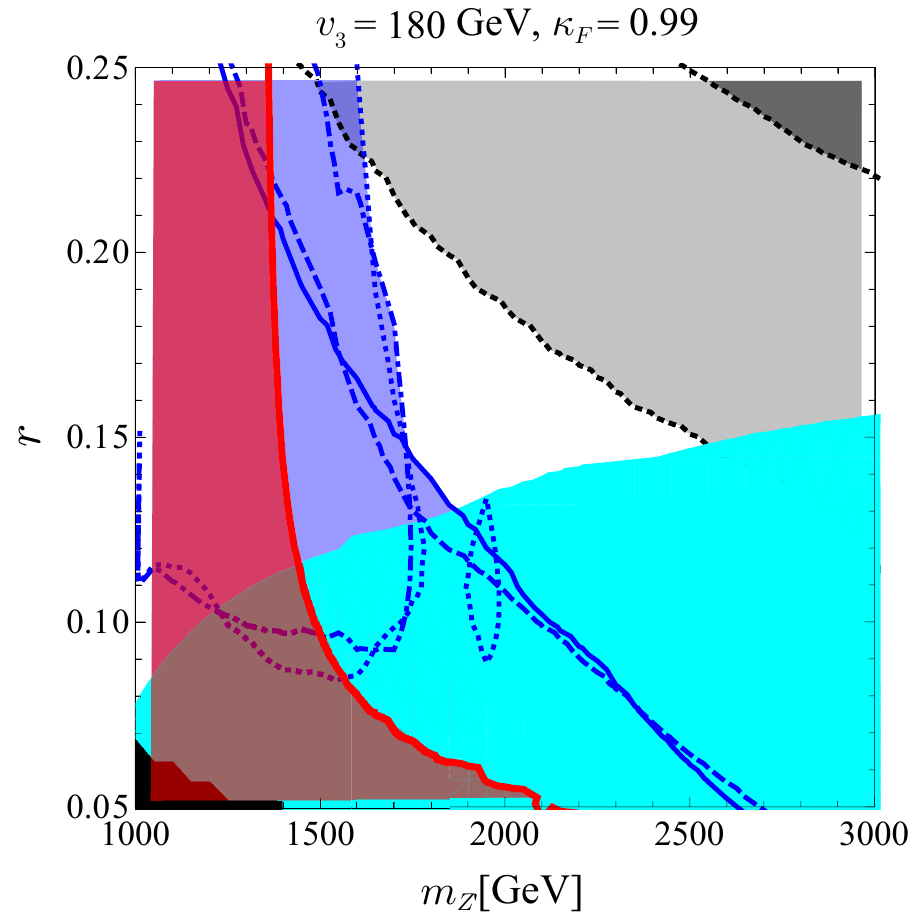}}
 \vspace{-.cm}
 \caption{The theoretical and experimental constraints in the
 $m_{Z'}$--$r$ planes.
 We take three different parameter choices for ($v_3$, $\kappa_F$), 
and the universal masses $m_{Z'} = m_A = m_{H'} = m_H$ are taken. A color
 notation is the same as in Fig.~\ref{fig:const}.
}
 \label{constriant2}
\end{center}
\end{figure}
Figure~\ref{constriant2} shows the constraints in $m_{Z'}$--$r$ plane
with the same color notation in Fig.~\ref{fig:const}.
We take the universal masses $m_{Z'} = m_A = m_{H'} = m_H$, 
and three different choices for $v_3$ and $\kappa_F$,
$(v_3, \kappa_F) = $(200~GeV, 1.00),
(180~GeV, 1.00), and (180~GeV, 0.99).
The perturbativity condition for the Higgs quartic couplings 
gives severe bounds in the heavy Higgs mass region. 
This is because the 
large mass differences in the CP-even Higgs mass spectra require the large Higgs quartic couplings,
and they become non-perturbative eventually. 
We also find in the panel (c)  that 
when  $\kappa_F$ deviates from $1$,  the  bounded below condition gives
stringent constraint.
This is because that the Higgs quartic couplings $\lambda_i$
are sensitive to  the small deviation of $\kappa_F$,
see Appendix~\ref{app:kappaz}.

\subsection{Current status: 8 TeV analyses}
\label{8TeVstatus}

In this subsection, we focus on the cross section times branching ratios of $V'$ at the LHC 8~TeV.
Inspired by the recent ATLAS diboson excess \cite{Aad:2015owa}, 
we concentrate on the case that the masses of the extra vector bosons are around 2 TeV.

\begin{figure}[tp]
\begin{center}
\includegraphics[width =10cm, bb=0 0 290 200]{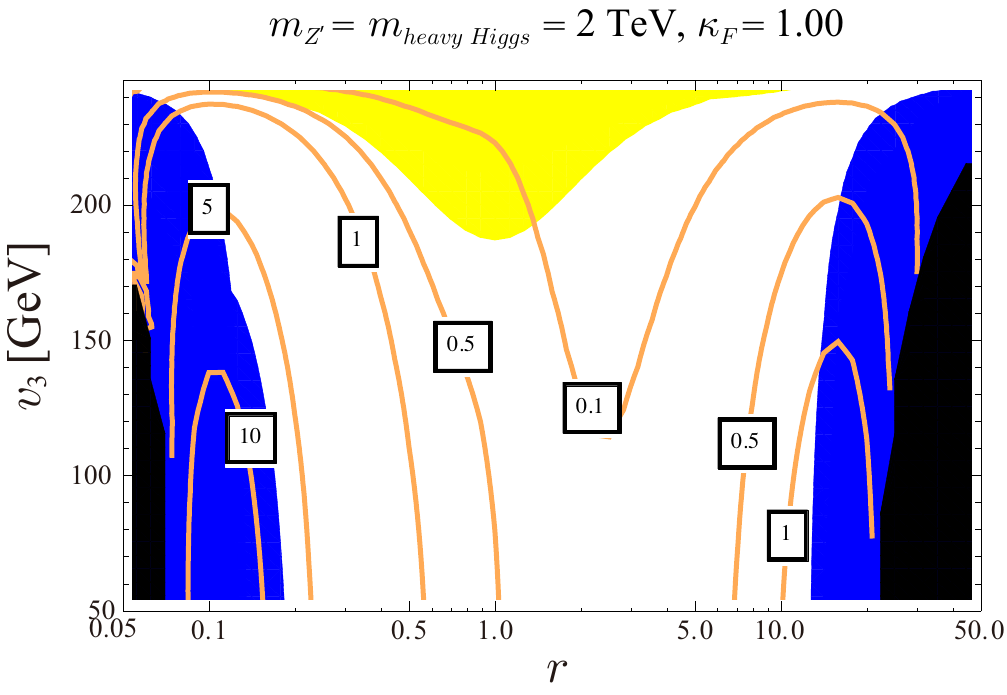}
 \vspace{-.cm}
\caption{ Contours of the cross section 
 $\sigma(pp \to W') \textrm{Br}(W' \to WZ) + \sigma(pp \to Z')
 \textrm{Br}(Z' \to WW)$ at the LHC
$\sqrt{s} = 8 \TeV$ in $fb$ unit. 
 We take $m_{Z'} = m_A = m_{H'} = m_H=$ 2 TeV, $\kappa_F = 1.00$. 
The blue regions are excluded by the current experimental bounds, and the
 regions of $g_1 (\mu = m_{Z'} ) >  4 \pi$ are filled with yellow. 
} 
\label{fig:8TeVprod}
\end{center}
\end{figure}
Since the separation of the $WW$, $WZ$, and $ZZ$ channels are not good and
there are significant overlap among them \cite{Aad:2015owa},
we investigate  the total cross section 
$\sigma(pp\to W') \textrm{Br}(W' \to WZ) 
+ \sigma(pp \to Z')\textrm{Br}(Z' \to WW)$ 
at $\sqrt{s} = 8 \TeV$  in Fig.~\ref{fig:8TeVprod}. 
In this figure,  $m_{Z'} = m_A = m_{H'} = m_H=$ 2 TeV, $\kappa_F = 1.00$ are taken.
The blue regions are excluded by the current experimental bounds
discussed in Sec.~\ref{sec:constraint}, 
and the regions of $g_1 (\mu = m_{Z'} ) >  4 \pi$ is filled with  yellow. 
Here we do not show the constraints from the Higgs sector.

We have to estimate the cross section value, $\sigma(pp \to V' \to VV)$,
required for the explanation of the diboson excess. 
We find $\sigma = 6$~fb with large error when we use the event numbers between 1.85 and 2.15~TeV bins, 
the estimated background,
and the efficiency given in
Ref.~\cite{Aad:2015owa}.
Hereafter we require $\sigma
= 6$~fb for the explanation of the diboson excess.

This cross section value can be achieved in the regions where $r$ is much smaller than 1.
This is because the production cross section is enhanced by   $r^{-2}$.
Hence, we focus on $r \ll 1$ regions in the rest of this paper.

Note that the strongest LHC constraint to the parameter regions where the ATLAS diboson excess can be explained comes from the hadronic channel of  $\sigma(pp \to V' \to 
Vh)$~\cite{Khachatryan:2015bma}, e.g. $\sigma(pp \to V' \to
Vh) \lesssim 7$ fb for  $m_{Z'} = 2$ TeV. 
This implies 
$\sigma (pp \to V' \to VV ) < 7$ fb  as we discussed in Sec.~\ref{sec31}.
If the coupling ratio $\kappa_Z$ deviates from 1, this LHC bound becomes weaker, because 
the relation between $\sigma (pp \to V' \to VV )$ and $\sigma (pp \to
 V' \to Vh)$ are modified.
However, in this model  $\kappa_Z$  is severely constrained close to 1
in the $m_A \gg m_h $ regime (see Appendix~\ref{app:kappaz}).

\begin{figure}[t]
\begin{center}
\includegraphics[width =5.1cm,bb=0 0 360 367]{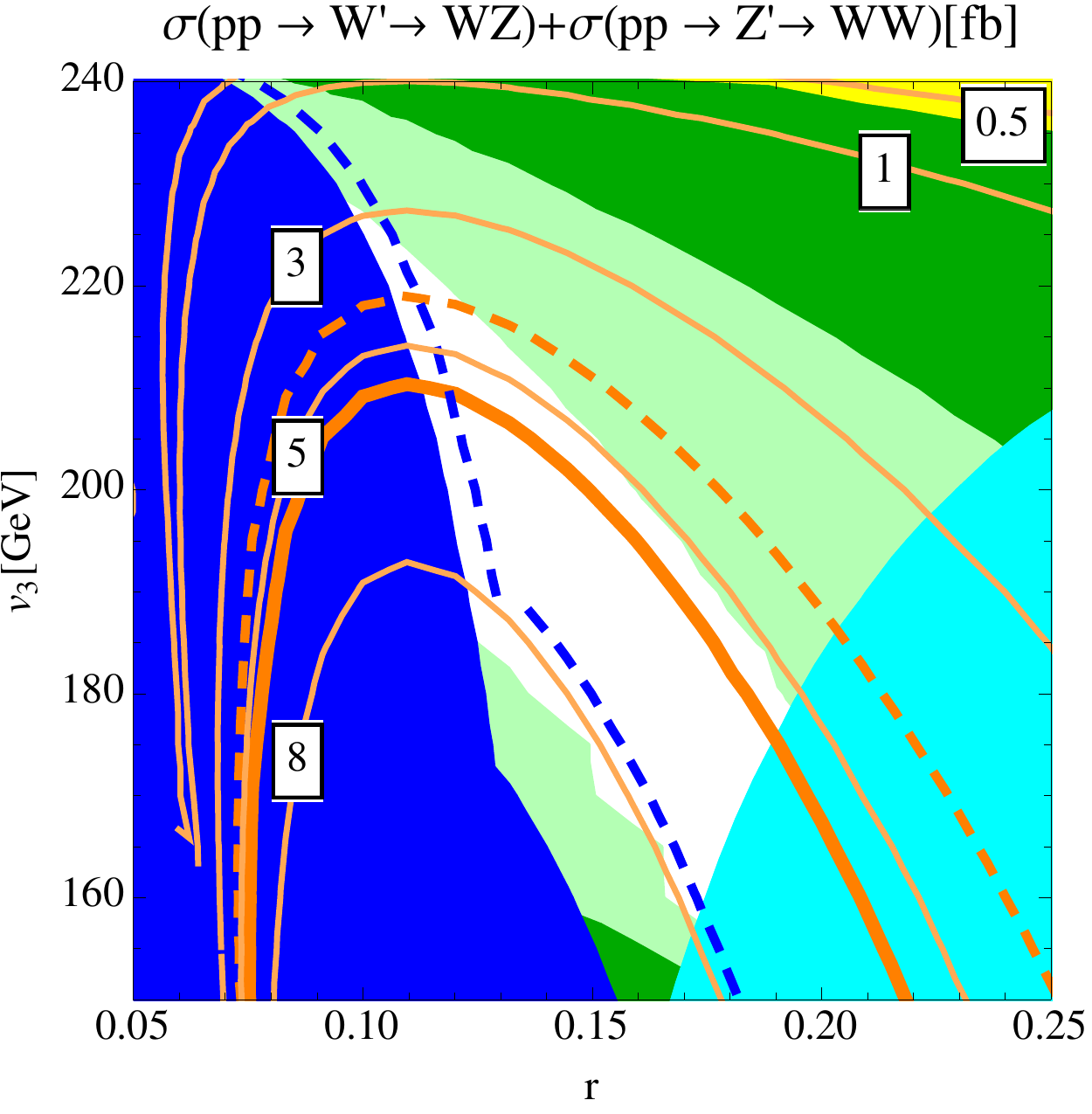}~~
\includegraphics[width =5.1cm,bb=0 0 360 367]{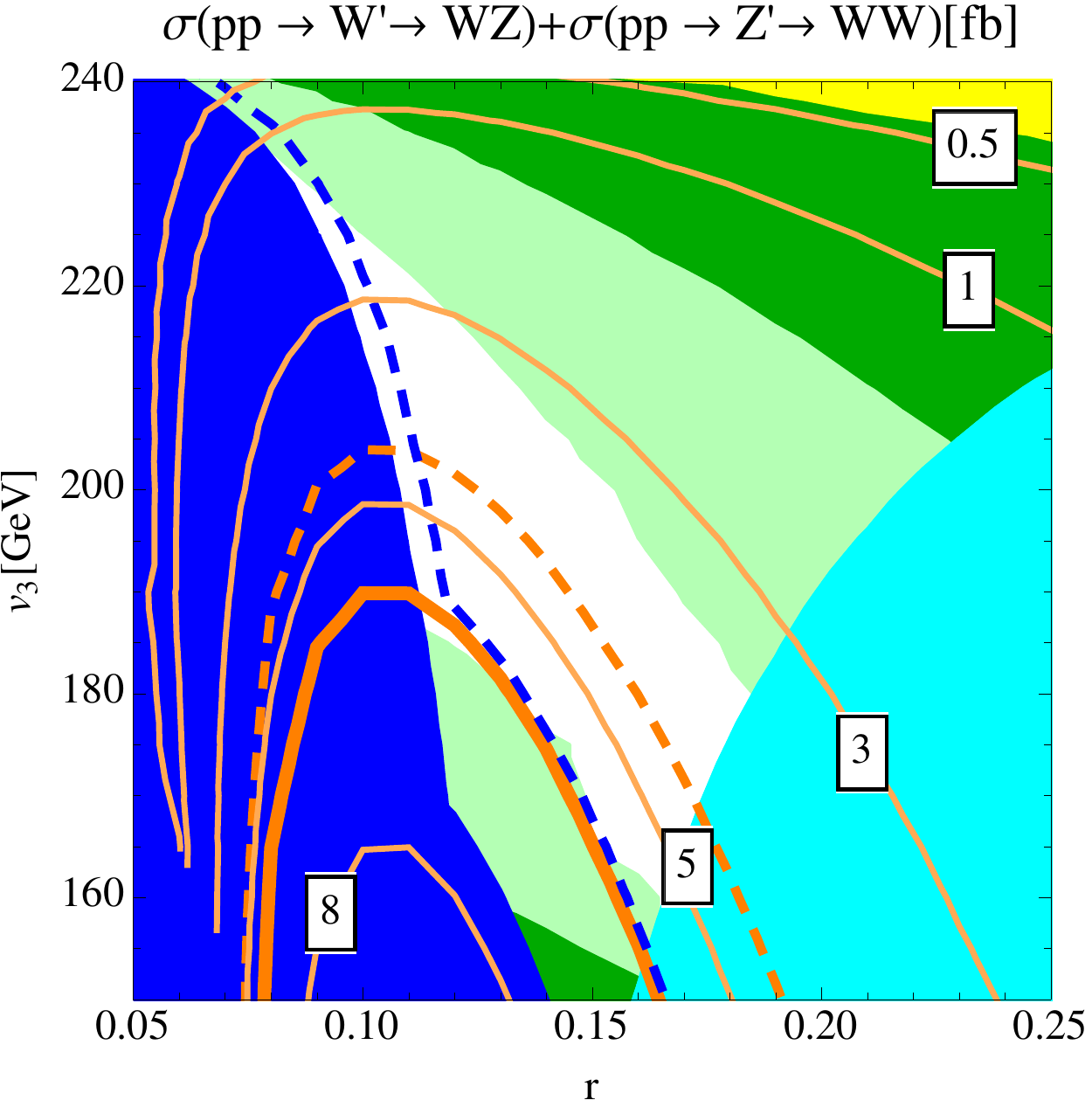}~~
\includegraphics[width =5.1cm,bb=0 0 360 367]{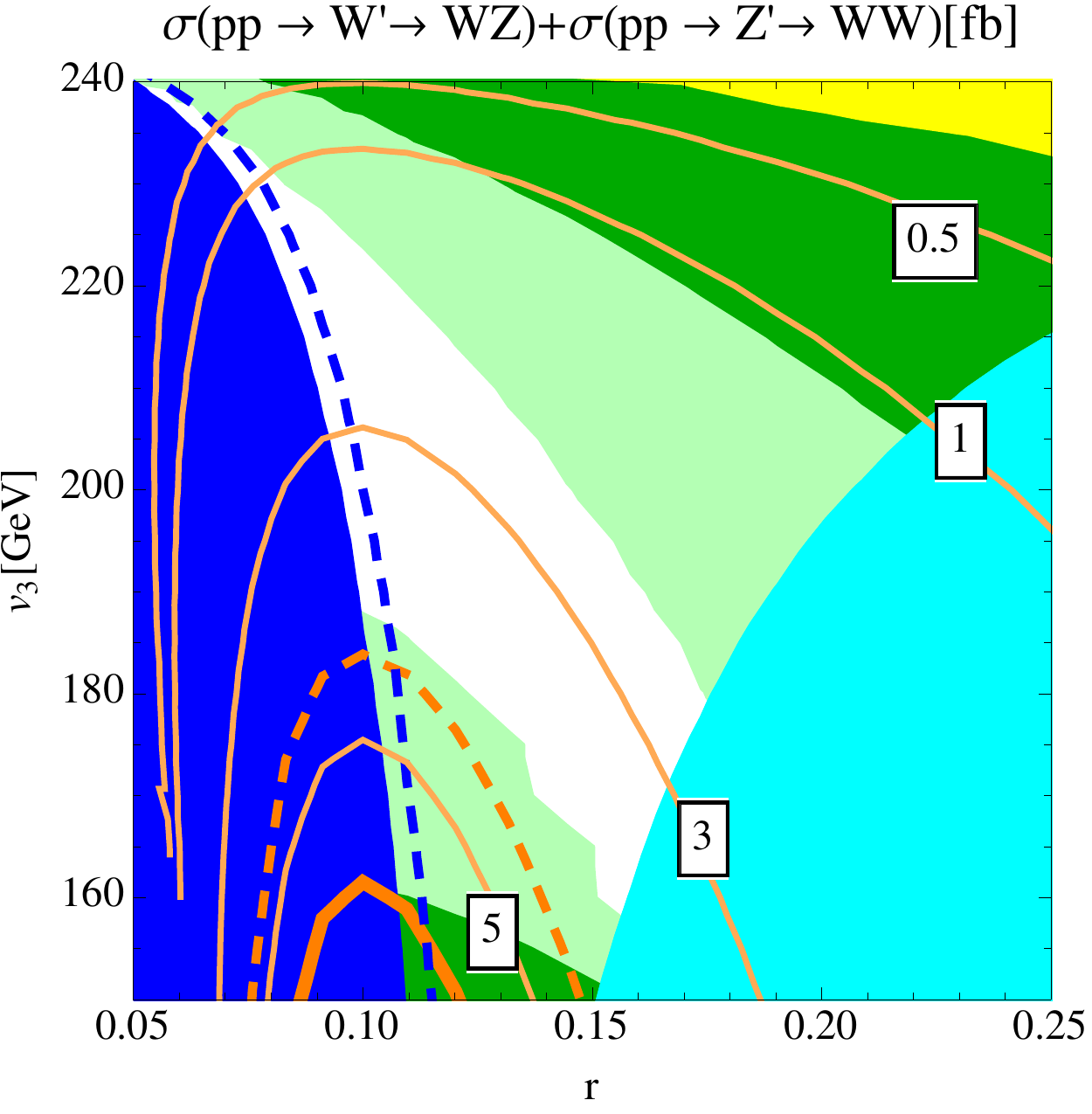}~~
 \vspace{.4cm}
 \caption{The cross sections of $W'$ and $Z'$ with two gauge bosons in
 their final states. We take $m_{Z'}$ = 1.9~TeV (the left panel), 2.0~TeV
 (the middle panel), and 2.1~TeV (the right panel). 
The color filled regions are excluded or constrained. 
The blue regions are excluded by the experimental bounds.
The regions of $g_1 (\mu = m_{Z'} ) >  4 \pi$ are represented by yellow.
The cyan regions are excluded by the bounded below condition. 
The green regions are constrained from the perturbativity and stability
conditions where their cut off  scale is 100 (10) TeV in the lighter
green (darker green) regions. 
Below the thick orange line  the cross section $\sigma(pp \to V' \to
VV)$ is consistent with the ATLAS result.
After taking account of the $K$-factor, 
the  boundaries of the regions  change into the dashed
orange lines, and  the regions below the dashed blue lines are excluded
 by the LHC bounds.
}
\label{fig:VV_1.9_2.0_2.1} 
\end{center}
\end{figure}
We show $\sigma(pp \to V' \to VV)$ in Fig.~\ref{fig:VV_1.9_2.0_2.1}, for
$m_{Z'}$ = 1.9, 2.0, and 2.1~TeV. We take $\kappa_F = 1.00$, and
all heavy scalar masses to be the same as $m_{Z'}$, $m_{Z'} = m_A = m_{H'}
= m_{H}$. The color filled regions are excluded or constrained.
The color notation is given in the caption.
We find that the cross section is sensitive to the mass. 
For example, by changing $m_{Z'} $ from  $ 1.9$ to $2 $ TeV ($5$ \% mass difference), 
the cross section is decreased by about 40\,\%.
This is because the PDF  rapidly changes in the heavier mass regions~(see~Fig.~\ref{fig:prop2}).

Below the thick orange line  the cross section $\sigma(pp \to V' \to VV)$ is consistent with the ATLAS result.
Here we apply the event selection efficiencies for the extended gauge
model (cf. $10$--$16$ \% at $m_{JJ} = $ 2 TeV~)~\cite{Aad:2015owa}.
Note that a part of these regions are constrained from  the stability
condition at $\bar{\Lambda} = 100$ TeV filled with light green. However,
once we take account of the higher order correction, these constraint
would be weaker as we discussed in Sec.~\ref{sec:cons}.
In the parameter regions shown in the figure,  $\sigma(pp\to W'\to
WZ)$/$\sigma(pp \to Z'\to WW) \simeq 2$. This is because the custodial
symmetry is enhanced in the small $r$ regime.

In the figure  we show the leading order (LO) production cross sections. 
Next-to-leading order and next-to-leading logarithmic (NLO$+$NLL)
corrections to the production cross sections of  $W'$ and $Z'$  are evaluated in
Ref~\cite{Fuks:2007gk,Jezo:2014wra}, and
the $K$-factor ($\sigma/\sigma_{\textrm{LO}}$) is about $1.3$.
This means that once we consider the QCD corrections,
the production cross sections
in the figures should be scaled by about 30 \%, and the LHC bounds
become severer, 
while the theoretical constraints do not change.
After taking account of the $K$-factor, 
the LHC diboson excess can be explained  for the regions below dashed
orange lines, and the regions below the dashed blue lines are excluded
by the LHC bounds.

\begin{figure}[tp]
\begin{center}
 \includegraphics[width =5.1cm,bb=0 0 360 367]{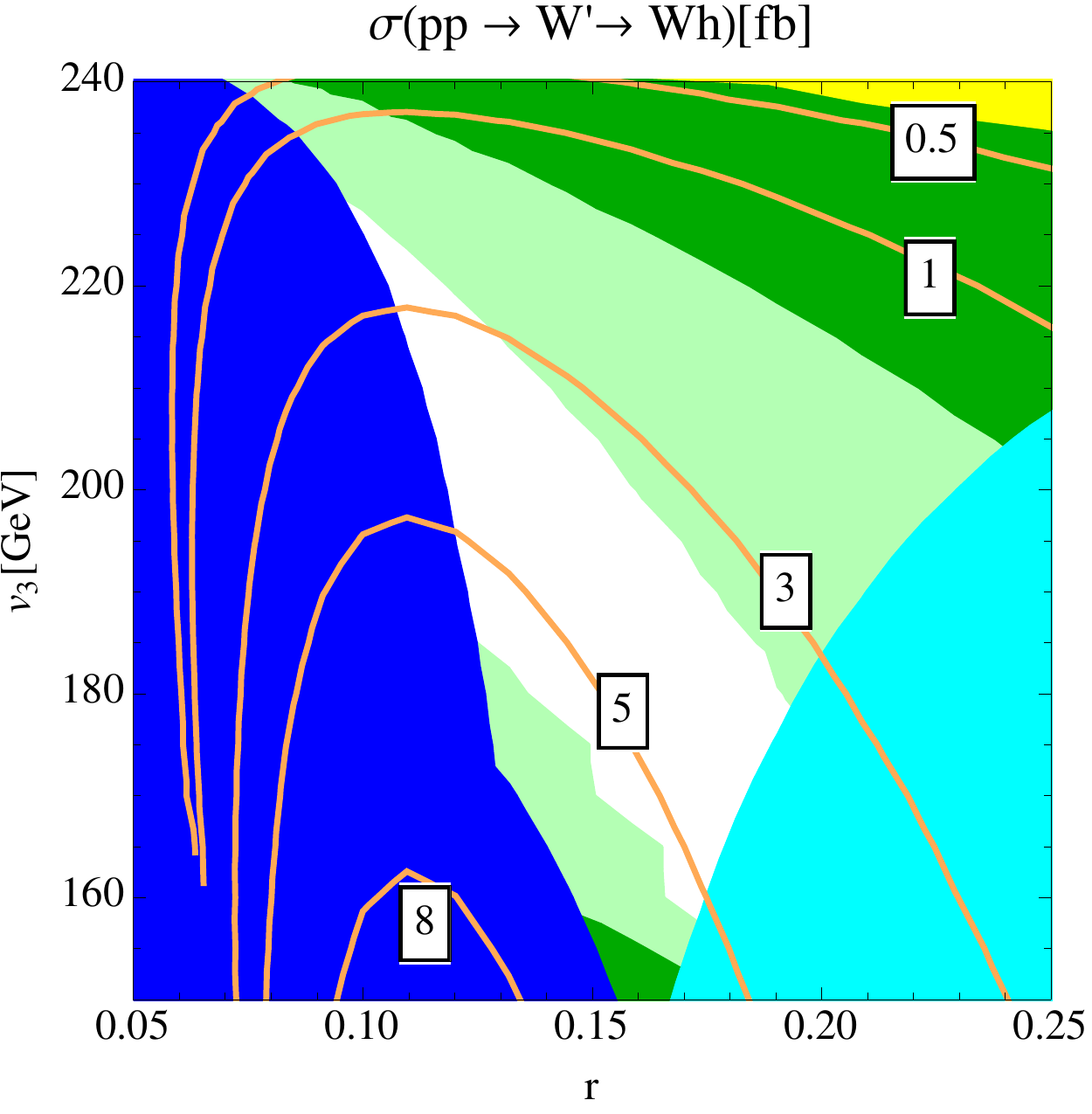}~~
 \includegraphics[width =5.1cm,bb=0 0 360 367]{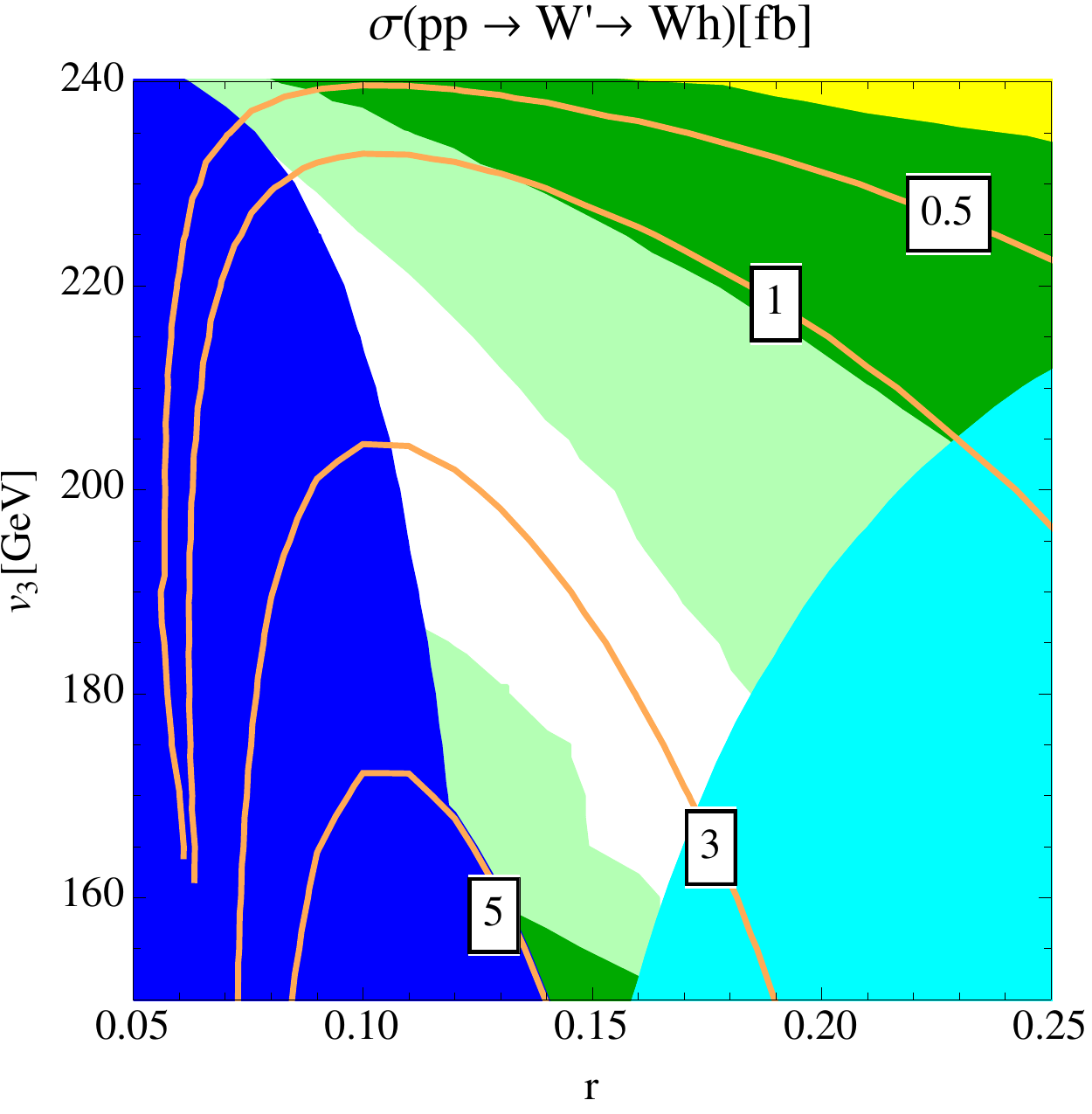}~~
 \includegraphics[width =5.1cm,bb=0 0 360 367]{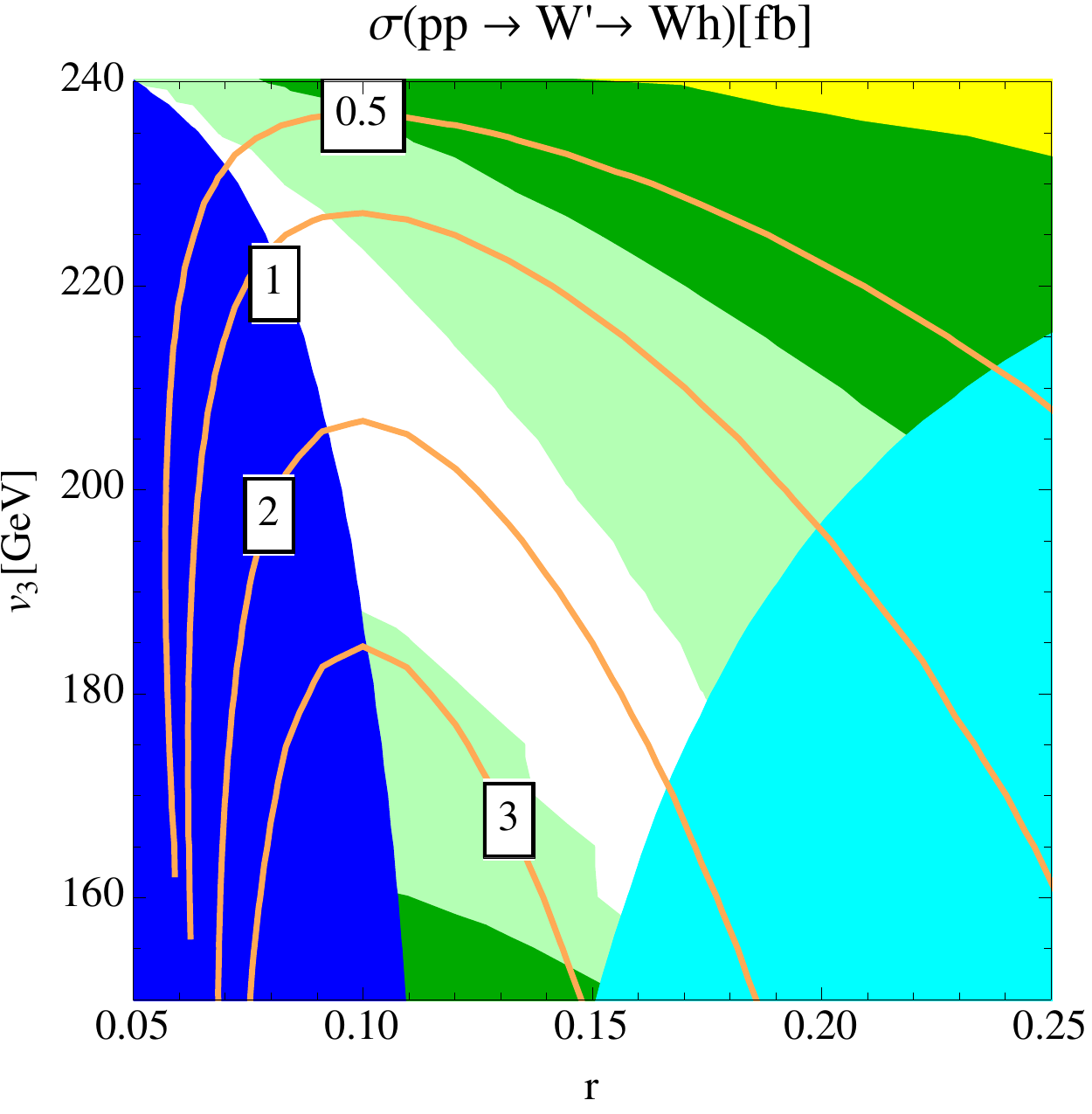}~~
 \vspace{.4cm}
 \caption{The cross sections of the diboson channels with $h$.
 We take $m_{W'} \simeq$  1.9~TeV (the left panel), 2.0~TeV
 (the middle panel), and 2.1~TeV (the right panel). 
The colored regions are  experimental and theoretical constraints. 
}
\label{fig:Vh_1.9_2.0_2.1} 
\end{center}
\end{figure}
Figure~\ref{fig:Vh_1.9_2.0_2.1} shows $\sigma(pp \to W' \to
Wh)$.\footnote{Detailed studies of this process are found in
Refs.~\cite{Zerwekh:2005wh, Hernandez:2015nga}.}
The parameter choices and the color notations are the same as
Fig.~\ref{fig:VV_1.9_2.0_2.1}. We find that $\sigma(pp \to W'\to
Wh)$/$\sigma( pp \to Z'\to Zh)
\simeq 2$ due to the enhancement of the custodial symmetry.
We find  $\sigma(pp \to W' \to
Wh)$ $\times 20$ fb$^{-1}$ $\times$  Br($h\to b\bar{b})$  Br($W\to e \nu
+ \mu \nu)$  $\sim$ 9 events
with $\sigma \sim 4$ fb in the regions where the ATLAS diboson excess can be explained.
This is consistent with the excess of the event for the $1.8$--$1.9$ TeV bins 
at the CMS with a local significance of 2.2$\sigma$ for $W' \to Wh \to \ell \nu
b\bar{b}$ search \cite{CMS:2015gla}, 
although the detail of the event selection has not been reported.

\begin{figure}[tb]
\begin{center}
\includegraphics[width =5.2cm,bb=0 0 360 364]{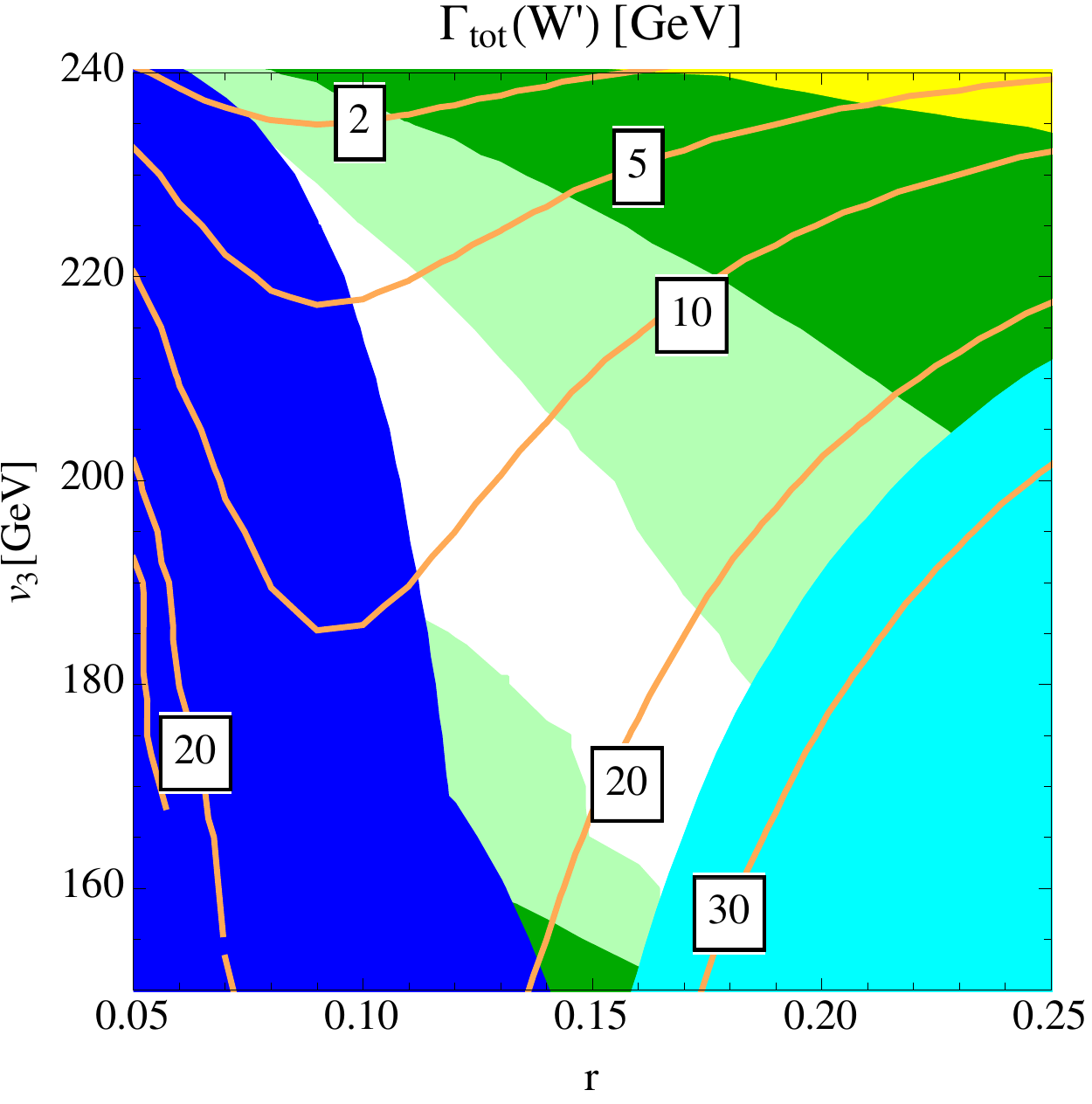}~~~~
\includegraphics[width =5.2cm,bb=0 0 360 364]{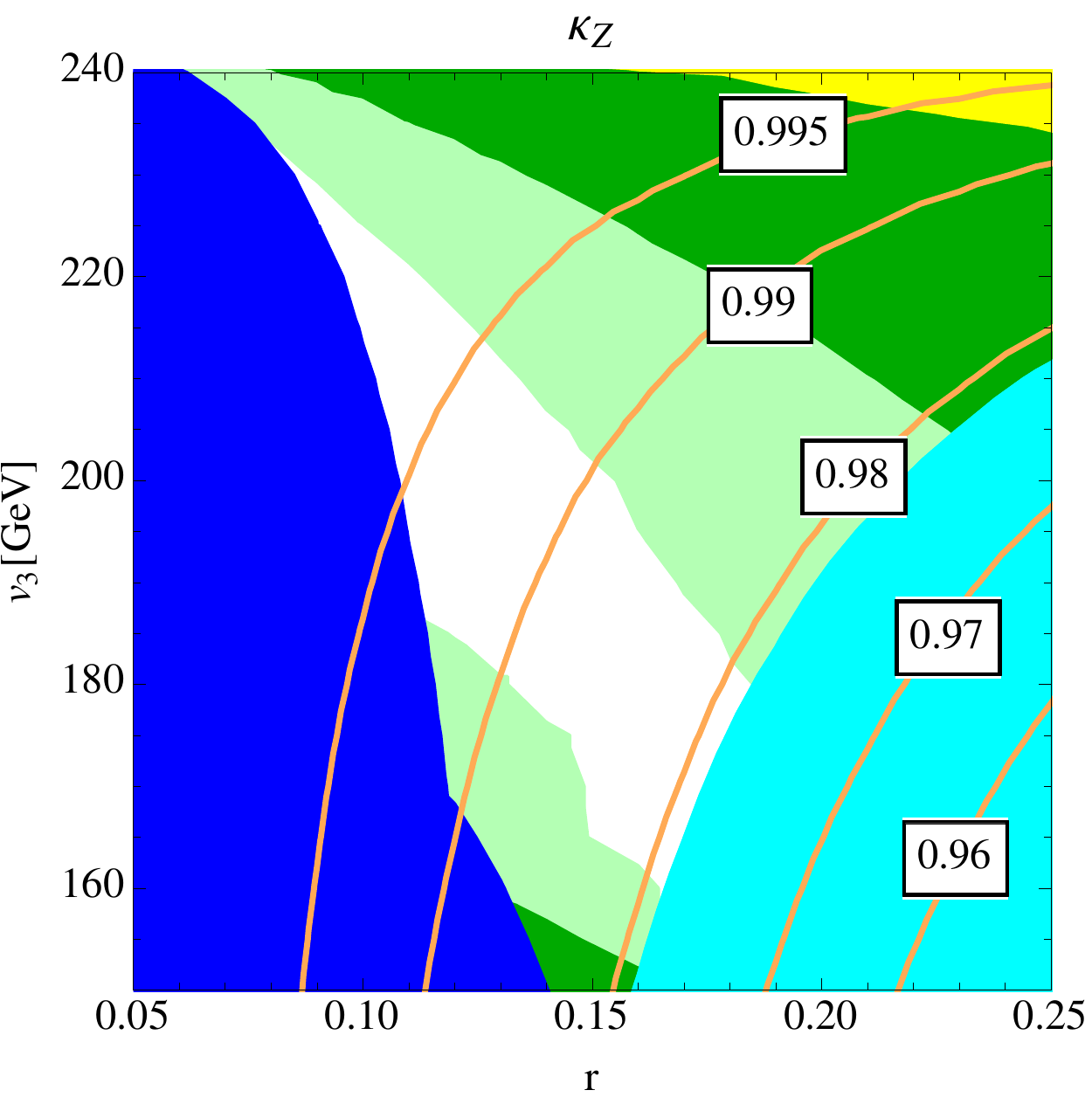}
\\
\includegraphics[width =5.2cm,bb=0 0 360 369]{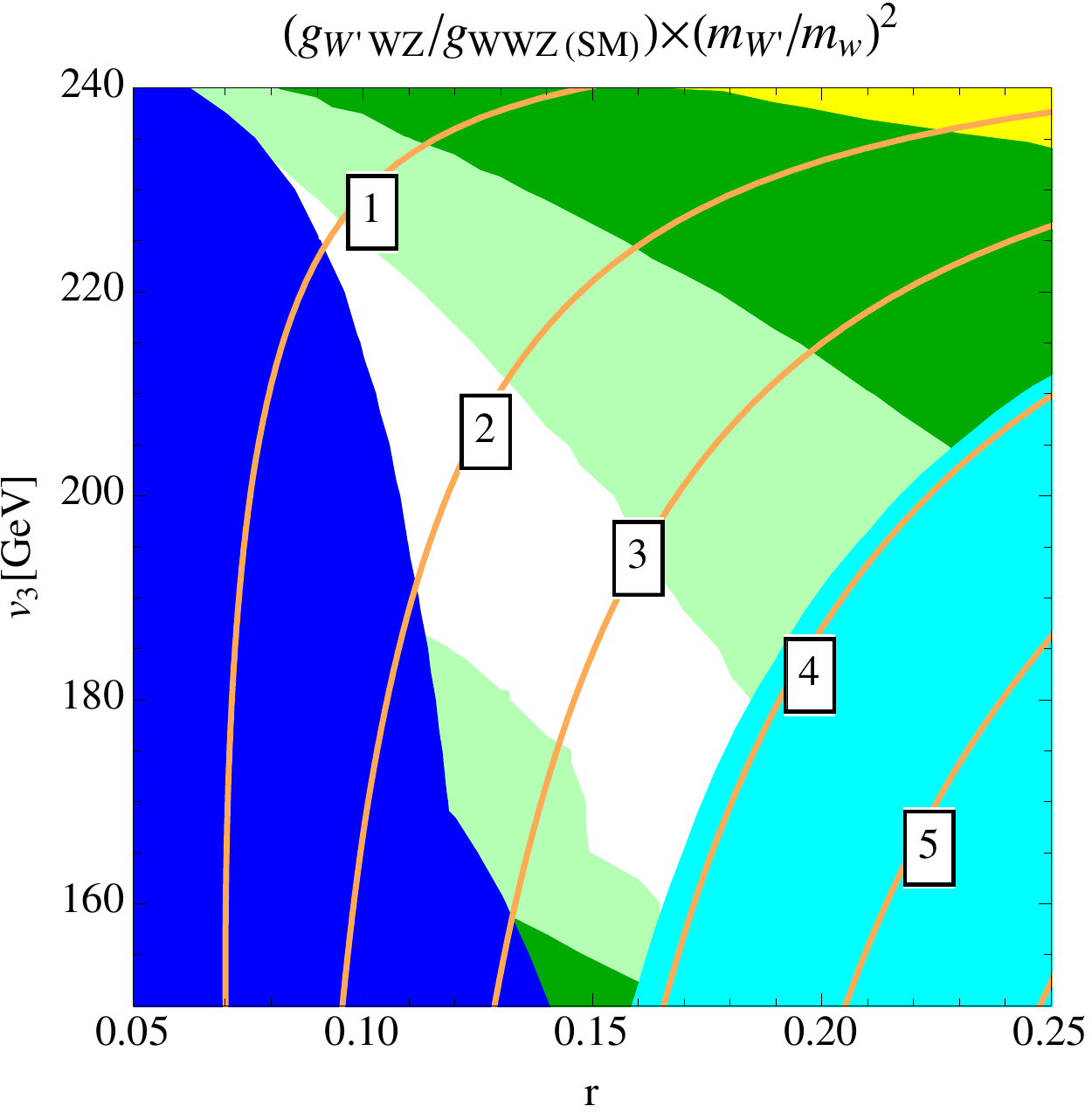}~~~~
\includegraphics[width =5.2cm,bb=0 0 360 367]{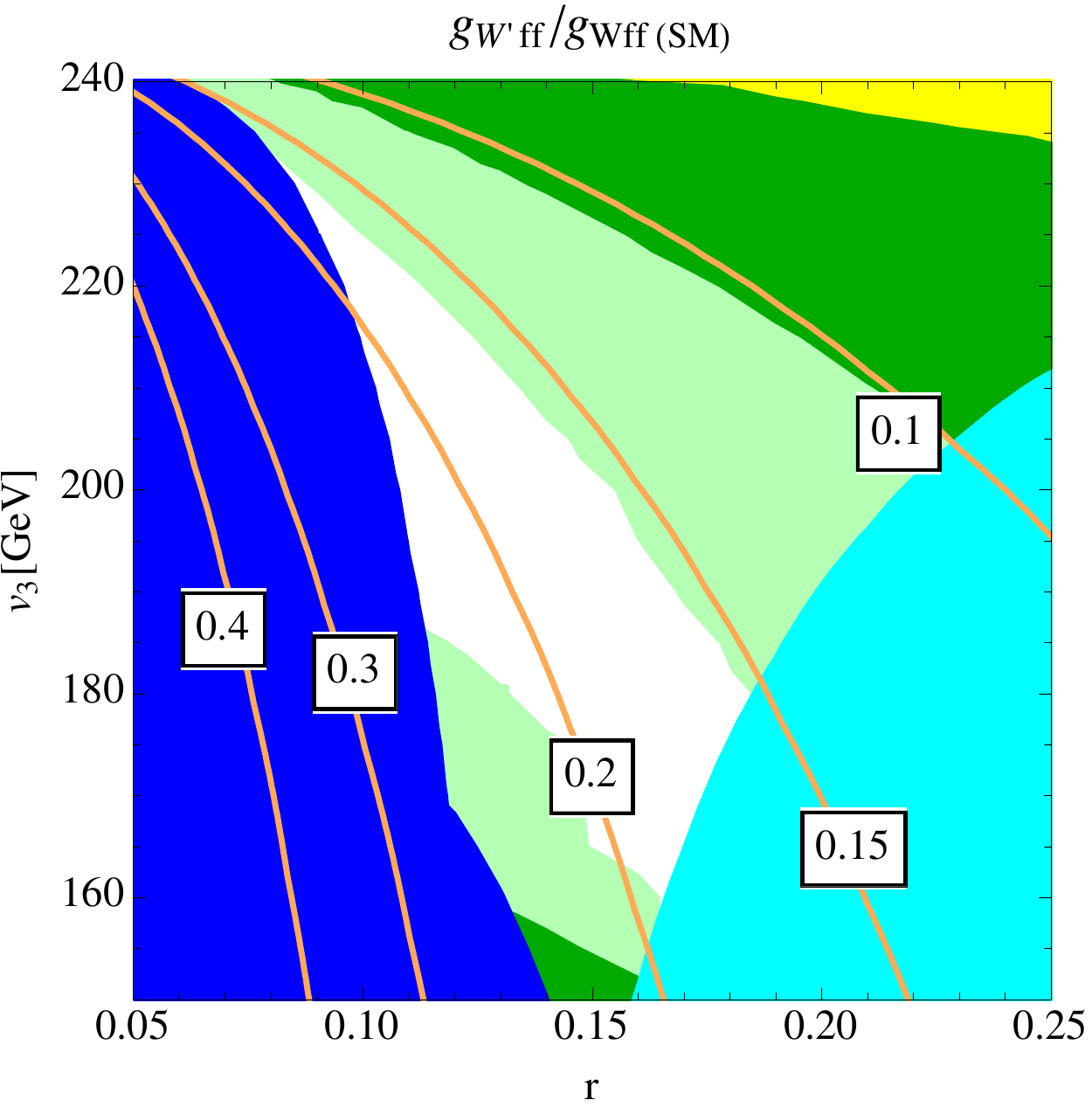}
 \vspace{.4cm}
 \caption{
$\Gamma_{\text{tot}}(W')$, $\kappa_Z$, 
$(g_{W'WZ} /  g_{WWZ}^{\textrm{SM}}) \times (m_{W'} /m_{W})^2$, 
and 
$g_{W'ff}/ g_{Wff}^{\textrm{SM}}$ 
for $m_{Z'} = 2$~TeV. 
They are insensitive to small difference of $m_{Z'}$.
The color notations are the same as in Fig.~\ref{fig:VV_1.9_2.0_2.1}. 
}
\label{fig:obs_1.9_2.0_2.1} 
\end{center}
\end{figure}
We show 
$\Gamma_{\textrm{tot}} (W')$, 
$\kappa_Z$,
$(g_{W'WZ} / g_{WWZ}^{\textrm{SM}}) \times (m_{W'} /m_{W})^2$
and 
$g_{W'ff}/ g_{Wff}^{\textrm{SM}}$ 
in Fig.~\ref{fig:obs_1.9_2.0_2.1}.
We fix $m_{Z'}=2$~TeV 
because they are not sensitive to the $Z'$ mass.
The choices of other parameters and the color notations are
the same as in Fig.~\ref{fig:VV_1.9_2.0_2.1}. 
The width of $W'$ is shown in the top-left panel in the figure. 
We find that it is narrow and is less than 1 \% of its mass,
 because the $W'$ couplings to the SM particles are suppressed by
powers of $m_{W}/m_{W'}$ and the
decay into the heavy scalars are suppressed kinematically.
The observable related to the Higgs couplings $\kappa_Z$
defined in Eq.~(\ref{eq:def_of_kappa})
is shown in the top-right panels.
The deviation from the SM prediction is small and the model is
consistent with the current LHC data \cite{Khachatryan:2014jba, ATLASHiggs}.
Since the International Linear Collider (ILC) can measure the
$\kappa_Z$ at 1\% level \cite{Asner:2013psa}, 
some parameter points are within the reach of the proposed ILC.
We show $g_{W'WZ}/g_{WWZ}^{\text{SM}}$ and
$g_{W'ff}/g_{Wff}^{\text{SM}}$ at the lower panels in the figure. In the
benchmark model used in Ref.~\cite{Aad:2015owa},
$g_{W'WZ}/g_{WWZ}^{\text{SM}} = m_W^2/m_{W'}^2$ and 
$g_{W'ff}/g_{Wff}^{\text{SM}} = 1$. 
In our model, we find,  due to the extra suppression by
$\sqrt{1 - v_3^2/v^2}$ and small $r$, $g_{W'WZ}/g_{WWZ}^{\text{SM}}$ is
numerically the same order as the
benchmark model although its $m_{W'}$ dependence is $m_W/m_{W'}$ (see
Eq.~(\ref{eq:WpWZcoupling})).
On the other hand, $g_{W'ff}/g_{Wff}^{\text{SM}}$ is about 10\% of the
benchmark model.

\section{MC simulation of $W' \to WZ$ at $\sqrt{s} = 13$ TeV }
\label{sec:LHC}

In this section, we perform a collider simulation of  $pp \to W' \to WZ$
at  $\sqrt{s} = 13$ TeV.
To study a discovery potential, we generate both QCD dijet background and $pp
\to W' \to WZ$ signal events.

We generate $1.73 \times 10^6$ QCD dijet events as the dominant
background by using {\sc Pythia 8.205}  \cite{Sjostrand:2007gs} 
with the generation cut so that the parton-parton center of mass energy 
 must exceed 1 TeV and $p_T > 400$ GeV  at $\sqrt{s}=13$ TeV. 
The tree level production  cross section is $350$ pb. 
Our sample therefore corresponds to roughly $ \int dt \mathcal{L} = $5 fb$^{-1}$. 
We use the Tune 4C for fragmentation and
hadronization \cite{Corke:2010yf}. 
We also generate $10^4$ signal events ($pp \to W' (W^{'+} +W^{'-})  \to
WZ$ ) for the mass between 1800  
and
3200 GeV, and take $\Gamma_{\textrm{tot}} (W')$ to be $25$ GeV in the simulation.
Note that the total width of $W'$ is less than about $30$ GeV 
in the allowed region of this model  (see Fig. \ref{fig:obs_1.9_2.0_2.1}).
We also generate the signal and background  events
at $\sqrt{s } =$ 8 TeV 
and compare them with the ATLAS plots~\cite{Aad:2015owa}.

The simple detector simulator {\sc Delphes3}  \cite{deFavereau:2013fsa} 
 is modified using {\sc FastJet3}~\cite{Cacciari:2005hq,Cacciari:2011ma}
 so that the mass drop and the grooming cuts used in
 the ATLAS study can be applied to the jets.
 We apply the cluster track matching algorithm  of {\sc Delphes3} so that
information of tracks inside  jets can be used,  otherwise the default ATLAS card is used.

Reconstruction of boosted objects
using jet substructure was originally proposed in Refs.~\cite{Butterworth:2008iy,Cacciari:2008gd}.  
See recent developments  in Refs.~\cite{Abdesselam:2010pt,Altheimer:2013yza}. 
In our simulation, we closely follow the ATLAS analysis. 
The Cambridge-Aachen algorithm with $R=1.2$ is used
\cite{Dokshitzer:1997in, Wobisch:1998wt}
for the jet clustering.
Then, $p_{T_1}> 600$~GeV, $p_{T_2}>540$~GeV, $ (p_{T_1}-p_{T_2})/(p_{T_1}+p_{T_2})<0.15$,
$\vert y_1-y_2 \vert <1.2$, $\vert \eta_1\vert < 2$ and $\vert
\eta_2\vert < 2$ are required for the jets. 
In addition, we require $E_{Tmiss}<350$ GeV, and  veto events with isolated electrons and muons with $p_T>20$ GeV.
For each jet, the pair of the subjets which satisfies
the subjet momentum balance criteria $\sqrt{y}>\sqrt{y_f}=0.45$ are selected, where
\begin{equation}
\sqrt{y}=\textrm{min}(p_{T_{j1}}, p_{T_{j2}})  \frac{\Delta R_{(j1, j2)}}{m_0}.
\end{equation}
\begin{figure}[t]
\begin{center}
\includegraphics[width =7cm,bb=0 0 468 324]{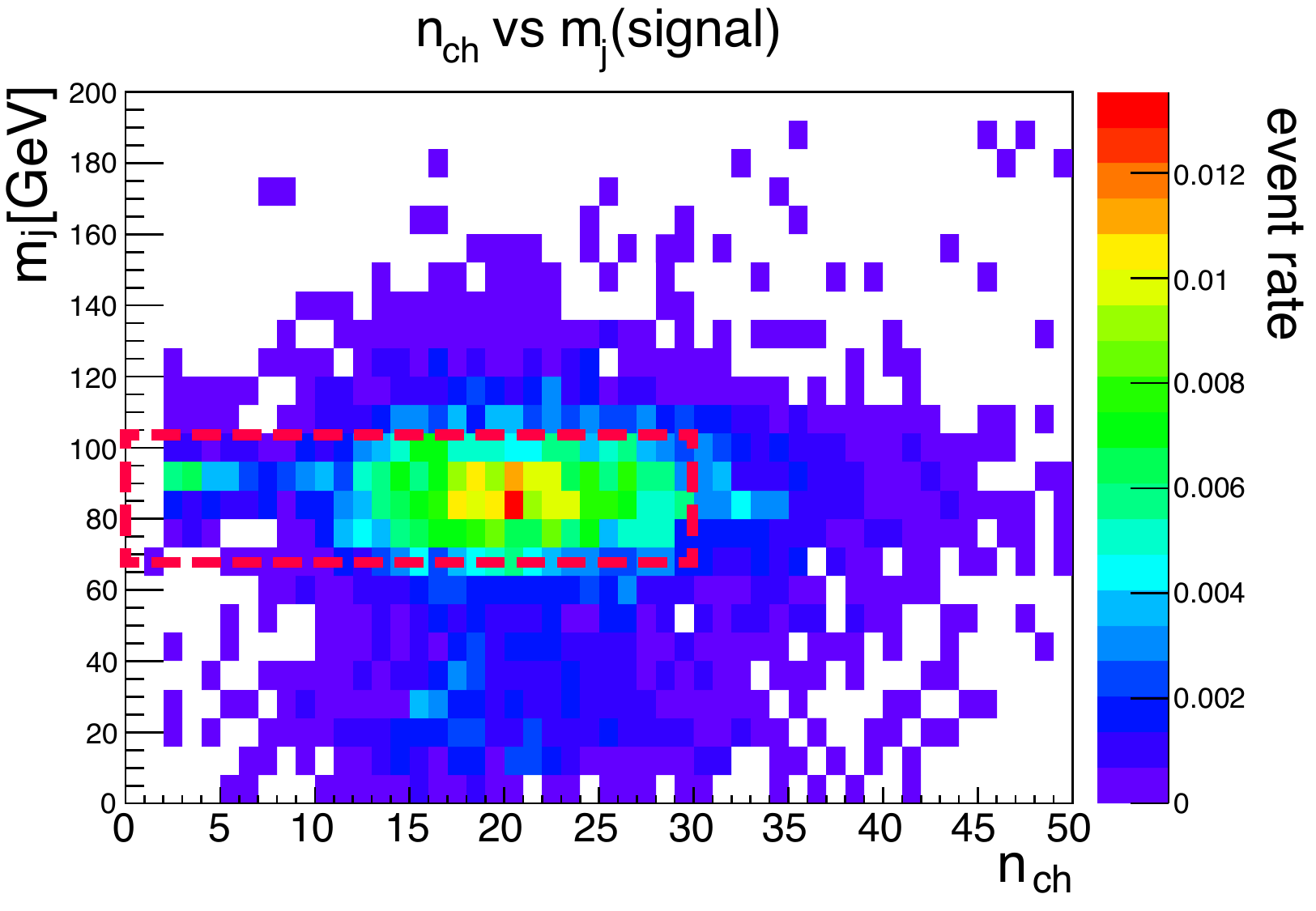}~~
\includegraphics[width =7cm,bb=0 0 468 328]{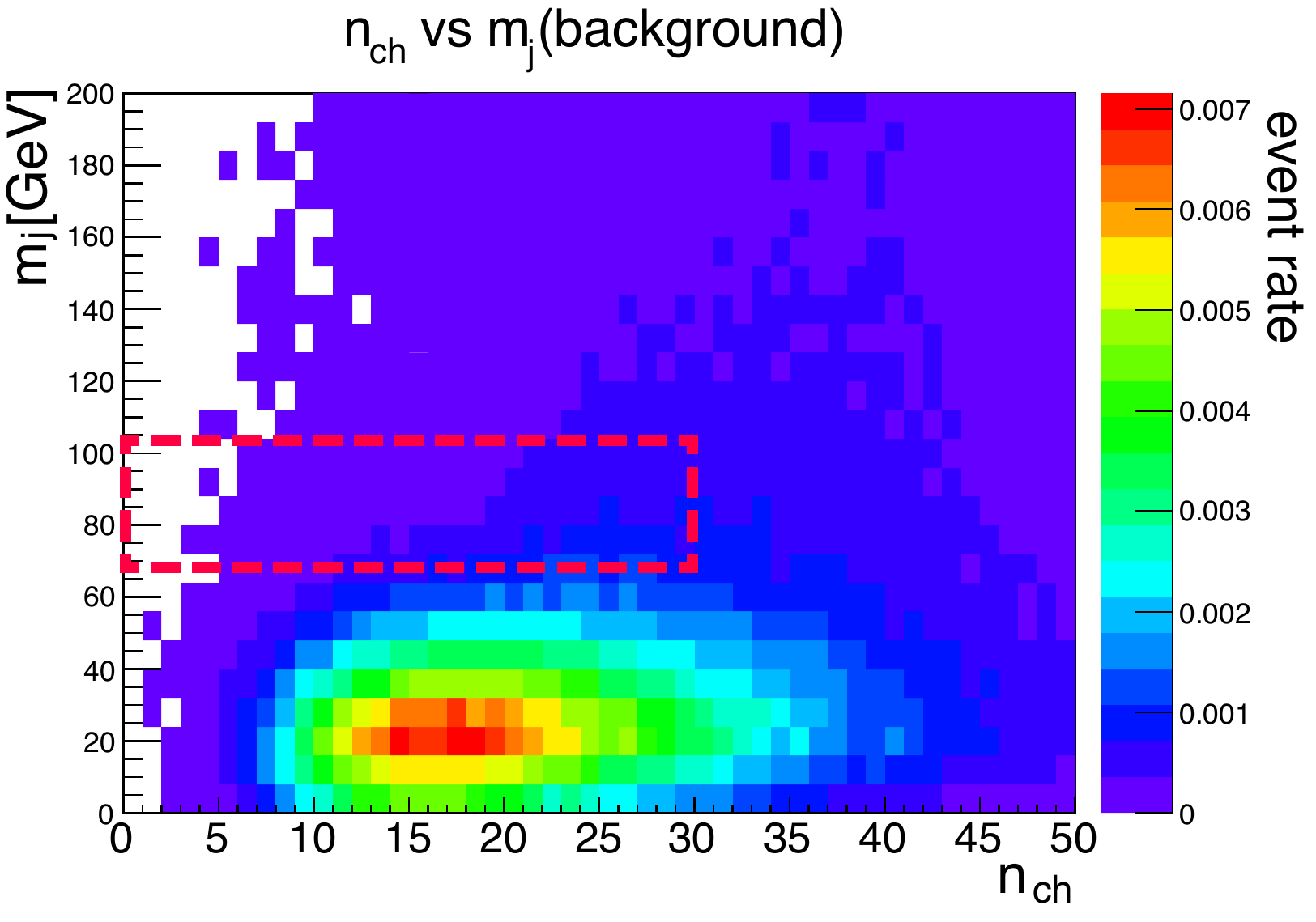}
 \vspace{.2cm}
\caption{
The distributions of  the signal ($m_{W'} = 1800 $ GeV and $\Gamma_{\textrm{tot}} (W') = 25$ GeV) 
and the dijet background in $n_{\textrm{ch}}$ and $m_j$ plane.
All the cut except for $n_{ch} $ and $m_j$ are applied, and we required   $m_{jj} >  1500$ GeV.  The ATLAS signal regions are marked by squares.   
}
\label{fig:charged1}
\end{center}
\end{figure}
Here, $p_{T_{j1}}$ and $p_{T_{j2}}$ are the transverse
momenta of subjets $j_1$ and $j_2$, $\Delta R_{(j1, j2)}$
 is the distance between the subjets $j_1$ and $ j_2$, and $m_0$
 is the mass of the parent jet.
Then the constituents of the selected pair of subjets are filtered. 
Namely the constituents are clustered with the radius parameter $R=0.3$, and  up to the highest 3 jets are taken to calculate the groomed jet mass and momentum. 
We require $\vert m_V-m_j\vert < 13$ GeV,
where $m_V$ is  $m_Z$ or $m_W$, and $m_j$ is an invariant
mass of the groomed jet.
Finally, the number of charged-particle tracks which are  associated with the jet is required to be $n_{ch} < 30$.
In Fig.~\ref{fig:charged1}, we show the distribution in $n_{ch}$ and
$m_j$ plane for the events with $m_{jj} > 1500$ GeV where all 
the cuts except for $n_{ch} $ and $m_j$ are applied.  
 The ATLAS signal regions are marked by squares.   
The figure shows very good separation between the signal and
the background events.

\begin{figure}[tp]
\begin{center}
\includegraphics[width =8cm,bb=0 0 258 169]{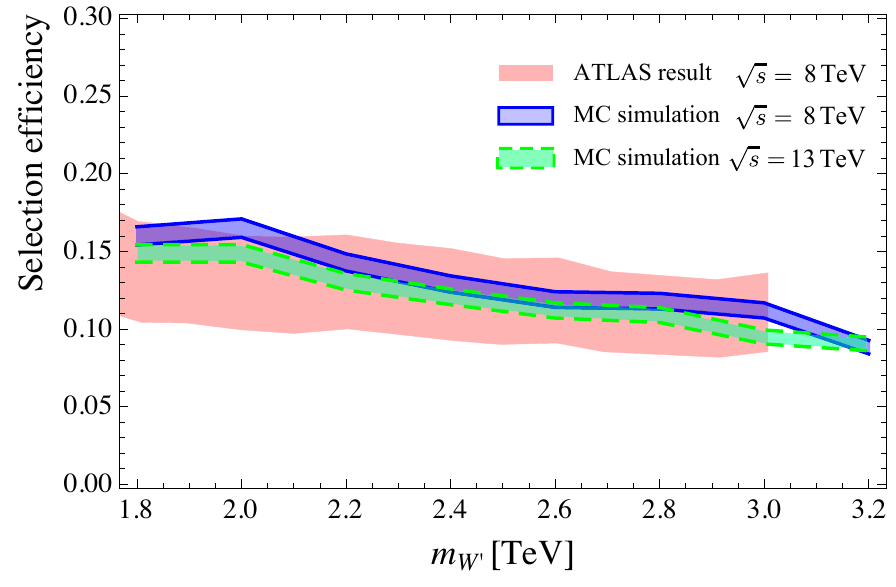}
\caption{
The event selection efficiencies for $W' \to WZ \to JJ $ simulated
 events generated by {\sc{Pythia 8}} as a function of the  $W'$ mass at
 $\sqrt{s}  = 8$ and $13$ TeV. The red band is the result of the ATLAS
 \cite{Aad:2015owa}, where the thickness  corresponds to   $\pm $ 1
 $\sigma$ statistical and systematical errors. 
Our
 result of MC simulation at $\sqrt{s}  = 8$ ($13$) TeV is represented in
 the blue (green) band where the thickness corresponds to
 $\pm $ 1 $\sigma$ statistical error. 
}
\label{fig:efficiency}
\end{center}
\end{figure}

To check  our simulation, we compare the distributions of our 
$\sqrt{s} = $ 8 TeV samples to the ATLAS ones.
The Fig.~\ref{fig:efficiency} shows the reconstruction efficiency of the signal  for various input $W'$ mass.
Our result at $\sqrt{s} = 13$ TeV is also shown.
We find that the signal selection efficiency agrees with
the ATLAS one.
Since the jets get narrower with increasing $p_T$, the
signal efficiency becomes lower for higher $W'$ mass.

\begin{figure}[th]
\begin{center}
\includegraphics[width =7cm,bb=0 0 567 427]{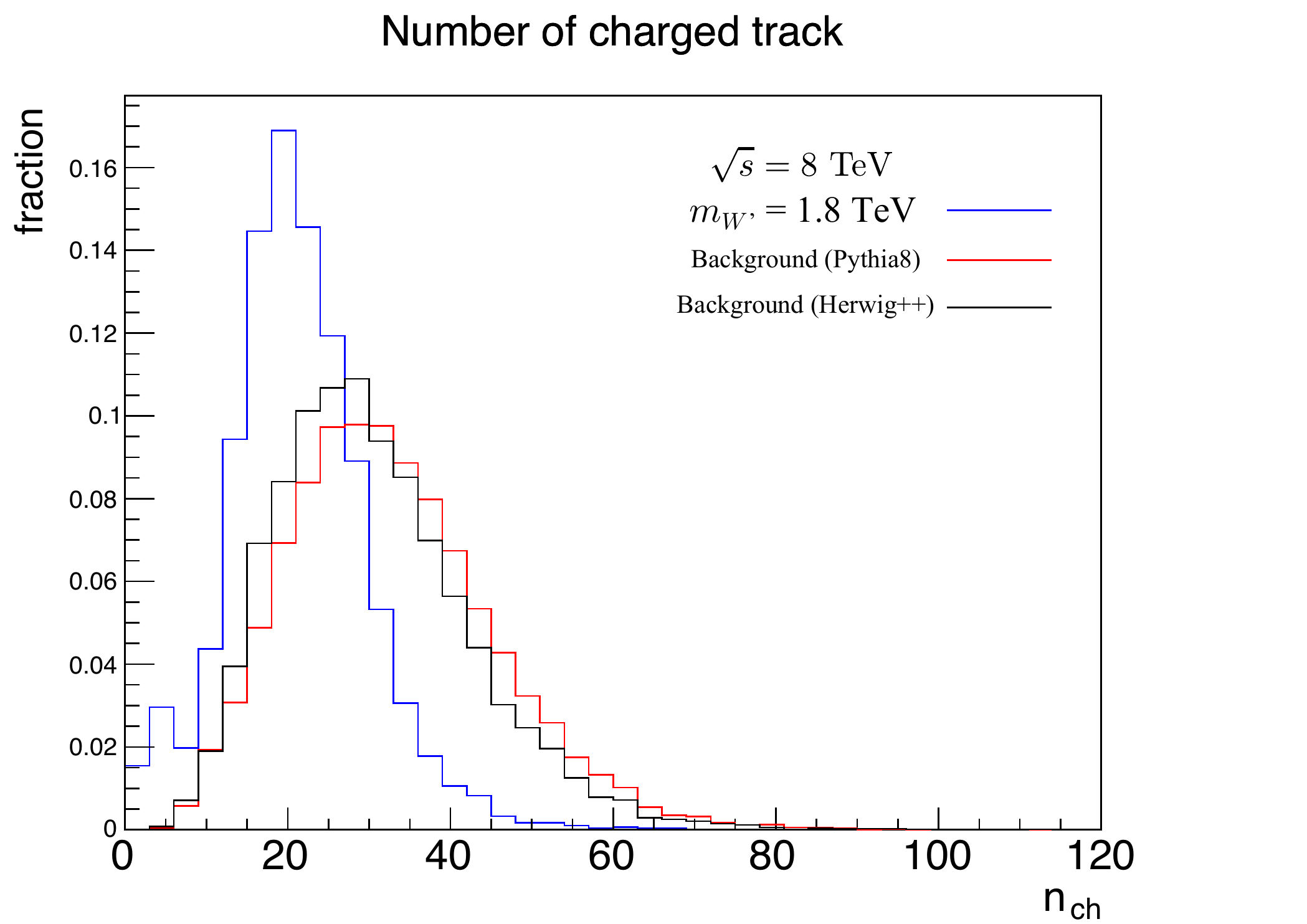}~~
\includegraphics[width =7cm,bb=0 0 567 427]{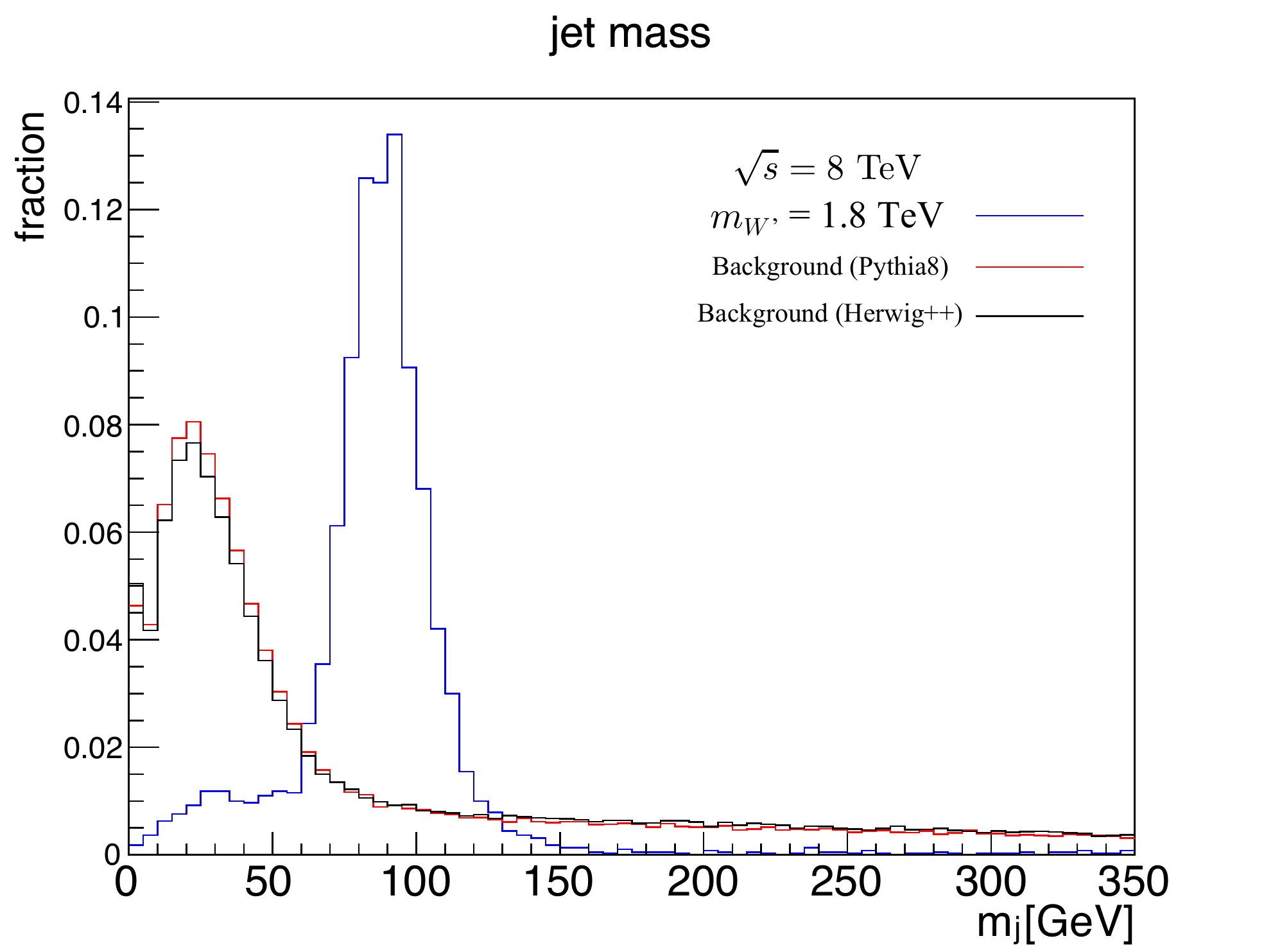} \vspace{.2cm}
\caption{
The distributions of the  number of
 charged tracks $n_{\textrm{ch}}$ (left) and  the jet mass
 $m_j$ (right) for the signal (Blue) and background
 events (Black and Red). 
We take $m_{W'} = 1800 $ GeV and $\Gamma_{\textrm{tot}} (W') = 25$ GeV.
We simulate the  background events
 using {\sc Pythia8} (Red) and {\sc Herwig++} (Black).  
}
\label{fig:charged}
\end{center}
\end{figure}
On the other hand, we find that the number of the background events
after all the selection cuts are twice as large as that of
the  ATLAS final result.
The discrepancy in the total background events
might arise from several sources. 
The number of charged tracks
of QCD jets are controlled by soft physics and varies significantly depending 
on  Monte Carlo (MC) generators and tunes of the parton shower parameters especially for gluon jets.
The distributions are  shown in the left panel of Fig.~\ref{fig:charged}, under the cut of
Fig. 1 of Ref.~\cite{Aad:2015owa}, $1.62$ TeV$< m_{jj}<1.98$ TeV and 60
GeV $<m_j<110$~GeV together with the selection cuts listed above except
that for $n_{ch}$, where $m_{jj} $ is  the dijet invariant mass. 
 The signal distribution which is represented in the
 blue line agrees quite well with the ATLAS ones.
 However, the average number of charged tracks of  dijet event is significantly
higher than  the ATLAS ones.
To see the MC dependence,  we also show the distribution of the MC sample generated by {\sc Herwig++} with default tunes by the black line 
\cite{Bahr:2008pv}, which predicts slightly  small $\langle n_{ch}
\rangle $ compared with {\sc Pythia8} Tune C4, but higher
than ATLAS CT10 Tune.\footnote{ In the study of quark-gluon separation \cite{Aad:2014gea},
{\sc Herwig++} reproduces  high gluon $p_T$ jet nature well. }

In the right panel of Fig.~\ref{fig:charged}, we also show the $m_j$
 distributions for the signal and background.  
 The signal distribution agrees quite well with the ATLAS MC results again. 
 For the background distribution,
we find that the number of events above $m_j > 50 $ GeV is smaller compared with Fig.~1 of Ref.~\cite{Aad:2015owa}.
This is because we do not generate underlying events 
together with the dijet events.

In the experimental side,  ATLAS  counts the  well reconstructed track inside the jet.
 The efficiencies are not implemented in our simulation. 
Naively speaking,  the efficiency
is expected to be lower for the jets with high charged track multiplicity.
For jet clustering, ATLAS selects calorimeter towers using Topocluster
algorithm and does not use cluster-track matching which is  only crudely
implemented in our simulation. 
 In any case, data driven approaches are adopted in Ref.~\cite{Aad:2015owa}
to estimate the selection efficiency, 
and reproducing the result precisely is beyond the scope of this paper.
It is probably worth doing more dedicated theoretical
and  experimental studies 
on jet nature relevant to the boosted $W$ and $Z$ bosons reconstruction in future.

\begin{figure}[tp]
\begin{center}
\includegraphics[width =7cm,bb=0 0 567 428]{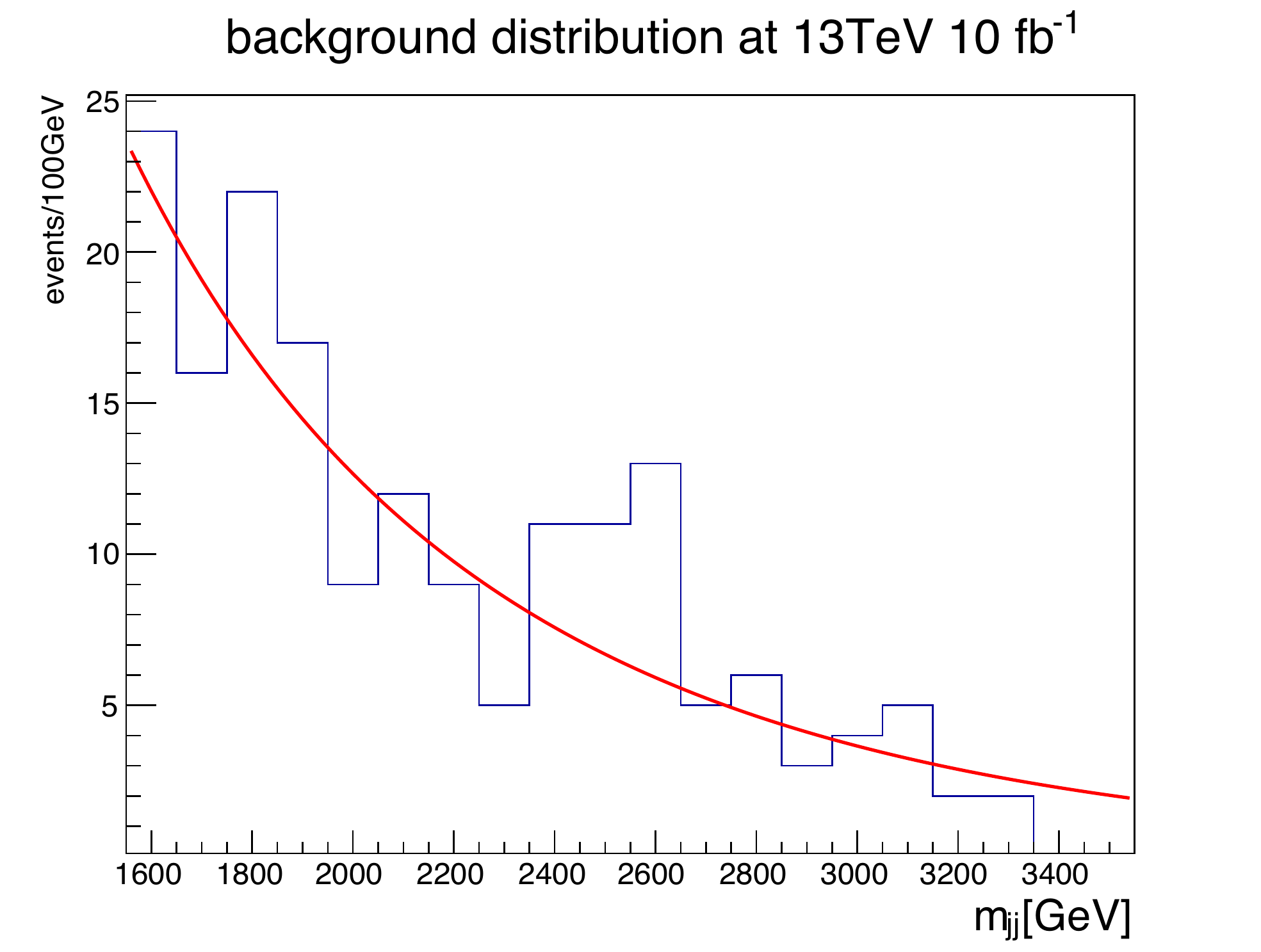}~~
\includegraphics[width =7cm,bb=0 0 567 427]{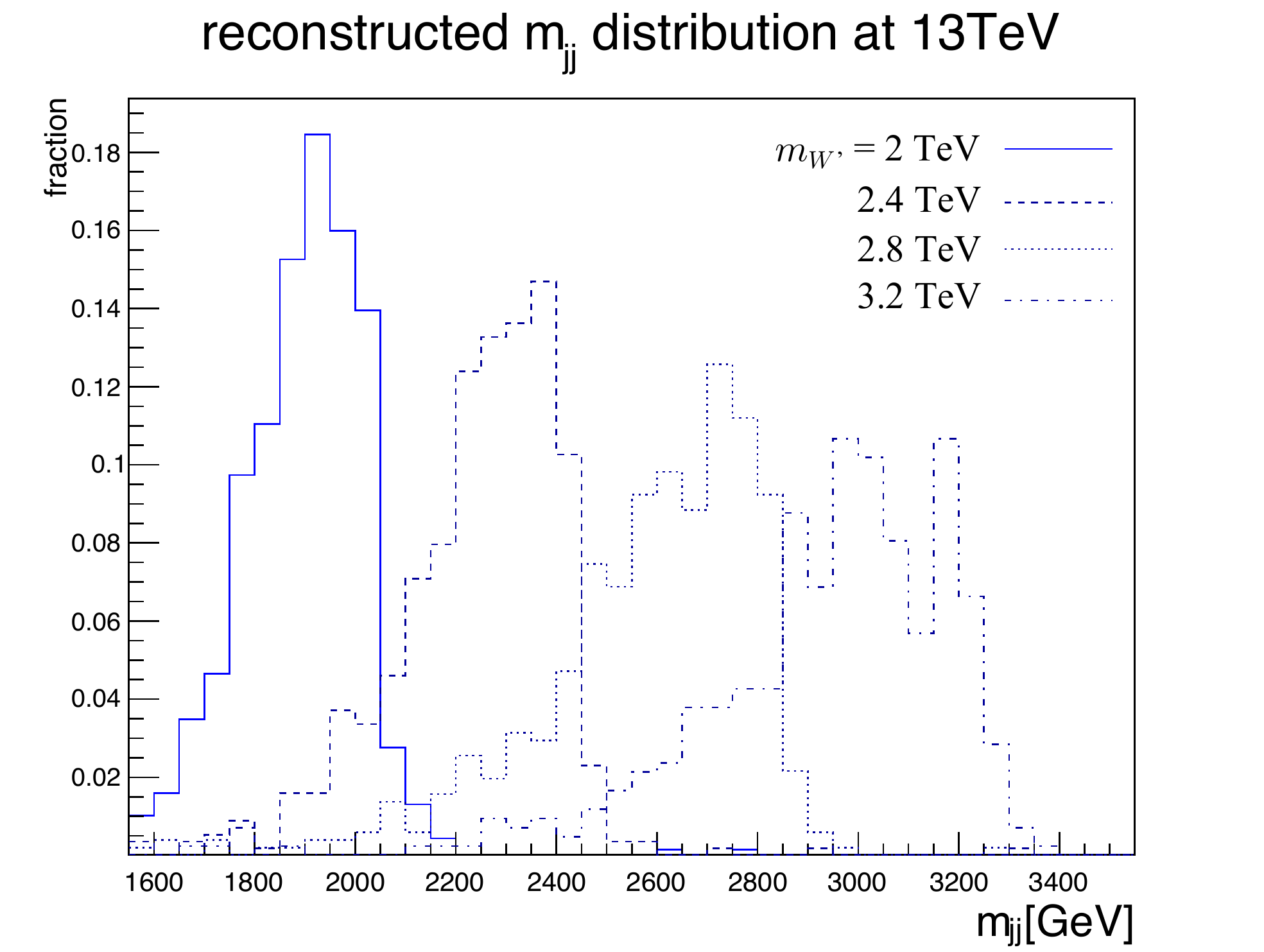}
 \vspace{.2cm}
\caption{The $m_{jj}$ distributions for the background (left)
 and the signal (right)  at $\sqrt{s} = 13$ TeV.
The background distribution is for $10 $ fb$^{-1}$, and the signal distributions are normalized to be 1 for various input $W'$ mass.
}
\label{fig:signal}
\end{center}
\end{figure}
\begin{figure}[hbp]
\begin{center}
\includegraphics[width =10cm,bb=0 0 360 251]{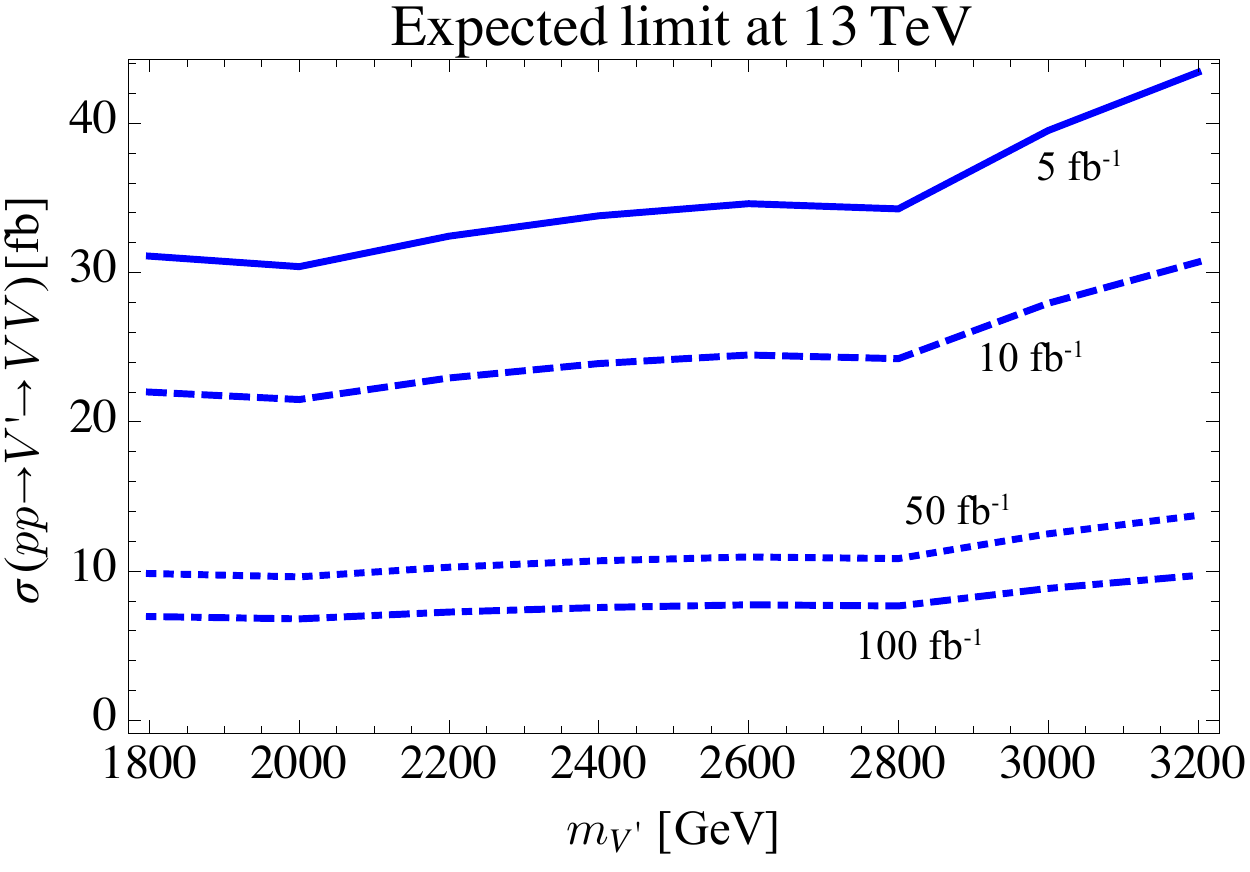}
 \vspace{.2cm}
\caption{
Expected limit on  $\sigma(pp\to V' \to VV)$
at $\sqrt{s} = 13$ TeV.
We take $\Gamma_{\textrm{tot}}(V') = 25$~GeV.
}
\label{fig:expected}
\end{center}
\end{figure}
Keeping crudeness of our simulation in mind, we  estimate
the
signal efficiency at $\sqrt{s} = $ 13 TeV using our signal MC and detector simulation
without rescaling, while  the number of the background
events obtained from
our MC is rescaled by factor of $1/2$ which is needed to
 reproduce the ATLAS results at $\sqrt{s} = $ 8 TeV.  
 The scaling approach comes from an assumption that the change 
in the center of mass energy from $8$ TeV to $13$ TeV 
does not alter the structure inside jets of the same $p_T$ jets.
 Note that gluon jets are involved in the QCD
background at $\sqrt{s} = $ 13 TeV but this effect is not taken into account.
Under this assumption, the background is fitted to estimate the
distribution, and the result is shown in the left panel of Fig.~\ref{fig:signal}.
We found 169 events/10 fb$^{-1}$ for 1550~GeV $ < m_{jj } < $  3550~GeV after rescaling.
We also show the dijet invariant mass distributions
of the signal for various input $W'$ mass in the right panel of Fig.~\ref{fig:signal}.

Figure~\ref{fig:expected} shows expected the exclusion limit at 95 \% C.L.  for 
$\sigma(pp\to V' \to VV)$ 
at $\sqrt{s} = 13$ TeV.\footnote{Prospects for the electroweak gauge boson scattering which can also probe $W'$  are discussed
in, for example, Ref.~\cite{Englert:2015oga}. }
We calculated $\Delta \chi^2 $ of the signal plus background
distributions to the background distribution.
For the signal, we generate $10^4$ signal events to obtain
the result for each $m_{W'}$, and 
take the number of events 
in the bins $i_{\textrm{max}} - 3 \leq i \leq i_{\textrm{max}} +3$,
where $i_{\textrm{max}}$  is the 
highest signal event bin and  the bin size is $50$ GeV.
Then, we  rescale them by the cross section and luminosity.
The selection efficiency of the signal event is shown in Fig.~\ref{fig:efficiency}.
For the background, we use our fit and rescaled it by the luminosity. 
Here y-axis means 
$\sigma(pp \to W' \to WZ)$ + $\sigma(pp \to Z' \to WW)$ with the mass at $m_{V'}$.
We find that the region where 
$\sigma(pp\to V' \to VV) $
is larger than 20 fb may be excluded at $\int dt  \mathcal{L}=10$ fb$^{-1}$ and
$\sqrt{s} =13$~TeV.

\section{Future prospects: 13 TeV analyses}
\label{sec13TeV}

In this section, we analyze
the future prospects of $W'$ and $Z'$ searches by applying result in Sec.~\ref{sec:LHC}.

We investigate the prospects for the $\sqrt{s} = 13$ TeV collision  in the case of
$m_{Z'} = 2$ TeV. 
\begin{figure}[tp]
\begin{center}
\includegraphics[width =6cm,bb=0 0 360 367]{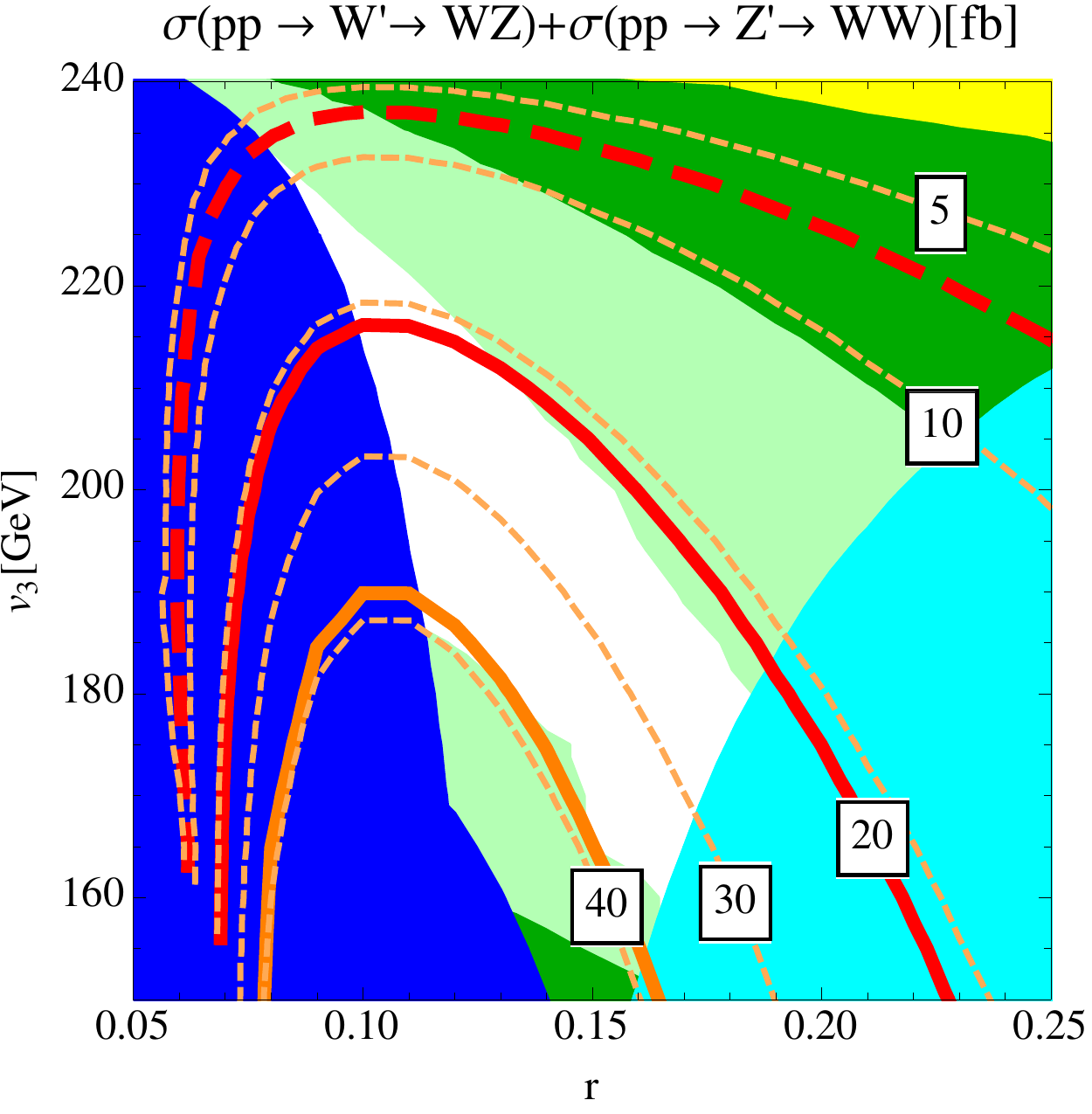}~~~~
\includegraphics[width =6cm,bb=0 0 360 367]{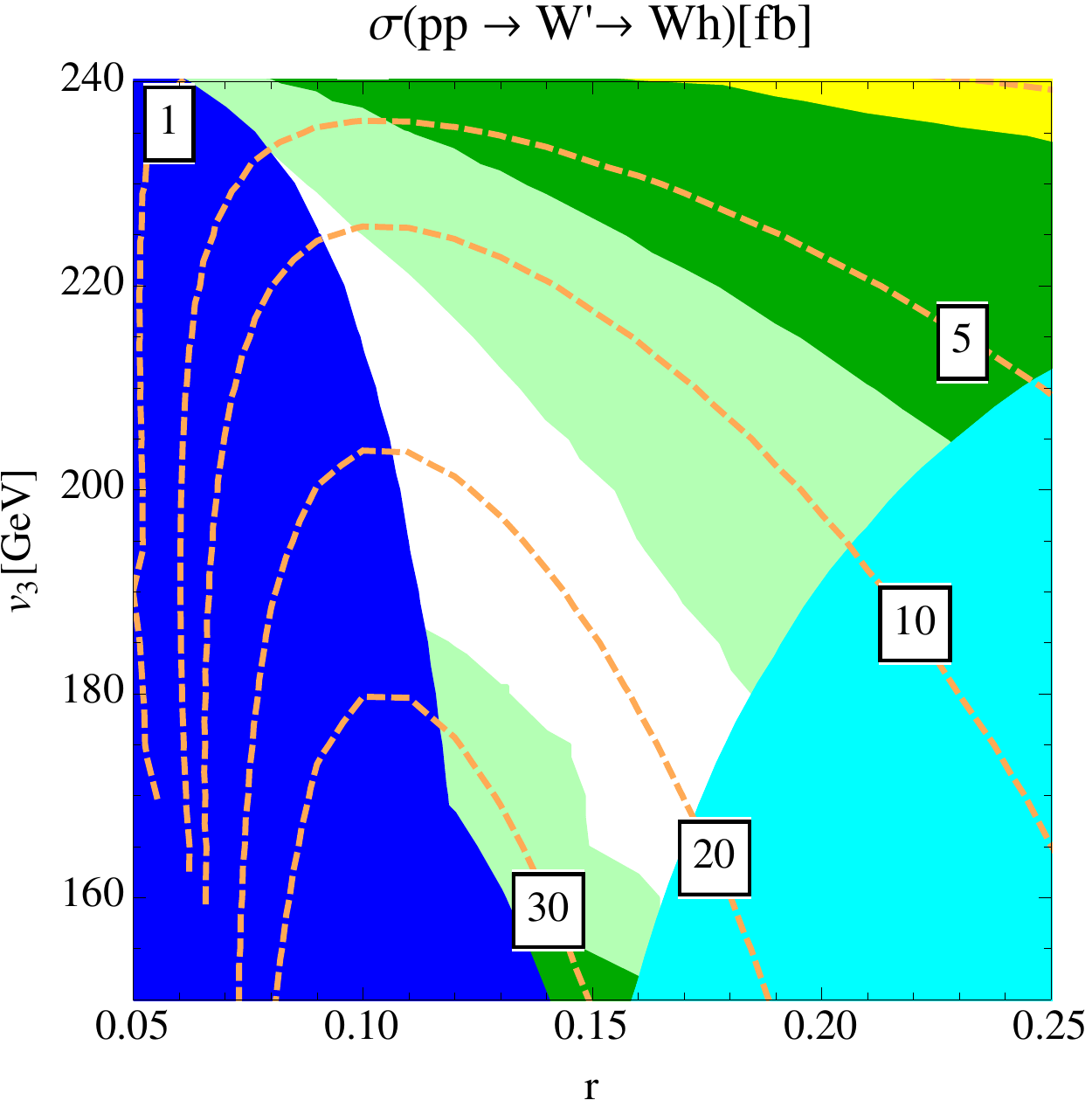}\\
\includegraphics[width =6cm,bb=0 0 360 367]{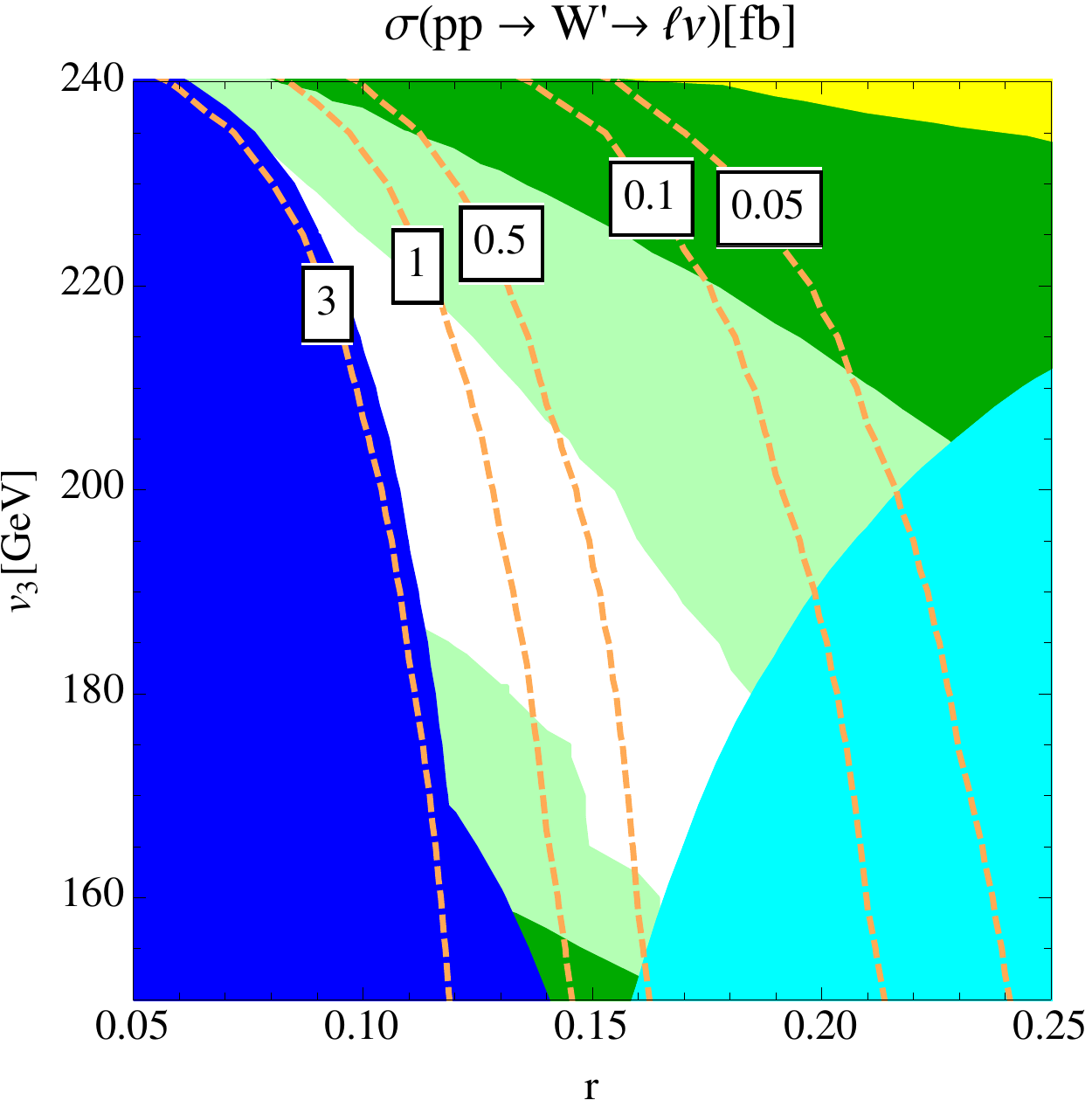}~~~~
\includegraphics[width =6cm,bb=0 0 360 367]{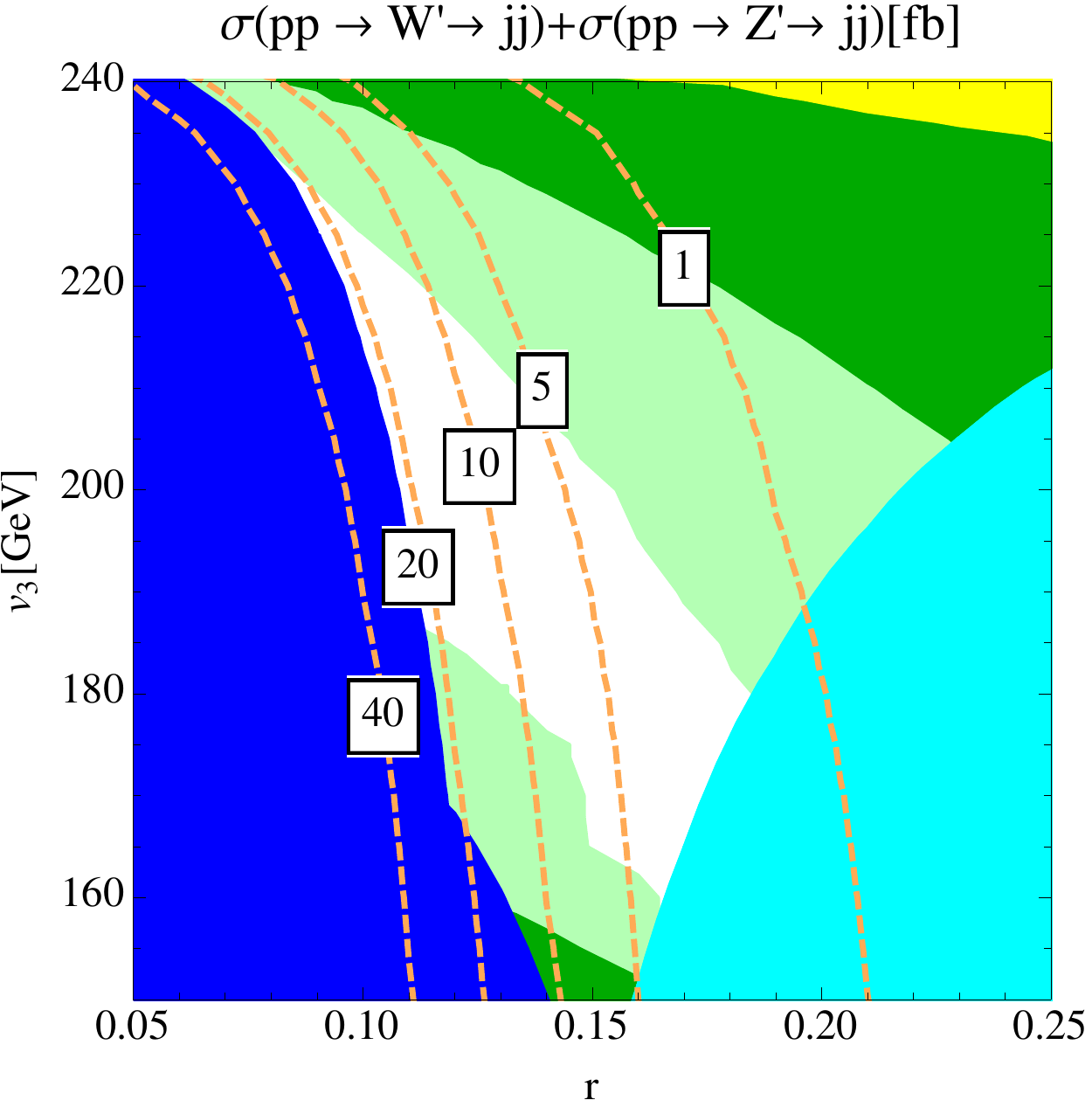}\\
\includegraphics[width =6cm,bb=0 0 360 367]{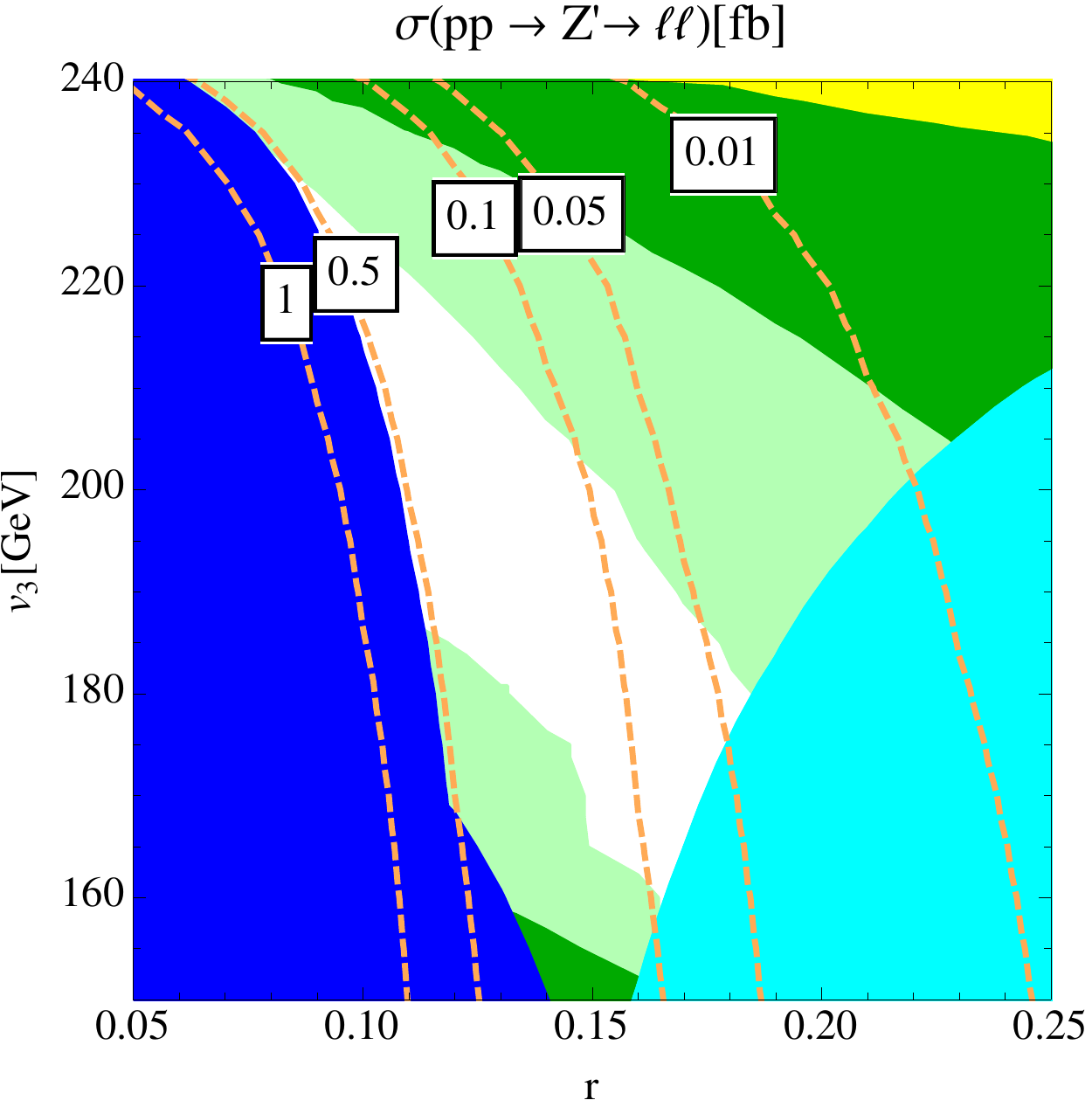}~~~~
\includegraphics[width =6cm,bb=0 0 360 373]{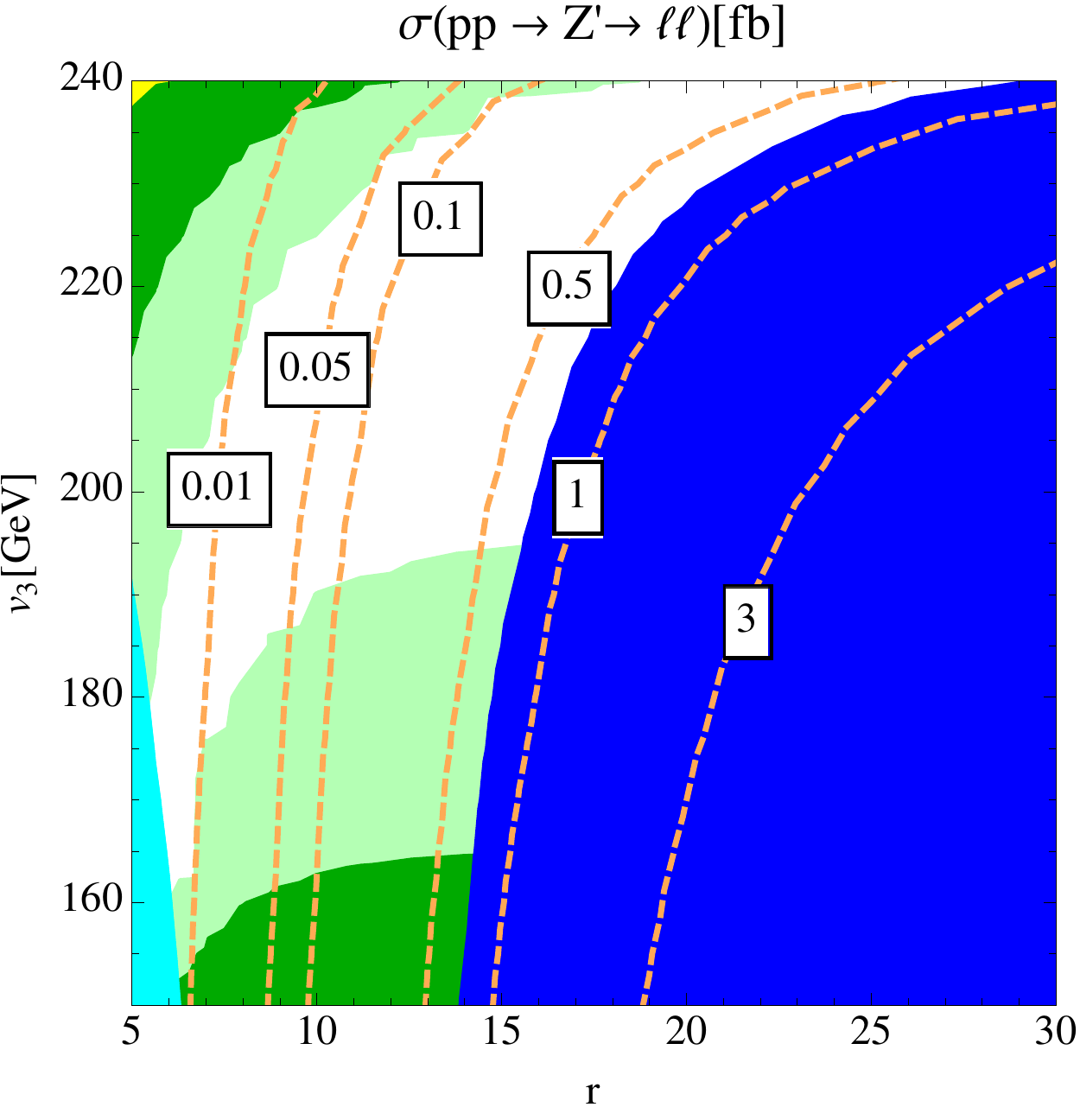}
 \vspace{.4cm}
 \caption{The prospects for the $\sqrt{s} = 13$ TeV collision.
We take $m_{Z'} = m_A = m_{H'} = m_H = 2000$ GeV, $\kappa_F =
 1.00$. 
  The color notation is the same as in Fig.~\ref{fig:VV_1.9_2.0_2.1}. 
 Below the thick orange line, the cross
 section $\sigma(pp \to  V' \to  VV )$ is enough to explain the diboson excess at ATLAS. 
Below the (dashed) red line is expected to be excluded with $\int dt
 \mathcal{L} =  10$ (100) fb$^{-1}$. 
}
 \label{2000_13TeV} 
\end{center}
\end{figure}
In Fig.~\ref{2000_13TeV}, 
we show  contours of the several  cross sections by orange dashed lines  in $r$--$v_3$ planes:  $\sigma(pp \to W') \textrm{Br}(W' \to WZ) + \sigma(pp \to Z') \textrm{Br}(Z' \to WW)$, $\sigma(pp \to W') \textrm{Br}(W' \to Wh)$,  $\sigma(pp \to W') \textrm{Br}(W' \to \ell \nu)$, $\sigma(pp \to W') \textrm{Br}(W' \to jj) + \sigma(pp \to Z') \textrm{Br}(Z' \to jj)$, and $\sigma(pp \to Z') \textrm{Br}(Z' \to \ell \ell)$.
The color filled regions are excluded and constrained as we discussed in Sec.~\ref{8TeVstatus}. 
The color notations are the same as in
Fig.~\ref{fig:VV_1.9_2.0_2.1}.

The expected exclusion limit at $\sqrt{s} = 13 $ TeV provided in
 Fig.~\ref{fig:expected} is shown by 
the red  (dashed) line for $\int dt \mathcal{L} = 10 $ (100) fb$^{-1}$.
The masses of $W'$ and $Z'$ are highly degenerate in the small $r$ regime,
 and we assume that the signal efficiency for $Z' \to WW$
 event is equivalent to  the one for  $W' \to WZ$ which is simulated in the previous section.
 Thus we can apply the prospect shown 
 in Fig.~\ref{fig:expected} to $\sigma (pp \to W' \to  WZ )+ \sigma ( pp \to Z' \to  WW )$.
The region where the ATLAS diboson excess can be
 explained is within the
 reach of the LHC at $\int dt  \mathcal{L}=10$ fb$^{-1}$ and $\sqrt{s}=
 13 $ TeV. In addition, the cross sections of the other channels are also large,
so that the spin-1 resonances could be probed  in the channels as well.

\begin{figure}[tp]
\begin{center}
\includegraphics[width =5.1cm,bb=0 0 360 362]{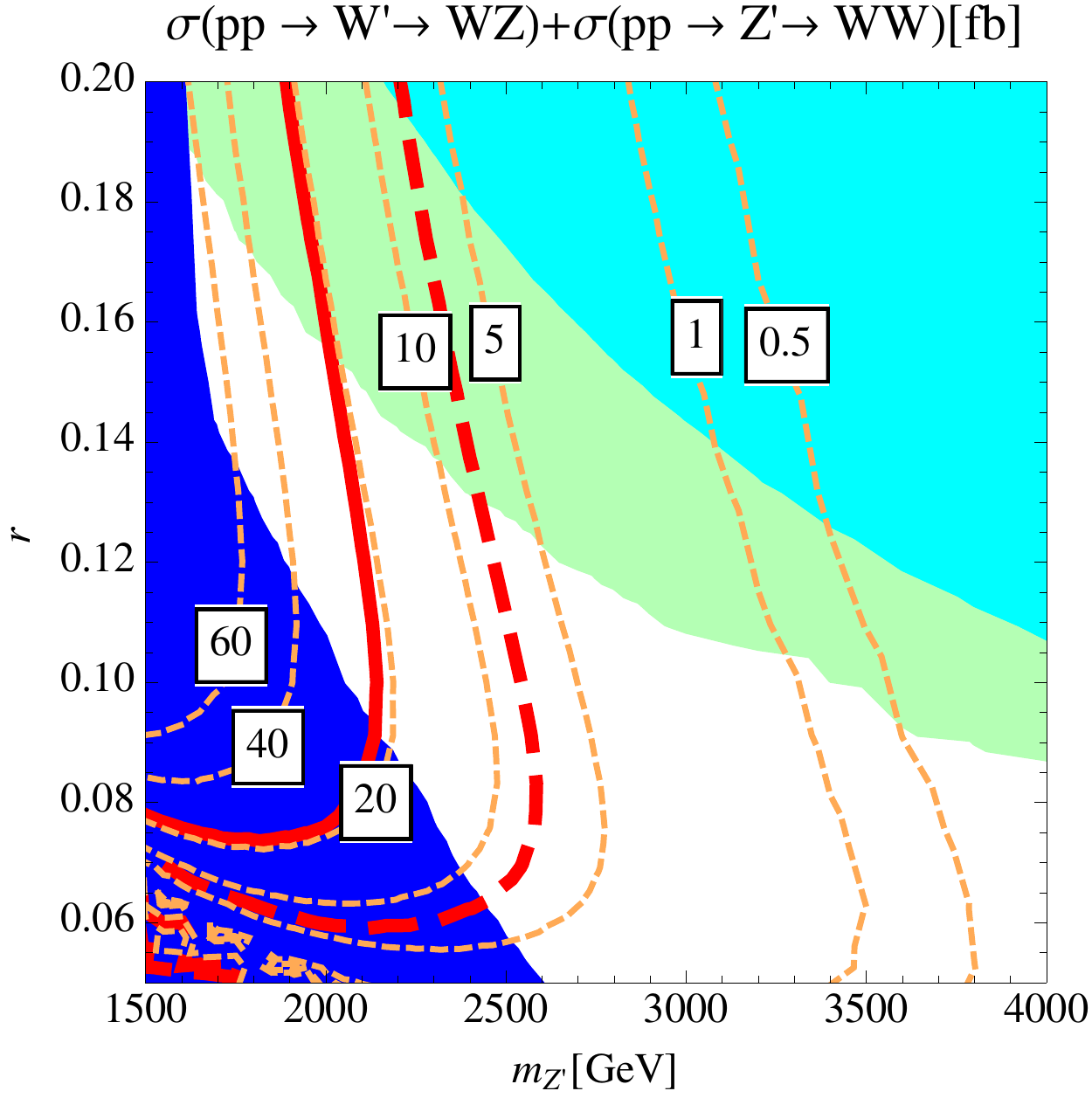}~~
\includegraphics[width =5.1cm,bb=0 0 360 362]{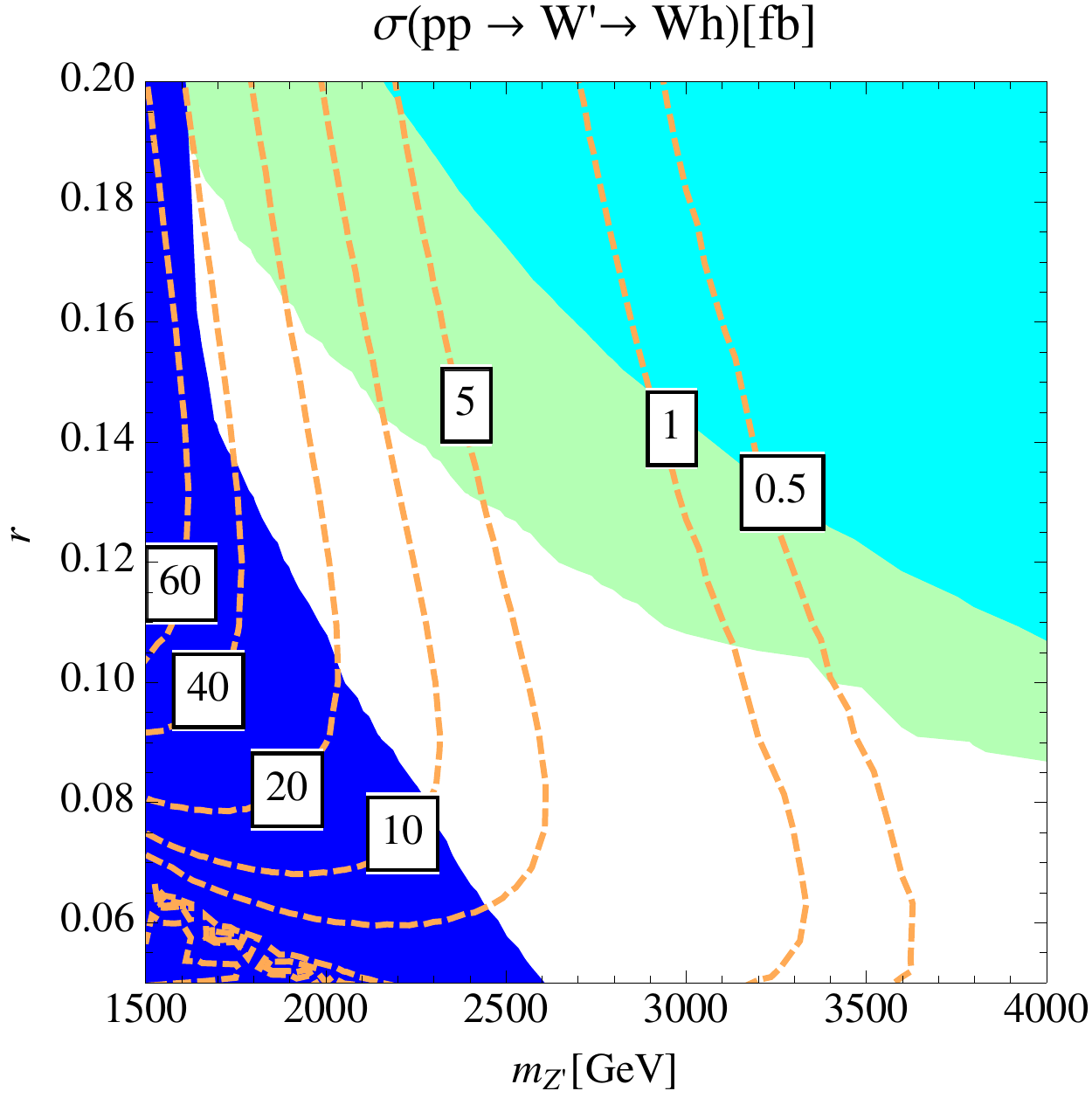}~~
\includegraphics[width =5.1cm,bb=0 0 360 362]{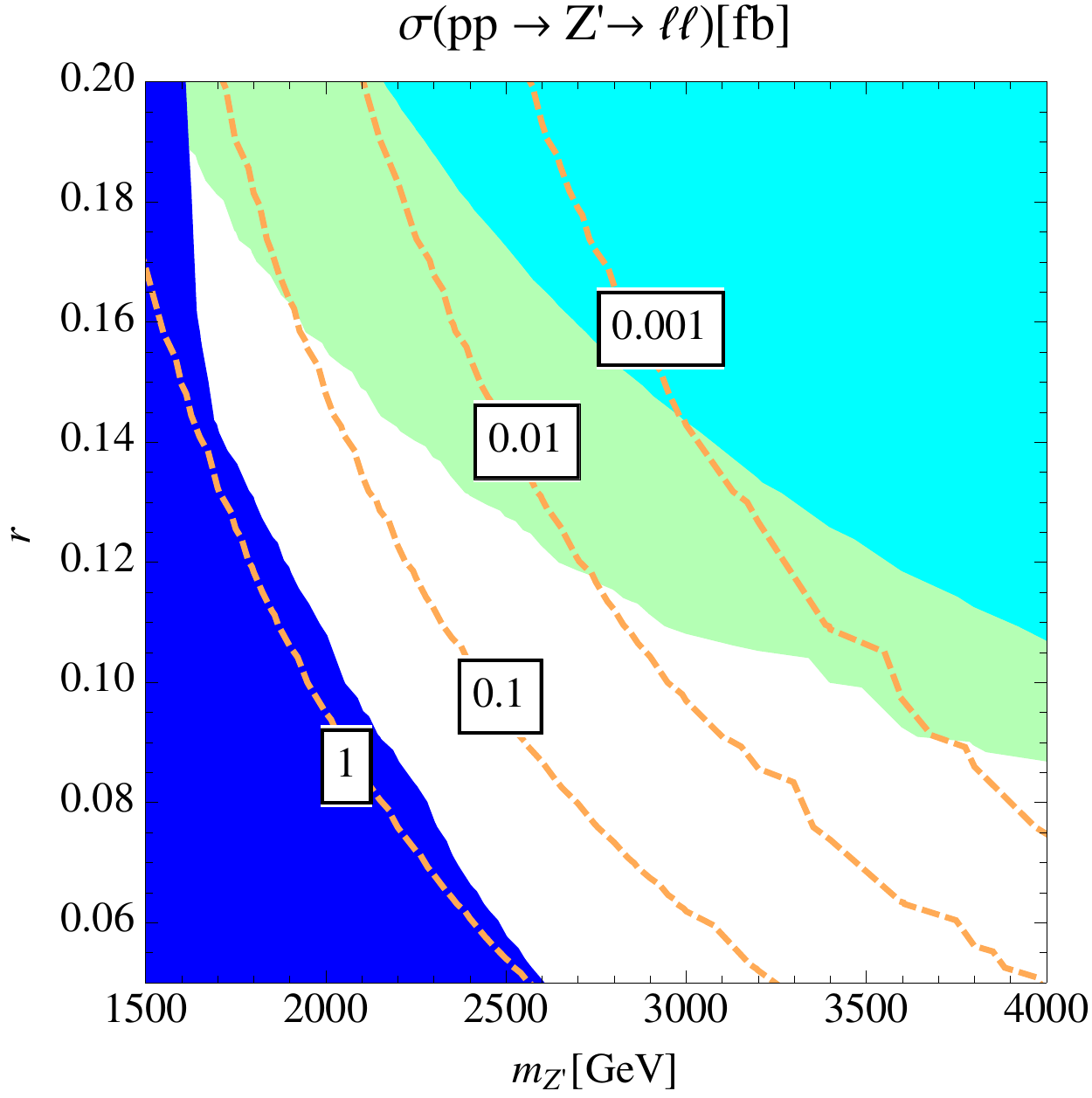}
 \vspace{.4cm}
\caption{The prospects of the discovery as a function of $m_{Z'} $
 and $r$ for the LHC Run-2.
We take $v_3 = 200 \GeV$ and $\kappa_F = 1.00$. 
The color notation is the same as in Fig.~\ref{fig:VV_1.9_2.0_2.1}.
The (dashed) red is the future expected 
 exclusion limit for $\int dt \mathcal{L} = 10$ (100) fb$^{-1}$. 
} 
\label{13TeV_v3200}
\end{center}
\begin{center}
\includegraphics[width =5.1cm,bb=0 0 360 362]{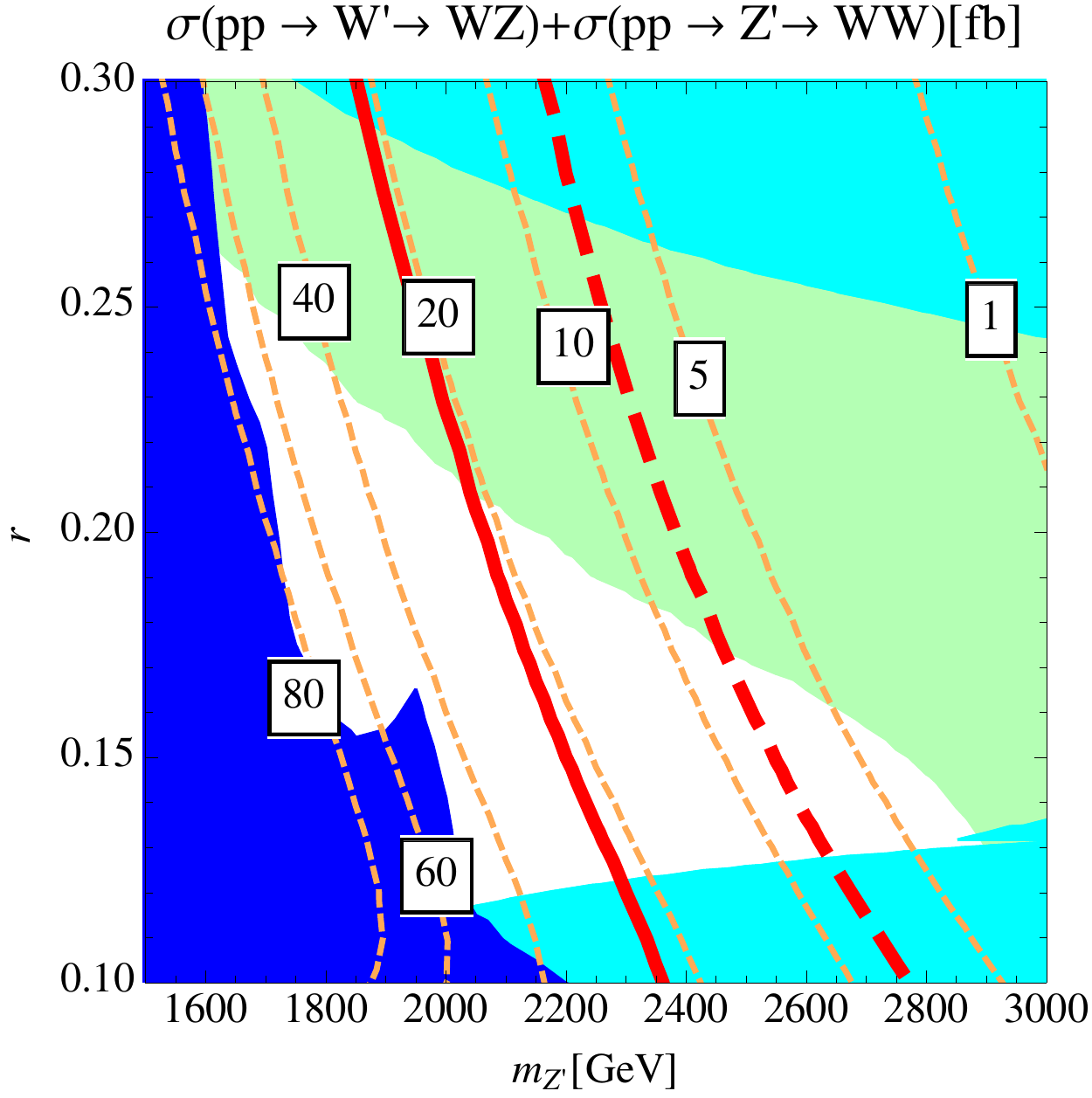}~~
\includegraphics[width =5.1cm,bb=0 0 360 362]{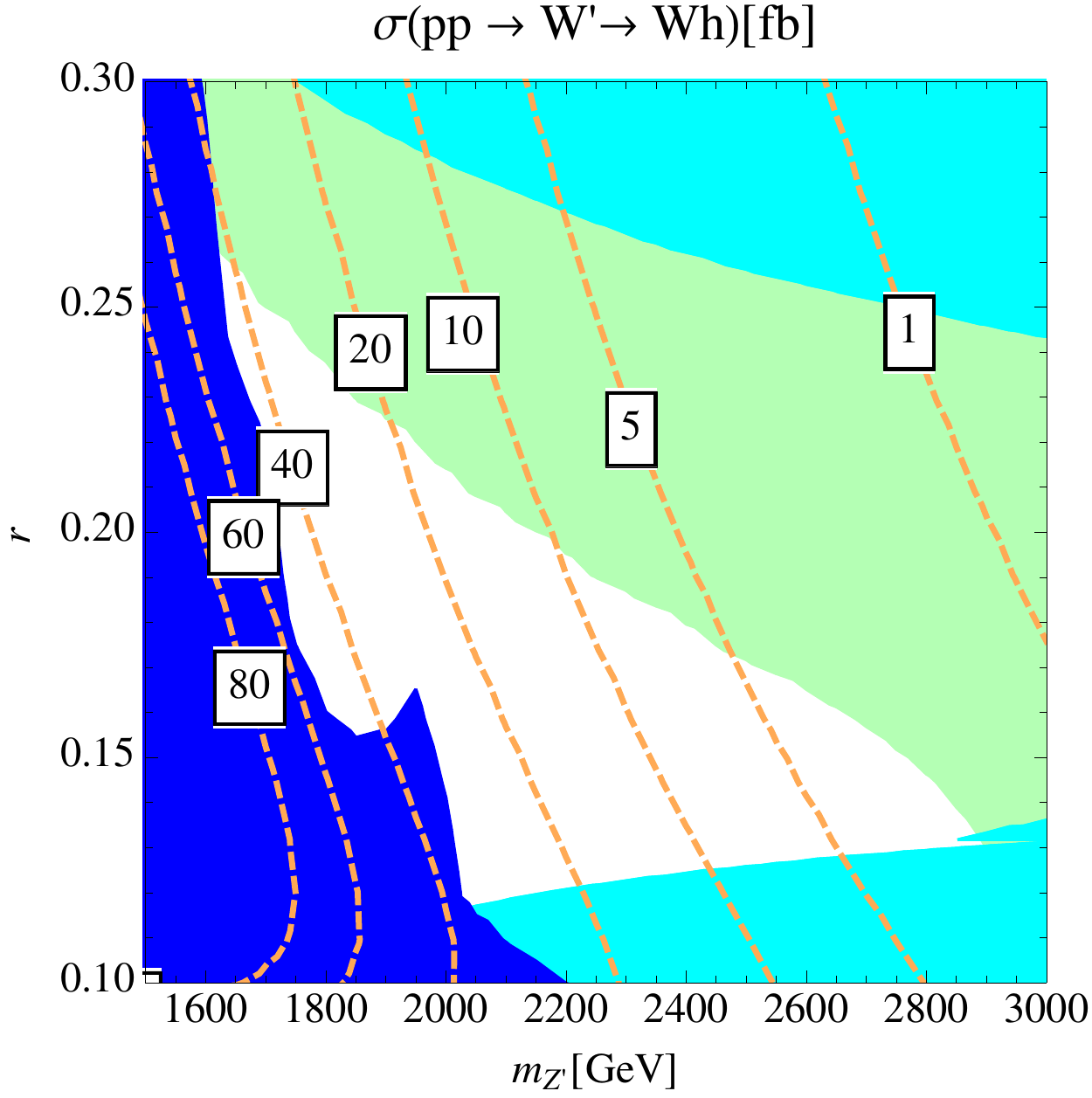}~~
\includegraphics[width =5.1cm,bb=0 0 360 362]{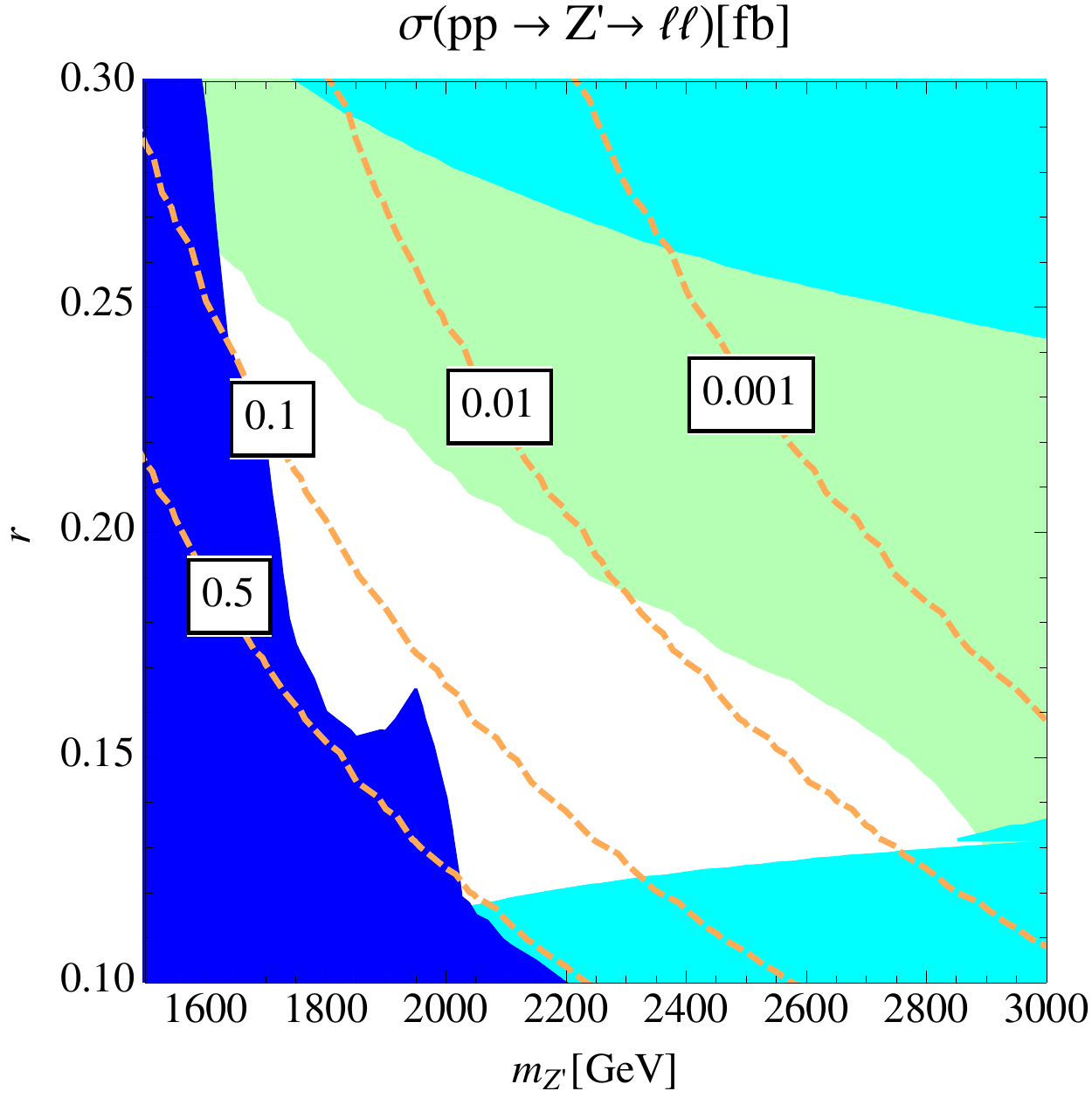}
 \vspace{.4cm}
\caption{The prospects of the discovery as a function of $m_{Z'}$
 and $r$ for the LHC Run-2.
We take $v_3 = 150 \GeV$ and $\kappa_F = 0.99$.
Below the (dashed) red line is expected to be excluded with $\int dt
 \mathcal{L} =  10$ (100) fb$^{-1}$. 
}
  \label{13TeV_v3150}
\end{center}
\end{figure}
We plot the $\sqrt{s} = 13$ TeV cross sections as a function of spin-1 resonance mass in Fig.~\ref{13TeV_v3200}.
We take the universal mass $m_{Z'} = m_A = m_{H'} = m_H$, $v_3 = 200 \GeV$, and $\kappa_F = 1.00$  in Fig.~\ref{13TeV_v3200}.
The color notation is the same as in Fig.~\ref{fig:VV_1.9_2.0_2.1}.
The mass of $W'$  is highly degenerate with the mass of $Z'$  in this figure. 
We find that the spin-1 resonances lighter than 2.1 (2.6)~TeV
can be excluded by the diboson search at the LHC 
with $\int dt \mathcal{L} =  10~(100)$ fb$^{-1}$.
Figure~\ref{13TeV_v3150} shows the same cross sections for another
parameter set ($v_3 = 150 \GeV$ and $\kappa_F = 0.99$). 
With the consideration of the bounded below condition, 
we find that the spin-1 resonances  
up to masses  of 2.3 (2.6) TeV can be excluded  with 
$\int dt \mathcal{L} =  10~(100)$ fb$^{-1}$.

Before closing this section, we briefly comment on the prospect for the
large $r$ region. 
Under the several theoretical bounds, this model has two allowed
regions, namely small $r$ and large $r$ regions (see
Fig.~\ref{fig:const}).
The cross section of $pp \to Z' \to \ell \ell$ is enhanced by $r^2$
factor in large $r$ regime (see Eq.~(\ref{prod:Zp})).
ATLAS study found $m_{Z'}$ around 3~TeV with the cross section
around 0.01 fb is accessible with the high luminosity LHC at $\sqrt{s} =
14$~TeV with integrated luminosity 3000 fb$^{-1}$ \cite{ATLASdilepton}. 
Thus we can expect that the cross section for the large $r$ region in
the model is accessible at the LHC Run-2 as we can see in Fig.~\ref{2000_13TeV}.
However, the $r$ regime would be constrained  from the electroweak
precision parameters at one-loop level due to the custodial symmetry breaking.

\section{Conclusion}
\label{sec:conclutions}

Motivated by the ATLAS diboson excess around 2 TeV, we have investigated
the phenomenology of the spin-1 resonances ($W'$ and $Z'$) in the
partially composite standard model.
In this model, $W'$ and $Z'$ couple to the SM fermions weakly
through the mixing to the elementary gauge bosons.
  We find  that the  main decay modes of  the resonances  are
  $V' \to VV$ and $V' \to Vh$, and the width is narrow enough so that
  the ATLAS diboson excess can be explained.
The couplings of the spin-1 resonances  with the SM sector can
be  controlled by the  ratio of the Higgs VEVs so that the ATLAS diboson
excess can be explained.

We have explored  not only the current bounds from the LHC and 
the precision measurements 
but also the theoretical constraints,
i.e. perturbativity condition, bounded below condition, global minimum
vacuum condition, and stability condition of the scalar potential.
The parameter regions where 
the diboson excess at the ATLAS can be explained are still allowed after
including  those constraints.

In order to  investigate future prospects of
the spin-1 resonance search, we  have performed the simulation at
$\sqrt{s} = 13$ TeV LHC, and estimated model independent exclusion
limit for $\sigma(pp \to V' \to VV \to JJ)$ 
shown in Fig.~\ref{fig:expected}.
Applying our simulation result, we find that 
the parameter regions 
consistent with the ATLAS diboson excess will be excluded at
$\int dt  \mathcal{L}=10$ fb$^{-1}$ and $\sqrt{s}= 13 $ TeV.

Finally, we have  investigated  future prospects of diboson resonance
search in our model. 
The spin-1 resonances  up to a mass of  $2.6$ TeV    can be probed  at
$\sqrt{s} = 13$ TeV and $\int dt \mathcal{L} = $ 100 fb$^{-1}$.

\section*{Acknowledgments}
The authors  would like to thank Koji Terashi and Ryuichiro Kitano for useful discussions.
This work is supported by JSPS KAKENHI No. 26287039 (M.M.N.). The work is supported by Grant-in-Aid
for Scientific research from the Ministry of Education, Science, Sports, and Culture
(MEXT), Japan, No. 23104006 (T.A. and M.M.N.) and No. 25105011 (T.K.), and also by World
Premier International Research Center Initiative (WPI Initiative), MEXT, Japan.

\appendix

\section{Viable range of the coupling ratio $\kappa_Z$}
\label{app:kappaz}

In this appendix, we show $\kappa_Z$ is very restricted to be one.
We first replace 
the Higgs quartic couplings  $\lambda_i$  by the other parameters Eq.~(\ref{eq:inputParams}). 
Here we take $m_A  = m_{H'} = m_{H}$ for simplicity,
and work in $m_A \gg m_{h}$ regime. 
For $r\ll 1$ regime, we find
\begin{align}
 \lambda_1 \simeq& 0, \\
\lambda_2 \simeq& f_1(\kappa_F, \kappa_Z), \\
\lambda_3 \simeq&
 \frac{m_h^2}{2 v^2} \kappa_F^2 + \frac{m_{A}^2}{2 v^2} (1-\kappa_F^2),\\
\lambda_{12} \simeq& 0,\\
\lambda_{23} \simeq& f_2(\kappa_F, \kappa_Z), \\
\lambda_{31} \simeq& 0,
\end{align} 
and for $r\gg 1$ regime,
\begin{align}
\lambda_1 \simeq& f_1(\kappa_F, \kappa_Z), \\
 \lambda_2 \simeq& 0, \\
\lambda_3 \simeq&
 \frac{m_h^2}{2 v^2} \kappa_F^2 + \frac{m_{A}^2}{2 v^2} (1-\kappa_F^2),\\
\lambda_{12} \simeq& 0,\\
\lambda_{23} \simeq& 0,\\
\lambda_{31} \simeq& f_2(\kappa_F, \kappa_Z), \\
\end{align} 
where
\begin{align}
 f_1(\kappa_F, \kappa_Z)
\simeq&
\frac{m_h^2}{2 v^2}
-
(1 - \kappa_F)
\frac{v_3^2}{v^2}
\frac{2 (m_A^2 - m_h^2)}{v^2 - v_3^2}
+
(1 - \kappa_Z)
\frac{m_A^2 - m_h^2}{v^2 - v_3^2}
,\label{f1}
\\
 f_2(\kappa_F, \kappa_Z)
\simeq&
\frac{m_h^2}{v^2}
+
 (1-\kappa_F)
\frac{v^2 - 3 v_3^2}{v^2} 
 \frac{m_A^2 -m_h^2}{v^2 - v_3^2}
 +
 (1 - \kappa_Z)
 \frac{m_A^2 - m_h^2}{v^2 - v_3^2}
.\label{f2}
\end{align}
In Eqs.~(\ref{f1}) and~(\ref{f2}), we make an expansion around
$\kappa_F \simeq 1$ and
$\kappa_Z \simeq 1$. 
Since both coefficients of $(1-\kappa_F)$ and
$(1-\kappa_Z)$ are large enough in $m_A \gg v$ regime, 
even the small deviations of $\kappa_F$ and $\kappa_Z$ from $1$ 
change the Higgs quartic couplings $\lambda_i$ drastically.

Let us consider the case of $\kappa_F = 1$.
The largest Higgs quartic coupling is  $\lambda_{23}$ ($\lambda_{31}$) in the $r \ll 1$ ($r \gg 1$) regime.
We further demand $\lambda_{23}$ ($\lambda_{31}$) $ < \left(4 \pi\right)^2$, namely
\beq
\kappa_Z &>& 1 - \frac{(16 \pi^2 v^2 - m_h^2)}{(m_A^2 - m_h^2)} \left( 1 - \frac{v_3^2}{v^2} \right).
\eeq
Due to the running effects, these quartic couplings can be even larger at the high scale, so that the lower bound on $\kappa_Z$ is actually severer than above estimation.

\begin{figure}[tp]
\begin{center}
\includegraphics[width =7cm,bb=0 0 270 285]{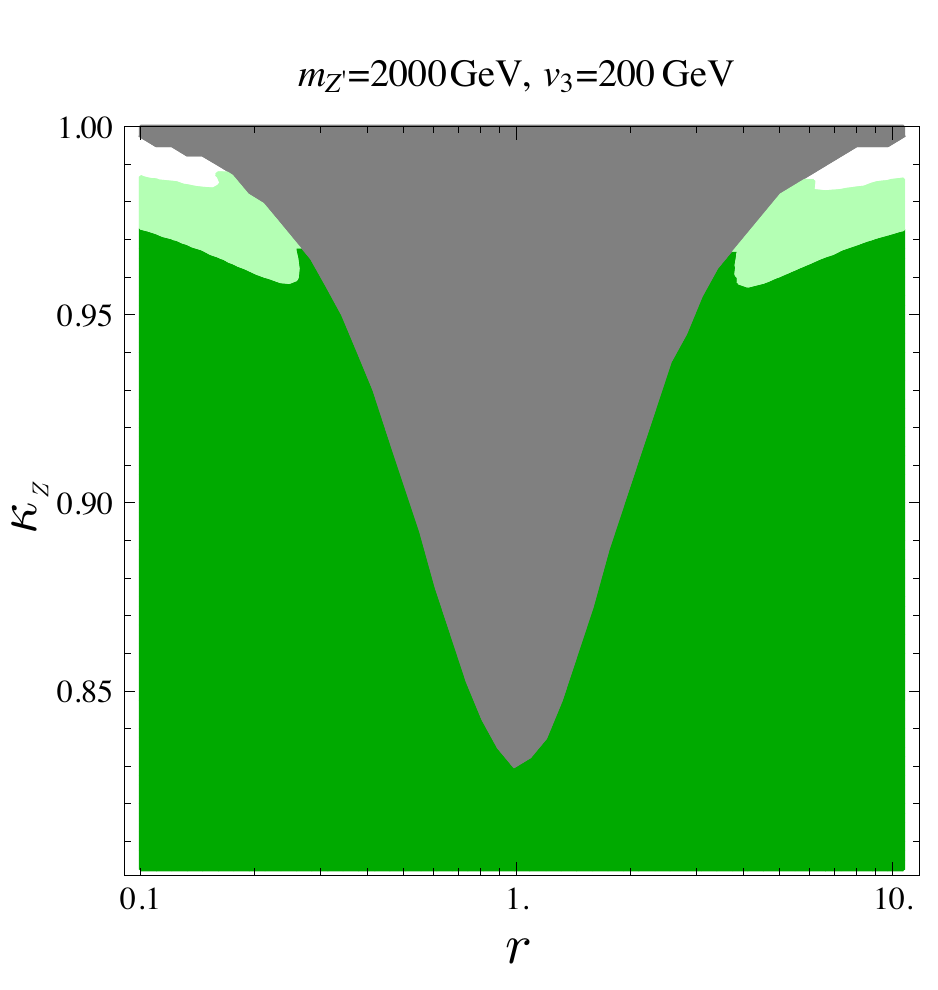}
 \vspace{-.2cm}
\caption{ The viable range of the coupling ratio $\kappa_Z$, 
 where we take $v_3 = 200 \GeV$, $m_{Z'} = m_A = m_{H'} = m_H =$ 2 TeV,  and $\kappa_F = 1.00 $.
The regions where $\lambda_i (\mu = 10 ~(100) \TeV) > \left( 4 \pi\right)^2$ are filled with
 green (light green).}
\label{fig:kappaz}
\end{center}
\end{figure}

In Fig.~\ref{fig:kappaz}, we show that 
the viable range of the coupling ratio $\kappa_Z$, where we take $v_3 = 200 \GeV$, $m_{Z'} = m_A = m_{H'} = m_H =$ 2 TeV,  and $\kappa_F = 1.00 $.
The regions where $\lambda_i (\mu = 10 ~(100) \TeV) > \left(4 \pi\right)^2$ are filled with
 green (light green), 
and  there is no physical solution  in the gray region.
The maximal value of $\kappa_Z$ is achieved at
a boundary of the gray region.
The  white area represents the allowed region of parameter space.
Thus we find $\kappa_Z$ is severely constrained close to its maximized value for $m_A \gg m_h $.

\section{Production cross sections of $W'$ and $Z'$}
\label{App:xsec}
The leading order production cross sections of $W'$ and $Z'$ are given as follows.
\beq
\sigma(pp \to W'^{\pm} X) &\simeq& \frac{\pi}{12 s }  \frac{m_W^2}{m_{W'}^2}  \frac{e^2}{s_W^2} \frac{1}{r^2} \left( 1 - \frac{v_3^2}{v^2} \right) \int^1_{m_{W'}^2 /s}  \frac{d x}{x} \non
& & \times \left( f_u (x,m_{W'}) f_{\bar{d}} \left(\frac{m_{W'}^2}{s x}, m_{W'} \right) + f_{\bar{d}} (x,m_{W'}) f_{u} \left(\frac{m_{W'}^2}{s x}, m_{W'}\right)  \right. \non
&& ~~\left. + f_d (x,m_{W'}) f_{\bar{u}} \left(\frac{m_{W'}^2}{s x}, m_{W'} \right)   + f_{\bar{u}} (x,m_{W'}) f_{d} \left(\frac{m_{W'}^2}{s x}, m_{W'}\right)  \right. \non
& & ~~+ (u \leftrightarrow c, d \leftrightarrow s) \Bigr), \label{prod:Wp} \\
\sigma(pp \to Z' X) &\simeq& \frac{\pi} {6 s} \frac{m_W^2}{m_{W'}^2}\frac{e^2}{s_W^2}  \frac{1}{r^2}
\left( 1 - \frac{v_3^2}{v^2}\right) \non
& & \times \left(  \left( \left( 1 - r^2 \frac{s_Z^2}{c_Z^2} \right) T^3_f + r^2 \frac{s_Z^2}{c_Z^2} Q_f \right)^2 + \left( r^2 \frac{s_Z^2}{c_Z^2} Q_f \right)^2  \right) \int^1_{m_{Z'}^2 /s}  \frac{d x}{x} \non
& & \times \left( f_u (x , m_{Z'}) f_{\bar{u}}\left( \frac{ m_{Z'}^2}{s x}, m_{Z'} \right) + f_{\bar{u}} (x , m_{Z'}) f_{u}\left( \frac{ m_{Z'}^2}{s x}, m_{Z'} \right) \right. \non
& & ~~+  (u \leftrightarrow d,~s,~c,~b) \Bigr), \label{prod:Zp}
\eeq
where $s$ is a square of the center of mass energy of the $pp$ collider and  $f_{q} (x ,~Q)$ is the parton distributions inside the $p$ at the factorization scale $Q$ for quark flavor $q$.
Note that \eq{prod:Wp} is a sum of production cross section of  $W'^{+}$ and $W'^{-}$.

\section{Renormalization group equations}
\label{App:RGE}

We derive the one-loop $\beta$ functions for this model \cite{Cheng:1973nv, Grimus:2004yh}, and obtain,
\beq
\beta_{g_0} &=& -3 g_0^3,\\
\beta_{g_1} &=& -7 g_1^3,\\
\beta_{g_2} &=& 7 g_2^3,\\
\beta_{g_s} &=& -7 g_s^3,\\
\beta_{\lambda_1} &=& 24 \lambda_1^2 + 2 \lambda_{12}^2 + 2 \lambda_{31}^2 -9 \lambda_1 (g_0^2 + g_1^2)  + \frac{9}{8} g_0^4 + \frac{9}{8} g_1^4 + \frac{9}{4} g_0^2 g_1^2,\\
\beta_{\lambda_2} &=& 24 \lambda_2^2 + 2 \lambda_{12}^2 + 2 \lambda_{23}^2 -3 \lambda_2 (3 g_1^2 + g_2^2)  + \frac{9}{8} g_1^4 + \frac{3}{8} g_2^4 + \frac{3}{4} g_1^2 g_2^2,\\
\beta_{\lambda_3} &=& 24 \lambda_3^2 + 2 \lambda_{23}^2 + 2 \lambda_{31}^2 -3 \lambda_3 (3 g_0^2 + g_2^2)  + \frac{9}{8} g_0^4 + \frac{3}{8} g_2^4 + \frac{3}{4} g_0^2 g_2^2 \non
& & +2 \lambda_3 ( 3 y_t^2 + 3 y_b^2 + y_{\tau}^2) -\frac{3}{2} y_t^4 -\frac{3}{2}y_b^4 - \frac{1}{2} y_{\tau}^4,  \label{RGElambda3}  \\
\beta_{\lambda_{12}} &=& 4 \lambda_{12}^2 + 12 \lambda_{12}( \lambda_1 + \lambda_2 ) + 4 \lambda_{23} \lambda_{31}  -\frac{3}{2} \lambda_{12} ( 3 g_0^2 + 6 g_1^2+ g_2^2 ) +\frac{9}{4} g_1^4, \\
\beta_{\lambda_{23}} &=& 4 \lambda_{23}^2 + 12 \lambda_{23}( \lambda_2 + \lambda_3 ) + 4 \lambda_{12} \lambda_{31}  -\frac{3}{2} \lambda_{23} ( 3 g_0^2 + 3 g_1^2+2 g_2^2 ) +\frac{3}{4} g_2^4 \non 
& &+ \lambda_{23} ( 3y_t^2 + 3 y_b^2 + y_{\tau}^2 ), \\
\beta_{\lambda_{31}} &=& 4 \lambda_{31}^2 + 12 \lambda_{31}( \lambda_1 + \lambda_3 ) + 4 \lambda_{12} \lambda_{23}  -\frac{3}{2} \lambda_{31} ( 6 g_0^2 + 3 g_1^2+ g_2^2 ) +\frac{9}{4} g_0^4 \non 
& &+ \lambda_{31} ( 3y_t^2 + 3 y_b^2 + y_{\tau}^2 ), \\
\beta_{y_t} &=& y_t \left( -\frac{9}{4} g_0^2 - \frac{17}{12} g_2^2 - 8 g_s^2 + \frac{9}{4} y_t^2 + \frac{3}{4} y_b^2 + \frac{1}{2} y_{\tau}^2 \right),\\
\beta_{y_b} &=& y_b \left( -\frac{9}{4} g_0^2 - \frac{5}{12} g_2^2 - 8 g_s^2 + \frac{3}{4} y_t^2 + \frac{9}{4} y_b^2 + \frac{1}{2} y_{\tau}^2 \right),\\
\beta_{y_{\tau}} &=& y_{\tau} \left( -\frac{9}{4} g_0^2 - \frac{15}{4} g_2^2 + \frac{3}{2} y_t^2 + \frac{3}{2} y_b^2 + \frac{5}{4} y_{\tau}^2 \right),
\eeq
where $\beta$ functions are defined in the following notation, 
\beq
\frac{d (\textrm{coupling})}{d \textrm{ln} \mu} = \frac{\beta_{(\textrm{coupling})}}{(4 \pi )^2}.
\eeq

\providecommand{\href}[2]{#2}
\begingroup\raggedright

\end{document}